\documentclass[prd,twocolumn,10pt,aps,longbibliography,nofootinbib]{revtex4-2}

\usepackage{amssymb}
\usepackage{amsmath}

\usepackage{graphicx}
\usepackage{dcolumn}
\usepackage{bm}
\usepackage{hyperref}

\usepackage{subfig, tabularx, afterpage,xifthen}
\usepackage{amsmath,amsfonts,amssymb,mathtools}
\usepackage{slashed}

\usepackage{xcolor}

\newcommand\scetone{SCET$_\mathrm{I}$} 
\newcommand\scettwo{SCET$_\mathrm{II}$} 

\newcommand\bra[1]{ \langle {#1} | }
\newcommand\ket[1]{ | {#1} \rangle}

\newcommand{\order}[1]{O({#1})}
\newcommand\Phat{P}
\newcommand\nbar{{\bar{n}}}
\newcommand\nb{{\bar{n}}}
\newcommand\n{n}

\newcommand\QLsqr{q_L^2}
\newcommand\pcorrec{(2)}

\newcommand{\overbar}[1]{\mkern 1.5mu\overline{\mkern-1.5mu#1\mkern-1.5mu}\mkern 1.5mu}
\newcommand\Deltabar{\overbar{\Delta}}

\newcommand\pquark{p_1}
\newcommand\panti{p_2}

\newcommand\pquarkslashed{\slashed{p}_1}
\newcommand\pantislashed{\slashed{p}_2}

\newcommand\figref[1]{{Fig.\ \ref{#1}}}
\newcommand\secref[1]{{Sec.\ \ref{#1}}}

\newcommand\initHad{N_{12}}
\newcommand\1{n}
\newcommand\2{\nb}

\newcommand\scale{\zeta}

\newcommand\alphabar{\overbar{\alpha}}

\newcommand\zbar{\overbar{z}_1}
\newcommand\zzbar{\overbar{z}_2}

\newcommand\zabar{\overbar{z}_a}
\newcommand\zbbar{\overbar{z}_b}

\newcommand\omegabar{\overbar{\omega}}

\newcommand\qqquad{\qquad\qquad}

\newcommand\myvec[1]{\boldsymbol{\mathbf{#1}}}

\newcommand\vecQ{\myvec{q}_T}
\newcommand\deltaT{\myvec{\delta}_T}
\newcommand\deltaP{\myvec{\delta}_\perp}

\newcommand\thisN{n}
\newcommand\thisNb{\nb}

\newcommand\wn{w_{\thisN}}
\newcommand\wnb{w_{\thisNb}}

\newcommand\upsn{\upsilon_{\n}}

\newcommand\nun{\nu_{\n}}
\newcommand\nunb{\nu_{\nb}}
\newcommand\nunnb{\nu_{\n,\nb}}

\newcommand\etan{\eta_{\thisN}}
\newcommand\etanb{\eta_{\thisNb}}

\newcommand\omegan{\omega_{1}}
\newcommand\omeganb{\omega_{2}}

\newcommand\omegabarn{\omegabar_{1}}
\newcommand\omegabarnb{\omegabar_{2}}

\newcommand\qcd{\mathrm{QCD}}

\newcommand\cusp{\mathrm{cusp}}
\newcommand\LL{\mathrm{LL}}

\newcommand\commonPrefac{{\sigma}_0}

\newcommand\etabar{\bar{\eta}}

\newcommand\chibar{\bar{\chi}}
\newcommand\that{\hat{t}}
\newcommand\thatt{\hat{t}_1}
\newcommand\thattt{\hat{t}_2}

\newcommand\Cff{C_{f\kern-2.4pt f}}

\newcommand\ggamma{\gamma\kern-8.2pt \gamma}
\newcommand\ff{f\kern-8.2ptf}

\DeclareSymbolFont{matha}{OML}{txmi}{m}{it}
\DeclareMathSymbol{\varv}{\mathord}{matha}{118}
\newcommand\vbar{\bar{v}}

\newcommand\jqcd{J_{\rm QCD}}
\newcommand\jscet{J_{\rm SCET}}
\def\OMIT#1{}

\def\xn{x_n}
\def\xnb{x_\nb}
\def\eqn#1{Eq.\ (\ref{#1})}
\def\lqcd{\Lambda_{\rm QCD}}

\def\pantisl{{\slashed{p}_2}}
\def\pquarksl{{\slashed{p}_1}}
\def\vvars{\{v\}}
\def\uvars{\{u\}}
\def\tvars{\{t\}}

\def\qtsq{q_T^2}

\def\rrgtimes{\ast}
\def\mmat#1{\mathcal{M}_{#1}}
\def\mmatn#1#2{\mmat{#1}^{#2}}
\def\amat#1{A_{#1}}
\def\amatn#1#2{\amat{#1}^{#2}}
\def\bmat#1{B_{#1}}
\def\bmatn#1#2{\bmat{#1}^{#2}}
\def\cmat#1{C_{#1}}
\def\cmatn#1#2{\cmat{#1}^{#2}}
\def\kmat#1#2#3#4{K^{(#1,#2)}_{(#3,#4)}}
\def\NLP{{\mathrm{NLP}}}

\def\Bigvert{\vphantom{\left(\left(\frac{zzbar^2}{zbar}\right)\right)_+}}
\def\bigvert{\vphantom{\left(\frac{zzbar}{zbar}\right)}}

\begin{document}

\title{Factorization of power corrections in the Drell-Yan process in EFT}

\author{Matthew Inglis-Whalen}
\email{minglis@physics.utoronto.ca}
\author{Michael Luke}
\email{luke@physics.utoronto.ca}
\author{Jyotirmoy Roy}
\email{jro1@physics.utoronto.ca}
\author{Aris Spourdalakis}
\email{aspourda@physics.utoronto.ca}

\affiliation{Department of Physics, University of Toronto, Toronto, Ontario, Canada M5S 1A7}

\begin{abstract}
We examine the quark-induced Drell-Yan process at next-to-leading power (NLP) in Soft-Collinear Effective Theory. Using an approach with no explicit soft or collinear modes, we discuss the factorization of the differential cross section in the small-$q_T$ hierarchy with $q^2\gg q_T^2\gg\Lambda_{\mathrm{QCD}}^2$. We show that the cross section may be written in terms of matrix elements of power-suppressed operators $T_{(i,j)}$, which contribute to $O(q_T^2/q^2)$ coefficients of the usual parton distribution functions.  We derive a factorization for this observable at NLP which allows the large logarithms in each of the relevant factors to be resummed. We discuss the cancellation of rapidity divergences and the overlap subtractions required to eliminate double counting at next-to-leading power.
\end{abstract}

\maketitle

%
\section{Introduction}
%

The Drell-Yan (DY) process $N_1(P_1) N_2(P_2) \rightarrow \gamma^*(q) + X \rightarrow (\ell \bar{\ell}) + X$ has been extensively studied in perturbative QCD \cite{Collins:1989gx,ellis_stirling_webber_1996,collins_2011}. In the limiting case that the transverse momentum $q_T$ of the lepton pair is parametrically larger than $\lqcd$ and smaller than its invariant mass $\sqrt{q^2}$, the cross section may be written as \cite{COLLINS1985}
\begin{equation}\begin{aligned}
\label{eq:Cff_factorization}
    \frac{1}{\sigma_0}\frac{d\sigma}{dq^2 dy d\qtsq} &= \sum_{a,b}\int \frac{dz_1}{z_1} \frac{dz_2}{z_2}  \Cff^{ab}(z_1,z_2,q^2,\qtsq)\\
    &\times 
    f_{a/N_1}\!\left( \frac{\xi_1}{z_1} \right) f_{b/N_2}\!\left( \frac{\xi_2}{z_2} \right)  +O\!\left( \frac{\lqcd^2}{q_T^2}\right)\,,
\end{aligned}\end{equation}
where $q^\mu$ is the four-momentum of the lepton pair, $\xi_{1}\equiv q^-/P_1^-$,  $\xi_{2} \equiv q^+ / P_2^+$, $y=\log(q^-/q^+)/2$, and $P_1^-$ and $P_2^+$ are the large light-cone components of the incoming hadron momenta. The sum is over parton types $a,b$, and the $f_i$ are the usual parton distribution functions (PDFs). In this paper we only study the quark-induced process for a single flavor of quark, so we define $\Cff\equiv\Cff^{q\bar{q}}$. 

The coefficient function $\Cff$ may be expanded in powers of $\qtsq/q^2$, 
\begin{equation}\begin{aligned}
\Cff(z_1,z_2,q^2,\qtsq) = &
\Cff^{(0)}(z_1,z_2,q^2,\qtsq) 
    \\& +\frac{1}{q^2}\Cff^{\pcorrec}(z_1,z_2,q^2,\qtsq) + \cdots\,.
\end{aligned}\end{equation}
where each subsequent term is suppressed by increasing powers of $\qtsq/q^2$. 
Since they depend on two parametrically different scales $q^2$ and $\qtsq$, the fixed-order perturbative expansions for each $\Cff^{(n)}$ contain large logarithms of $\qtsq/q^2$ which can spoil the behavior of perturbation theory and need to be resummed. The resummation of the leading power (LP) term $\Cff^{(0)}$ has been extensively studied in the literature, using both  perturbative QCD techniques \cite{Dokshitzer:1978hw,Dokshitzer:1978yd,Parisi:1979se,Curci:1979bg,Ellis:1997ii,Frixione:1998dw,Davies:1984hs,Davies:1984sp,COLLINS1985,deFlorian:2000pr,Bozzi:2010xn,Catani:2015vma} and effective field theory methods \cite{Becher:2006mr,Becher:2010tm,GarciaEchevarria:2011rb,Ebert:2016gcn}. Factorization theorems allow $\Cff^{(0)}$ to be written as a product of separate terms depending on distinct scales, each of which may be resummed to arbitrary order using a variety of renormalization group (RG) or related techniques.
The most recent analyses achieve a resummation up to N$^3$LL+NNLO order \cite{Bizon:2017rah,Bertone:2019nxa,Bacchetta:2019sam,Ebert:2020dfc,Becher:2020ugp,Camarda:2021ict,Re:2021con}. 
However, much less is known about the factorization and resummation properties of the first power correction $\Cff^{\pcorrec}$. $\Cff$\, has been computed in QCD at fixed order in perturbation theory up to N$^2$LO \cite{Ellis:1981hk,Gonsalves:1989ar},
but an all-orders RG resummation at next-to-leading power (NLP) has not been performed.

Soft-Collinear Effective Theory (SCET) \cite{Bauer:2000ew,Bauer:2000yr,Bauer:2001ct,Bauer:2001yt,Bauer:2002nz,Beneke:2002ph,Beneke:2002ni} is an effective field theory (EFT) that provides a systematic framework in which to study power corrections in hard QCD processes. There has been much recent work studying power corrections to various processes, with applications including beam thrust \cite{Moult:2018jjd}, Drell-Yan production near threshold \cite{Beneke:2018gvs}, threshold Higgs production from gluon fusion \cite{Beneke:2019mua}, Higgs production and decay \cite{Bhattacharya:2018hmss}, the energy-energy correlator in $\mathcal{N}=4$ Supersymmetric Yang-Mills \cite{Moult:2019vou} and Higgs to diphoton decays \cite{Liu:2019oav,Liu:2020tzd,Liu:2020wbn}. Power corrections have also been studied using non-EFT QCD techniques \cite{Kramer:1996iq,Penin:2015hel,Bahjat-Abbas:2019fqa, Cieri:2019hopc, vanbeekveld:2021hhv, Oleari:2020wvt, Boughezal:2020vjp}.

The DY process at small $\qtsq$ is typically referred to as a \scettwo\ process, characterized by collinear and soft modes in the EFT, and exhibiting rapidity logarithms in matrix elements. Rapidity logarithms are large logarithms in matrix elements which arise in SCET due to divergences in individual diagrams at large values of the rapidity of one of the particles. These divergences cancel between graphs with different modes, but the final result contains large finite logarithms of the hard scale of the scattering which cannot be resummed using usual RG techniques. Rapidity divergences require an additional regulator beyond dimensional regularization, and various techniques have been successfully employed to handle the rapidity resummation, including off-the-light-cone techniques \cite{COLLINS1985}, the rapidity renormalization group \cite{Chiu:2012ir},  the collinear anomaly framework \cite{Becher:2010tm}, the exponential regulator  \cite{Li:2016axz} and the recently proposed pure rapidity regulator \cite{Ebert:2018gsn,Moult:2019vou}. The latter regulator has recently been used \cite{Ebert:2018gsn} to calculate the small-$q_T$ DY cross section by expanding the QCD graphs in the soft and collinear limits, where it correctly treats the power-law rapidity divergences arising at NLP. The connection between rapidity renormalization in \scettwo\ and the usual renormalization group equation (RGE) in \scetone\ was discussed in \cite{Bauer:2020npd}.

In this paper we study power corrections to DY production using the version of SCET developed in \cite{Goerke:2017ioi,Goerke:2017lei,Inglis:2020rpi}. In this approach the degrees of freedom in the EFT are not analyzed using the method of regions \cite{Beneke:1997zp} in which they are explicitly separated into soft, collinear, ultrasoft, and possibly additional modes. Instead, states are separated into distinct sectors, where the relative invariant mass of particles within each sector is less than the renormalization scale $\mu$ of the EFT, but the relative invariant mass of different sectors is larger than the renormalization scale.  
As with the mode expansion, particles of the same type but in different sectors are described by different fields; however, interactions within a sector are described by QCD, while interactions between sectors are mediated via the external current, which is expanded in inverse powers of the hard matching scale. Factorization of different modes (soft-collinear, ultrasoft-collinear, and others) does not occur explicitly in the Lagrangian since different modes in a given sector are described by the same fields, but instead arises through the usual EFT process of integrating out degrees of freedom and matching onto a new EFT at appropriate threshold scales.

This reduces the number of separate fields in the Lagrangian and therefore simplifies the formalism, both conceptually and practically. One immediate feature is that subleading terms in the effective Lagrangian coupling different modes and violating manifest factorization are not present in this approach. In addition, rather than deriving a factorization theorem in terms of jet and soft functions which are individually well-defined and renormalized at the appropriate scale, the rate is simply expressed in terms of bilocal products of operators in the EFT which may be run both in the renormalization scale $\mu$ as well as in the rapidity scale $\nu$. Similar to the situation at LP discussed in \cite{Inglis:2020rpi}, we show here that the DY cross section naturally factorizes into hard matching coefficients, rapidity evolution factors, soft matching coefficients and parton distribution functions, and give expressions for the first three quantities up to NLP at one loop. The complete resummation of rapidity logarithms is left for a future work.

Consistency of this theory requires that double counting of degrees of freedom between the two sectors is consistently subtracted, similar to the usual zero-bin subtraction \cite{Manohar:2006nz} in SCET. This procedure of overlap subtraction is necessary for the theory to be well-defined and is implicit in all matrix elements. Furthermore, as discussed in detail in \cite{Inglis:2020rpi}, the scheme dependence of this subtraction allows rapidity logarithms to be summed using techniques similar to \cite{Chiu:2012ir,Ebert:2018gsn} without having manifest factorization of soft and collinear modes in the effective Lagrangian. 
At subleading powers this subtraction is nontrivial, requiring contributions from multiple operators as well as subleading corrections to the leading power subtraction. While these subtractions vanish using an appropriately chosen regulator, the interplay of these subtraction terms explains patterns of rapidity divergence cancellation between different operators, similar to the nontrivial cancellations of rapidity and endpoint divergences at NLP seen in other approaches \cite{Liu:2019oav,Liu:2020tzd,Liu:2020wbn}. 

QCD proofs of factorization in hard scattering processes require that the effects of the exchange of soft gluons in the Glauber regime relevant to small angle parton scattering cancel in the relevant observable \cite{Collins:1989gx,ellis_stirling_webber_1996,collins_2011,Collins:2004nx,Diehl:2015bca}. Glauber modes have been the subject of much recent interest in SCET \cite{Rothstein:2016bsq}, and a consistent treatment of gluons in the Glauber regime has been shown to be necessary to ensure that operator statements in SCET are independent of the external states \cite{Bauer:2010cc}. Investigation of these effects in the formalism presented here are beyond the scope of this paper, but we will assume that gluons in the Glauber regime do not introduce factorization-violating effects in the context of this calculation.

In \secref{subsec:fact} we sketch the ingredients of the calculation and the approach to factorization in this formalism. We present the one-loop calculations of the various pieces in \secref{sec:fierzed_basis}, and compare our fixed-order results with the unsummed QCD result. In \secref{sec:overlap_subtractions} we consider the cross section with no rapidity regulator to demonstrate the cancellation of rapidity divergences between different operators and their respective overlap subtractions across different regions of phase space. We present our conclusions in \secref{sec:conc}. A few details of plus distributions used here are given in the appendixes, as well as a comparison to a recent one-loop analysis \cite{Ebert:2018gsn} of power corrections to the DY process.

%
%
\subsection{Factorization}\label{subsec:fact}
%
%

In the SCET formalism introduced in \cite{Goerke:2017ioi,Goerke:2017lei,Inglis:2020rpi} there are no explicit soft, collinear or ultrasoft modes, so factorization does not arise explicitly from a Lagrangian mode expansion, but instead by integrating out  ultraviolet degrees of freedom at the relevant matching scales. In this section we briefly review the approach of \cite{Inglis:2020rpi} to DY scattering and introduce its extension to subleading power. Precise definitions of quantities appearing in this section will be given in \secref{sec:fierzed_basis}.

The cross section for the electromagnetic Drell-Yan production process, $N_1 N_2 \rightarrow \gamma^* +X \rightarrow (\ell \bar{\ell}) + X$ is given in QCD by 
\begin{equation} \begin{aligned} \label{eq:cross_section_definition}
d \sigma = \frac{4\pi \alpha^2}{3 q^2 s}  \frac{d^4 q}{(2\pi)^4}  
\int & d^d x \, e^{-iq\cdot x} \, (-g_{\mu\nu}) \\
&\times\bra{\initHad}  \jqcd^{\mu\dagger}(x) \jqcd^{\nu}(0) \ket{\initHad }\,,
\end{aligned} \end{equation}
where $q^2=(p_\ell+p_{\bar{\ell}})^2$ is the invariant mass of the lepton pair, $s=(P_1+P_2)^2$ is the invariant mass of the incoming hadrons, the initial hadronic state is $\ket{\initHad}=\ket{N_1(P_1) N_2(P_2) }$, and the vector QCD current $\jqcd^\mu$ is
\begin{equation} \begin{aligned}  \label{eq:current_defn}
\jqcd^\mu(x) =  \bar{\psi}(x) \gamma^\mu \psi(x) \,
\end{aligned} \end{equation}
for a single flavor of light quark.
The extension to electroweak currents is straightforward \cite{Becher:2010tm,COLLINS1985}.

For $\qtsq\ll q^2$, perturbative corrections to the cross section in \eqn{eq:cross_section_definition} contain powers of logarithms of $\qtsq/q^2$, which can spoil the apparent convergence of perturbation theory.  SCET provides a systematic approach to resumming these terms. At the renormalization scale $\mu=\mu_H\sim\sqrt{q^2}\gg q_T$, hard interactions are integrated out of the theory and QCD is matched onto SCET.
In the formalism used here, SCET consists of two decoupled QCD sectors, denoted by the lightlike vectors $n^\mu$ and $\nb^\mu$, with total momenta $p_n^\mu$ and $p_\nb^\mu$; the sectors are distinguished by the power counting
\begin{equation}
    p_n^2, p_\nb^2\ll q^2,\ p_n\cdot p_\nb\sim q^2 \,.
\end{equation}
Interactions between the sectors are mediated by the external current $J^\mu_{\rm SCET}$, which is written as a sum of operators of increasing dimension\footnote{Subleading operators are also labeled by continuous indices, so the discrete sums over operators also include integrals, which we neglect for simplicity in this section.}
\begin{equation} 
\begin{aligned} \label{eq:SSLO_expansion}
\jscet^\mu(x) =   \sum_i \frac{1}{q_L^{[i]}} C_2^{(i)}\left(\mu\right) O_2^{(i)\mu}(x,\mu)\,, 
\end{aligned} \end{equation}
where an operator $O_2^{(i)}$ has mass dimension $[i]$ in excess of the leading-power operator $O_{2}^{(0)}$. We have defined
$q_L^2\equiv q^+ q^-$, and for brevity we will not explicitly include the $\mu$ dependence of operators in subsequent equations unless required for clarity. It is convenient to expand in inverse powers of $q_L^2$ rather than $q^2=q_L^2-\qtsq$ so that the hard scale of the EFT is independent of the infrared (IR) scale $\qtsq$. This expansion has been performed up to $O(1/q_L^2)$ \cite{Manohar:2003vb,Freedman:2014uta,Goerke:2017ioi,Lee:2004bsg,Hill:2004nlp}, the details of which are summarized in \secref{sec:hard_scale_matching}.  
The SCET expansion for the differential cross section is then given in SCET by
\begin{equation} \begin{aligned}  \label{eq:factorization_derivation_part2}
\frac{d\sigma}{dq^2 dy d\qtsq} 
=&\,\frac{4\pi \alpha^2}{3 q^2 s}  (-g_{\mu\nu})\int\! \frac{d\Omega_T}{2} \!\int\!\! \frac{d^d x }{2(2\pi)^d} \sum_{ij}  
\frac{H_{\left(i,j\right)}(\mu) }{ q_L^{[i]+[j]}}  \\& \times e^{-iq\cdot x} \bra{\initHad}  O_{2}^{(i)\mu\dagger}(x) O_2^{(j)\nu}(0) \ket{\initHad }\,,
\end{aligned} \end{equation}
where $H_{(i,j)}(\mu) \equiv C_{2}^{(i)\dagger}(\mu) C_2^{(j)}(\mu)$
and the final angular integral $d\Omega_T$ corresponds to the angular integral in the transverse momentum $\vecQ$.
Since we have not subdivided the degrees of freedom of SCET into separate soft and collinear modes, there is no expansion of the SCET Lagrangian beyond that in \eqn{eq:SSLO_expansion}; in particular, there are no power corrections arising from soft-collinear mixing terms in the Lagrangian \cite{Pirjol:2002km,Chay:2002vy,Manohar:2002RPI,Bauer:2003mga,Beneke:2002ni,Beneke:2002ph}. This simplifies the analysis of power corrections considerably. 

While matrix elements of the operator products in \eqn{eq:factorization_derivation_part2} may be directly evaluated between partons in perturbation theory, it is convenient to perform a Fierz rearrangement to write the operator product as a convolution of 
transverse momentum dependent distribution operators (whose hadronic matrix elements are generally referred to as TMDPDFs), one in the $n$ sector and one in the $\nbar$ sector.
This is a standard procedure at leading power \cite{ellis_stirling_webber_1996,Becher:2010tm}; at subleading powers a similar procedure may be used to express the basis of operator products as convolutions of power-suppressed 
distribution operators,
\begin{equation} \begin{aligned} \label{eq:FierzingTransformation}
\int \! \frac{d^d x }{2(2\pi)^d}& \, (-g_{\mu\nu})e^{-iq\cdot x}   O_{2}^{(i)\mu\dagger}(x) O_{2}^{(j)\nu}(0)  \\&= \sum_{k,\ell}\frac{1}{N_c}\kmat{i}{j}{k}{\ell} T_{(k,\ell)}(q^-,q^+,\vecQ)
\\&\quad+\textnormal{spin\,dependent} \,,
\end{aligned} \end{equation}
where each $T_{(i,j)}$ relevant to this calculation will be defined explicitly in \secref{sec:fierzed_basis}. Rewriting the operator products in terms of the operators $T_{(i,j)}$ is simply a change of operator basis, and not a matching condition or expansion in SCET, and so introduces no new perturbative corrections. Typically in SCET this Fierz rearrangement is performed to write the operator product in a form that manifestly factorizes into jet and soft functions; since this factorization is not needed here this change of basis is not strictly necessary, but it is included here for easier comparison with other approaches.

At $O(\alpha_s)$ matrix elements of the $T_{(i,j)}$'s at small $q_T$ are insensitive to the cutoff scale $q_L$ and so
running the scattering operators $O_{2,(i)}$ from $\mu_H\sim q_L$ to $\mu_S\sim q_T$ sums the usual renormalization group logarithms of $q_L^2/\qtsq$ in the rate. If $q_T\sim \lqcd$, matrix elements of each $T_{(i,j)}$ are nonperturbative quantities which would have to be either modeled or extracted from experiment. In the scaling of interest here, $q_T\gg \lqcd$, each $T_{(i,j)}$ may be further expanded in powers $\lqcd^2/\qtsq$, allowing the operator product in \eqn{eq:factorization_derivation_part2} to be matched onto the usual light-cone distribution operators
whose hadronic matrix elements are the parton distribution functions. This expansion corresponds to matching SCET onto a soft theory of completely decoupled sectors of QCD at the scale $\mu_S\sim q_T$, and at leading twist takes the form
\begin{equation} \begin{aligned} \label{eq:soft_matching}
 T_{(k,\ell)} \! \left(q^-, q^+, \vecQ,\mu_S\right)
\to &
\int \! \frac{dz_1}{z_1} \frac{dz_2}{z_2}\, C_{S,(k,\ell)}\left(z_1,z_2,\vecQ,\mu_S\right) \\
& \times O_q \left( \frac{q^-}{z_1} , \mu_S \right)  O_{\bar{q}}\left( \frac{q^+}{z_2} , \mu_S\right)\,,
\end{aligned} \end{equation}
where the various $C_{S,(k,\ell)}$ are matching coefficients and the hadronic matrix elements of the light-cone quark and antiquark distribution operators $O_{q,\bar q}$ are the usual spin-averaged parton distribution functions
\begin{equation} \begin{aligned} 
f_{q/N_1}(\zeta_1)&=  \bra{N_1(P_1)}O_q(\zeta_1 P_1^-) \ket{N_1(P_1)}  \\
f_{\bar{q}/N_2}(\zeta_2)&=  \bra{N_2(P_2)}O_{\bar{q}}(\zeta_2 P_2^+) \ket{N_2(P_2)} \,,
\end{aligned} \end{equation}
with $P_i^-\equiv P_i \cdot \nb$ and $P_i^+\equiv P_i\cdot n$.
Combining these matching steps gives an expression for the DY cross section for a single quark flavor of the form
\begin{equation}\begin{aligned}\label{eq:factorization_x_sec}
   \frac{d\sigma}{dq^2 dy d\qtsq} =& \sigma_0\int \frac{dz_1}{z_1} \frac{dz_2}{z_2} \Cff(z_1,z_2,q_L^2,\qtsq)\\
   &\times 
    f_{q/N_1}\left( \frac{\xi_1}{z_1} \right) f_{\bar q/N_2}\left( \frac{\xi_2}{z_2} \right) + \dots
\end{aligned}\end{equation}
where $\sigma_0 = 4\pi \alpha^2 /(3N_c\, q^2 s)$, $\xi_1 = q^-/P_1^-$, $\xi_2=q^+/P_2^+$, and $\Cff$ has the partially factorized form
\begin{equation}\begin{aligned} \label{eq:Cff_defn}
\Cff(z_1, z_2,q_L^2,\qtsq)=& \int\! \frac{d\Omega_T}{2} \sum_{ijk\ell}  \kmat ijk\ell    \left( \frac{1}{q_L} \right)^{[i]+[j]} 
\\&\times H_{(i,j)}\left(\mu_S \right) 
C_{S,(k,\ell)}\left(z_1,z_2,\vecQ,\mu_S \right)  .
\end{aligned} \end{equation}
However, in this form the matching coefficients $C_S$ still contain large logarithms of $\qtsq/q_L^2$ which are not resummed by the usual renormalization group evolution.  These rapidity logarithms arise because the graphs renormalizing matrix elements of $T_{(i,j)}$ in SCET are separately divergent in each sector, even in $d$ dimensions, and the divergences only cancel in the sum. These graphs therefore require the introduction of an additional regulator beyond dimensional regularization, and the rapidity divergences are reflected in logarithms of the (scheme-dependent) rapidity scale. 
While a number of regulators have been used at leading power \cite{Chiu:2009yx,Smirnov:1997gx,Li:2016axz,Chiu:2012ir,Ebert:2018gsn}, the ``pure rapidity regulator" introduced in \cite{Ebert:2018gsn} is particularly convenient for studying power corrections, as it properly regulates the power divergences in phase space integrals arising at NLP.

In this paper we use a version of the pure rapidity regulator appropriate for our formalism which introduces separate scheme dependence for the $n$ and $\nb$ sectors, denoted by the
parameters $\nun$ and $\nunb$.  Rapidity logarithms are summed by running the operators $T_{(i,j)}$ from $\nunnb^H=q_L$ to $\nunnb^S=\mu\sim q_T$. Under rapidity renormalization the operators $T_{(i,j)}$ can mix, leading generically to rapidity renormalization group running of the form
\begin{equation} \begin{aligned}  \label{eq:scheme_matching}
&T_{(i,j)}\left(q^-,q^+,\vecQ,\mu_S,\nunnb^H \right) = 
\sum_{k,\ell}\int 
\frac{d\omegan}{\omegan} \frac{d\omeganb}{\omeganb} d^2\myvec{p}_T 
\\&\qqquad\times
V_{(i,j),(k,\ell)} 
    \left( \omegan,\omeganb, \myvec{p}_T,\mu_S, \nunnb^H, \nunnb^S \right) 
\\&\qqquad\times  
T_{(k,\ell)}
    \left( \frac{q^-}{\omegan},\frac{q^+}{\omeganb}, \vecQ-\myvec{p}_T,\mu_S, \nunnb^S \right) \,, 
\end{aligned} \end{equation}
where by $\nu_{n,\nb}$ we denote depends on both $\nu_n$ and $\nu_\nb$ separately, and the large logarithms of $\nu_{n,\nb}^H/\nu_{n,\nb}^S$ have been resummed in the rapidity evolution factors $V_{(i,j),(k,\ell)}$.
Combining all these steps gives the DY cross section in \eqn{eq:factorization_x_sec}, where $\Cff$ now has the fully factorized form
%
\begin{equation}\begin{aligned} \label{eq:final_Cff_factorization}
\Cff&(z_1, z_2,q_L^2,\qtsq)= \\ & \int \frac{d\Omega_T}{2} \int  \frac{d\omegan}{\omegan} \frac{d\omeganb}{\omeganb}\sum_{ijkk^\prime\ell\ell^\prime}     
\frac{H_{(i,j)}\left(\mu_S\right) \kmat ijk\ell}{q_L^{[i]+[j]}} \\
& \times  
\int d^2\myvec{p}_T \
V_{(k,\ell),(k^\prime,\ell^\prime)}
\left( 
    \omegan,\omeganb,\myvec{p}_T,\mu_S,\nunnb^H,\nunnb^S
\right)  \\
& \quad\times C_{S,(k^\prime,\ell^\prime)}\left( \frac{z_1}{\omegan},\frac{z_2}{\omeganb},\vecQ-\myvec{p}_T,\mu_S,\nunnb^S\right) \,.
\end{aligned} \end{equation}

%
In this paper the fixed-order $\order{\alpha_s}$ contributions to each of the factors in \eqn{eq:final_Cff_factorization} which are required to determine the fixed-order cross section at NLP are calculated. The $\order{\alpha_s}$ anomalous dimensions of the relevant hard matching coefficients may be found in the literature \cite{Freedman:2014uta,Goerke:2017lei}, and here we also calculate the $\order{\alpha_s}$ off-diagonal entries for the rapidity evolution kernels $\gamma_{(k,\ell),(0,0)}$ which mix the various subleading operators $T_{(k,\ell)}$ into the leading operator $T_{(0,0)}$ with an NLP coefficient. The calculation of the one-loop entries which mix the subleading operators among themselves is left for future work. Additionally, in most phenomenological applications, $q_T$ resummation is performed for the Fourier conjugate of $q_T$ ($b$ space); here we will work in $q_T$ space, where the SCET operators we are using are defined. Fourier transforming our results to $b$ space may be useful for future applications.
%
%
\section{NLP operator products in SCET} \label{sec:fierzed_basis}
%
%
In this section the $\order{\alpha_s}$ ingredients that contribute to $\Cff$ at next to leading power in SCET are calculated. We begin by summarizing the hard-scale matching of the QCD current onto SCET scattering operators, and then proceed by Fierz rearranging products of these scattering operators into a smaller basis of operators. The matrix elements of these operators are calculated using the pure rapidity regulator, and the final result is compared to the corresponding fixed order result from QCD. 

\subsection{Hard-scale matching}\label{sec:hard_scale_matching}

The invariant mass of the lepton pair is $q^2=q^+q^--\qtsq\equiv q_L^2-\qtsq$, where $q_L^2\gg \qtsq \gg\lqcd^2$ and $q^+\equiv q\cdot n$ and $q^-=q\cdot \nb$ are the large light-cone components of the external current defined in terms of the lightlike vectors $n^\mu=(1,0,0,1)$ and $\nb^\mu=(1,0,0,-1)$. This defines the relevant scales for this process.  
The incoming state consists of two hadrons; the invariant mass of partons in the same hadron is of order $\lqcd$, while the invariant mass of partons in different hadrons is of order $q_L$. Therefore, at a hard scale $\mu_H\sim q_L$ partons in different hadrons are above one another's cutoff, and QCD is matched onto an EFT in which direct interactions between the sectors have been integrated out. In the SCET formalism used in this paper, SCET consists of decoupled copies of QCD for each sector 
which only mutually interact via the external electromagnetic current \eqn{eq:SSLO_expansion}. Only quark and antiquark PDFs are considered in this paper. Gluon PDFs may be included in the same formalism, and the relevant hard-scattering operators are listed in Appendix \ref{sec:appendix_matrix_elements}, but the calculation for incoming gluons is beyond the scope of this work. We work in a reference frame where the incoming hard quark is in the $n$ sector and the antiquark is in the $\nb$ sector.

The matching of the external vector current from QCD to SCET at subleading power has been considered in a number of papers \cite{Goerke:2017ioi,Goerke:2017lei,Beneke:2019mua,Moult:2019mog,Moult:2019vou, Beneke:2018gvs} and is obtained by expanding QCD amplitudes in powers of $p_i\cdot\n/q\cdot\n$ for particles in the $n$ sector, and $p_i \cdot \nb/q\cdot \nb$ for particles in the $\nb$ sector. In addition to the analogs of operators considered in \cite{Goerke:2017lei} for two incoming partons, there are also operators suppressed by single powers of the net transverse  momentum $p_{n_i,T}$ in either sector (which were eliminated  by a choice of reference frame in  \cite{Goerke:2017lei}) as well as corrections to the multipole expansion of the energy-momentum conserving delta functions. 

The SCET current has the expansion \eqn{eq:SSLO_expansion}.  The corresponding scattering operators are constructed from the field building blocks  \cite{Goerke:2017lei,Kolodrubetz:2016slbb}
\begin{equation} \begin{aligned} \label{eq:building_blocks}
\bar{\chi}_\nb(x) &= \bar{\psi}_\nb(x) \overline{W}_\nb(x) P_n \\
\chi_n(x) &= \overline{W}_n^\dagger(x) P_n \psi_n(x) \\
\mathcal{B}_\nb^{\mu_1 \cdots \mu_N}(x)&= \overline{W}_\nb^\dagger(x) iD_\nb^{\mu_1}(x) \cdots iD_\nb^{\mu_N}(x) \overline{W}_\nb(x) \\
\mathcal{B}_n^{\dagger\mu_1 \cdots \mu_N}(x)&= (-1)^N\, \overline{W}_n^\dagger(x) i\overleftarrow{D}_n^{\mu_1}(x) \cdots i\overleftarrow{D}_n^{\mu_N}(x) \overline{W}_n(x)
\end{aligned} \end{equation}
where we note that $(\mathcal{B}^{\mu_1\cdots\mu_N})^\dagger = \mathcal{B}^{\dagger \mu_N\cdots\mu_1}$. The incoming Wilson lines $\overline{W}$ are defined as 
\begin{equation} \begin{aligned} \label{eq:Wline_defns}
\overline{W}_n^\dagger(x) &= \overline{\mathcal{P}} \exp \left( -ig \int_{-\infty}^0 \!\!\! ds\,\nb\cdot A_n(x+\nbar s) e^{s0^+ } \right) \\
\overline{W}_\nb(x) &= \mathcal{P} \exp \left( \ \ ig \int_{-\infty}^{0} \!\!\! ds\,n\cdot A_\nb(x+n s) e^{s0^+ } \right) \,.
\end{aligned} \end{equation}
We use the conventions 
\begin{equation} \begin{aligned} \label{eq:covariant_defns}
iD_\nb^{\mu}(x) = i\partial^\mu + gA_\nb^\mu(x) \ , \quad i\overleftarrow{D}_n^{\mu}(x) = i\overleftarrow{\partial}^\mu - gA_n^\mu(x) 
\end{aligned} \end{equation}
and it is convenient to define the four-vectors introduced in \cite{Goerke:2017ioi}
\begin{equation} \begin{aligned} \label{eq:etaVectors}
\eta^\mu = \sqrt{ \frac{\nbar \cdot q}{n \cdot q} } n^\mu, \quad \etabar^\mu = \sqrt{ \frac{n \cdot q}{\nbar \cdot q} } \nbar^\mu \, 
\end{aligned} \end{equation}
which are invariant under the boost reparametrization $n^\mu\to e^y n^\mu$, $\nb^\mu\to e^{-y}\nb^\mu$.

At leading power there is a single scattering operator,
\begin{equation} \label{O20}
O_{2}^{(0)\mu}(x)=[\bar{\chi}_\nb(x_\nbar)] \gamma^\mu [\chi_{n}(x_n)]\,, 
\end{equation}
where 
\begin{equation}\label{eq:xndef} 
\begin{aligned} x_n^\mu &\equiv x^+ \frac{\nb^\mu}{2} + x_\perp^\mu, \ \mbox{and} \\ x_\nbar^\mu &\equiv x^- \frac{n^\mu}{2} + x_\perp^\mu.
\end{aligned}\end{equation}  
Note that the fields in the operator are multipole expanded; this is necessary for the energy-momentum conserving delta functions to preserve the correct power counting. For example, if $p_n^\mu$ and $p_\nb^\mu$ are momenta in the $n$ and  $\nb$ sectors respectively, we have the expansion
\begin{equation}\begin{aligned}
\delta(p_n^- +p_\nb^- -q^-)=&\,\delta\left(p_n^- -q^-\right)\!+p_\nb^-\, \delta^\prime\left(p_n^- -q^-\right)\!+\dots
\end{aligned}
\end{equation}
and similarly for the $n$ components.
Performing this expansion up to $O(1/q_L^2)$ gives
\begin{equation} \begin{aligned} \label{eq:multipole_expansion}
[\bar{\chi}_\nb(x)] \gamma^\mu [\chi_{n}(x)] =& O_{2}^{(0)\mu}(x) \\& + \frac{1}{\QLsqr}\left(O_{2}^{(2\delta^+)\mu}(x)  +  O_{2}^{(2\delta^-)\mu}(x)\right)\,,  
\end{aligned} \end{equation}
where 
\begin{equation} \begin{aligned} \label{eq:explicit_Odelta_Defns}
O_{2}^{(2\delta^+)\mu}(x) &=\frac{1}{2} q^- q^+ x^- [\chibar_\nb(x_\nb)] \gamma^\mu [n\cdot \partial \chi_{n}(x_n)] \\ 
O_{2}^{(2\delta^-)\mu}(x) &= \frac{1}{2} q^- q^+ x^+ [\nb\cdot \partial \bar{\chi}_\nb(x_\nb)] \gamma^\mu [ \chi_{n}(x_n)] \,.
 \end{aligned} \end{equation}
Power counting the multipole-expanded operators is not immediately obvious. In $O_{2}^{(2\delta^+)}$, for example, $q^+ x^-$ is of order 1 since $x^-\sim \partial/\partial q^+$, whereas $q^- p_n^+ \sim O(p_n^- p_n^+)\sim O(p_{n\perp}^2)$; thus, matrix elements of the operators in \eqn{eq:explicit_Odelta_Defns} are $O(1/\QLsqr)$ relative to leading power.  
Since we are working up to $1/\QLsqr$ suppression, the contributions from higher multipole expansions in the fields are only included for the leading power operator $O_2^{(0)}$.

At $O(1/q_L)$, there are two operators suppressed by a single perpendicular derivative, 
\begin{equation}\label{eq:perp_defs}
\begin{aligned}
O_2^{(1_{\perp\1})\mu}(x)&= [ \bar{\chi}_\nb(\xnb)]  \gamma^\mu  \frac{\slashed{\etabar}}{2} \gamma_\alpha^\perp  [-i\partial^\alpha \chi_{n}(\xn)] \\
O_2^{(1_{\perp\2})\mu}(x)&= [ -i\partial^\alpha \bar{\chi}_\nb(\xnb)]  \gamma_\alpha^\perp \frac{\slashed{\eta}}{2} \gamma^\mu [\chi_{n}(\xn)] .
\end{aligned}
\end{equation}
These were not required in \cite{Goerke:2017ioi,Goerke:2017lei} since they could be removed by a suitable choice of reference frame, while here the presence of initial-state radiation prevents such a choice.

Finally, there are several operators containing  factors of ${\mathcal B}_{n,\nb}$ whose matrix elements begin at $O(g_s)$. These operators are labeled by a continuous parameter $t$ which parametrizes the separation of fields along the light-cone \cite{Hill:2004nlp}. We define the dimensionless parameter $\hat t\equiv q^- t$ if the shift occurs in the $n$ sector, and by $\hat t \equiv  q^+ t$ if the shift occurs in the $\nb$ sector. We define the $A$-type operators in which a gluon is emitted at leading order in the $n$ sector,
\begin{equation} \begin{aligned} \label{eq:AtypeDefns}
O_2^{(1A_1)\mu}(x,\hat{t}\,)=& \, [ \bar{\chi}_\nb (x_\nbar)  ] \gamma_\alpha^\perp \frac{\slashed{\eta}}{2} \gamma^\mu     [\mathcal{B}_n^{\dagger\alpha}(x_n-\nb t) \chi_{n}(x_n)] \\
O_2^{(1A_2)\mu}(x,\hat{t}\,)=& - [ \bar{\chi}_{\nb}(x_\nbar)  ]  \gamma^\mu \frac{\slashed{\etabar}}{2} \gamma_\alpha^\perp [\mathcal{B}_n^{\dagger\alpha}(x_n-\nb t) \chi_{n}(x_n)] \\
O_2^{(2A_1)\mu}(x,\hat{t}\,)=& -2\pi  i\,\theta(\hat{t}) \otimes [ \bar{\chi}_\nb (x_\nb) ]  \gamma_\alpha^\perp \gamma_\beta^\perp \gamma^\mu    \\ &\times [\mathcal{B}_n^{\dagger\alpha\beta}(x_n-\nb t)  \chi_{n}(x_n)] \,,
\end{aligned} \end{equation}
and the corresponding $B$-type operators where the gluon is emitted in the $\nb$ sector,
\begin{equation} \begin{aligned} \label{eq:BtypeDefns}
O_2^{(1B_1)\mu}(x,\hat{t}\,)=& - [ \bar{\chi}_\nb (x_\nb) \mathcal{B}_\nb^{\alpha}(x_\nb- n t) ]  \gamma^\mu \frac{\slashed{\etabar}}{2}  \gamma_\alpha^\perp    [ \chi_{n}(x_n)] \\
O_2^{(1B_2)\mu}(x,\hat{t}\,)=& \, [ \bar{\chi}_{\nb}(x_\nb) \mathcal{B}_\nb^{\alpha}(x_\nb- n t)  ]   \gamma_\alpha^\perp  \frac{\slashed{\eta}}{2}\gamma^\mu  [\chi_{n}(x_n)] \\
O_2^{(2B_1)\mu}(x,\hat{t}\,)=& -2\pi  i\,\theta(\hat{t}) \otimes [ \bar{\chi}_\nb (x_\nb) \mathcal{B}_\nb^{\alpha\beta}(x_\nb- n t)  ]\\&   \times\gamma^\mu  \gamma_\alpha^\perp \gamma_\beta^\perp  [ \chi_{n}(x_n)] \,.
\end{aligned} \end{equation}
Following \cite{Hill:2004nlp}, it is convenient to work with the Fourier-transformed operators
\begin{equation} \begin{aligned} \label{eq:fourier_conventions}
O_2^{(i)}(x,u)&= \int \! \frac{d\hat{t}}{2\pi} e^{-iu\hat{t}} 
O_2^{(i)}(x,\hat{t}\,) \\
C_2^{(i)}(x,u)&= \int  d\hat{t}\ e^{iu\hat{t}} \ C_2^{(i)}(x,\hat{t}\,).
\end{aligned} \end{equation}
We have also defined the convolutions in $\hat{t}$ space in these definitions as 
\begin{equation} \begin{aligned} \label{eq:convolution}
f(\hat{t}) \otimes g(\hat{t}) = \int\! \frac{dx\,dy}{2\pi}\, f(x) g(y) \, \delta(\hat{t}-x-y) \ .
\end{aligned} \end{equation}
Note that the one-gluon matrix element of $O_2^{(2A_1)}(x,u)$ is proportional to $\delta(u+k^-)/u$, where $k^\mu$ is the gluon momentum. If the convolution with $\theta(\hat t)$ had not been included in its definition (as was the case in \cite{Goerke:2017lei}), the matrix element of the operator would instead be proportional to $\delta(u+k^-)$, and the operator would have a factor of $1/u$ in its Wilson coefficient. This is inconvenient because in the DY process studied here, this factor of $1/u\sim 1/k^-$ corresponds to  a rapidity divergence, and rapidity renormalizing operator products such as $O_2^{(2A1)\dagger} O_2^{(0)}$ without the factor of $1/u$ in its matrix element would then give rise to an unregulated rapidity divergence in the final integral over $u$.\footnote{The importance of having a finite integral over convolution variables was stressed in \cite{Moult:2019mog,Moult:2019uhz,Beneke:2019oqx}.} These are similar to the endpoint divergences which have been previously noted at NLP in SCET, in particular in $b$-mediated $h\rightarrow \gamma\gamma$ decay \cite{Liu:2019oav,Liu:2020tzd,Liu:2020wbn}. With the definition given here -- which is similar to the modification of SCET operators proposed in \cite{Beneke:2019kgv} -- the $u$ integral does not introduce any additional singularities and thus all rapidity divergences are correctly regulated by the pure rapidity regulator. We illustrate this with an example in \secref{sec:T23example}. 

Since SCET currents and their products contain operators with zero, one, or two factors of $u$ at this order, we use the notation $\{u\}$ to denote the dependence of a quantity on any number of $u$'s, as well as $\int d\{u\}$ to indicate integration over any number of $u$'s (including zero). The expansion of the SCET current may therefore be written
\begin{equation}\label{eq:J_SCET}
\begin{aligned}
J^\mu_{\rm SCET}(x) &= \sum_i \int \! d\{ u \} \frac{1}{q_L^{[i]}} C_2^{(i)}(\{u\}) O_{2}^{(i)\mu}(x,\{u\}) \\
&=
C_{2}^{(0)}
\left[ 
    O_{2}^{(0)\mu}(x) \bigvert\right.
    \\&\qqquad
    \left.
    +\frac{1}{q_L} 
    \left(
        O_2^{(1_{\perp\1})\mu}(x)+O_2^{(1_{\perp\2})\mu}(x)
    \right)
    \right.
\\ &\qqquad+
\left.
    \frac{1}{q_L^2}
    \left(
       O_2^{(2\delta^+)\mu}(x)+O_2^{(2\delta^-)\mu}(x)
    \right)
\right]\\
&+\frac{1}{q_L}\sum_i \int \!du \, C_2^{(1i)}(u) O_2^{(1i)\mu}(x,u) 
\\ &+
    \frac{1}{q_L^2}\sum_i \int\!du\, C_2^{(2i)}(u) O_2^{(2i)\mu}(x,u)
+O\left(\frac{1}{q_L^3}\right)\,
\end{aligned}
\end{equation}
where on the first line the sum is over all operators $i=0,1_{\perp n}, ..., 2B_1$, while in the last two lines the sums are over the operators of the appropriate dimension whose coefficients are not fixed by reparametrization or translation invariance.
The operators $O_2^{(1_{\perp\n,\nb})}$ are related to $O_2^{(0)}$ through reparametrization invariance (RPI) \cite{Manohar:2002RPI,Freedman:2014uta}, and so to all orders in $\alpha_s$ we have the equalities $C_2^{(0)}=C_2^{(1\perp_\1)}=C_2^{(1\perp_\2)}$, while translation invariance of QCD ensures that $C_2^{(0)}=C_2^{(2\delta^+)}=C_2^{(2\delta^-)}$. The normalizations of all operators have been chosen so that their tree-level matching coefficient is unity in $u$ space,
\begin{equation}
    C_2^{(i)}(\mu,\{u\})=1+O(\alpha_s)\,.
\end{equation}
The one-gluon matrix elements of the operators $O_2^{(i)}(\mu,\{u\})$ are given in Appendix \ref{sec:appendix_matrix_elements}.

There are additional operators not included in \eqn{eq:AtypeDefns} and \eqn{eq:BtypeDefns} that are part of the general SCET current expansion \cite{Goerke:2017lei}, but which do not contribute to $\Cff$ at the order (in $\alpha_s$, $q_T^2/q^2$, or $\lqcd^2/q^2$) to which we are working, or which contribute only to the gluon-initiated Drell-Yan subprocess. These operators do not mix under renormalization at one loop with the operators considered here, and so are not included in this analysis, though we list them in Appendix \ref{sec:appendix_matrix_elements} for completeness.

%
\subsection{Renormalization group running}
%

The anomalous dimensions of all the required matching coefficients $C_2^{(i)}(\{ u \})$ have been calculated previously in \cite{Manohar:2003vb,Freedman:2014uta,Goerke:2017lei}. They obey the integro-differential equation
\begin{equation} \begin{aligned} \label{eq:C2i_running}
\frac{d}{d\log\mu} C_2^{(i)}(\mu,\{ u\}) = \int d\{ v\}  \gamma_2^{\left(i\right)}(\{u\},\{v \})  C_2^{(i)}(\mu ,\{ v\})\,,
\end{aligned} \end{equation}
where the kernels $\gamma_2^{(i)}$ have the form
\begin{equation} \begin{aligned} \label{eq:C2_anom_dim_decomp}
\gamma_{2}^{(i)}(\{u\},\{v\}) =&   \Gamma_\cusp^{(i)}[\alpha_s] \log \frac{-\QLsqr-i0^+}{\mu^2} \delta(\{u\}-\{v\}) \\&+ \gamma_\mathrm{non-cusp}^{(i)}(\{u\},\{v\}) \,.
\end{aligned} \end{equation}
Working in the leading-log (LL) approximation only the cusp anomalous dimension is required. The one-loop cusp anomalous dimension is universal,
\begin{equation}  \label{eq:C2_LL_anom_dim}
\Gamma_\cusp^{(0)} = \Gamma_\cusp^{(1A_i)} = \Gamma_\cusp^{(2A_i)} = \Gamma_\cusp^{(1B_i)} = \Gamma_\cusp^{(2B_i)}= \frac{\alpha_s C_F}{\pi}   \,.
\end{equation}
With the definition $H_{(i,j)}(\{u \})=C_2^{(i)\dagger}( \{u\}) C_2^{(j)}( \{u\})$, 
the leading-log running of the hard functions is determined by the RGE 
\begin{equation}  \begin{aligned} \label{eq:H_running}
\frac{d}{d\log\mu} H_{(i,j)}(\mu,\{ u \}) &= \left( 2\Gamma_\cusp \log\frac{q_L^2}{\mu^2}\right) H_{(i,j)}(\mu,\{ u\})\,.
\end{aligned}\end{equation}
This gives the LL unitary evolution for all $H_{(i,j)}$
\begin{equation} \begin{aligned} \label{eq:H_unitary}
H_{(i,j)}(\mu,\{ u\}) = U_{H}^\LL(\mu,\mu_H) H_{(i,j)}(\mu_H,\{ u\})\,,
\end{aligned} \end{equation}
where, with $\beta[\alpha_s] \equiv d\alpha_s/d\log\mu = -\beta_0\alpha_s^2/2\pi + \cdots$\ ,
\begin{equation} \begin{aligned} \label{eq:UH_implicit}
\log &U_{H}^\LL(\mu,\mu_H)= -4 
\int_{\alpha_s(\mu_H)}^{\alpha_s(\mu)} \frac{d\alpha}{\beta[\alpha]} 
\Gamma_\cusp[\alpha] 
\int_{\alpha(q_L)}^{\alpha} \frac{d\alpha^\prime}{\beta[\alpha^\prime]} \\
&=\frac{16\pi C_F}{\beta_0^2} \left( 
\frac{1}{\alpha_s(\mu_H)}   -\frac{1}{\alpha(\mu)}
-\frac{1}{\alpha(q_L)} \log\frac{\alpha(\mu)}{\alpha(\mu_H)}   \right) \, .
\end{aligned} \end{equation}
Beyond LL there will be operator mixing, and the solution to the RGE will be more involved.
This sums the RG logarithms of $\mu_H/\mu$ in the hard functions.

\subsection{
\texorpdfstring{$T_{(i,j)}$}{T\_(i,j)} definitions} \label{sec:generalized_TMDPDF_defns}

The differential cross section for DY production is given in terms of hadronic matrix elements of products of operators $O_2^{(i)\dagger}(x)O_2^{(j)}(0)$ in \eqn{eq:factorization_derivation_part2}.
Matrix elements of these operator products may be evaluated between partons in perturbation theory to calculate the matching conditions onto 
light-cone distribution operators (whose matrix elements are the usual PDFs); however, it is convenient to perform a Fierz rearrangement for each operator product to write it as the product of factors in the $n$ and $\nb$ sectors, corresponding to the convolution of generalized transverse momentum dependent
distribution operators. 
At leading power, this gives
\begin{equation} \begin{aligned}\label{LOT00}
\int \frac{d^d x}{2(2\pi)^d}& (-g_{\mu\nu}) e^{-iq\cdot x} O_{2}^{(0)\mu\dagger}(x) O_{2}^{(0)\nu}(0) 
\\&= \frac{1}{N_c} \int \frac{d^d x}{2(2\pi)^d} e^{-iq\cdot x} 
\Phi_\1^{(0)}(x_n) \Phi_\2^{(0)}(x_\nb) 
\\
&\equiv \frac{1}{N_c}  T_{(0,0)}(q^-,q^+,\vecQ)\,.
\end{aligned} \end{equation}
The leading power position space 
distribution operators
are defined as
\begin{equation}
\Phi_\1(x_n)\equiv \chibar_n(x_n) \frac{\slashed{\nb}}{2} \chi_n(0), \ \Phi_\2(x_\nb)\equiv \chibar_\nb(0)\frac{\slashed{n}}{2} \chi_\nb(x_\nb)\,,
\end{equation} 
where $x_n$ and $x_\nb$ are defined in \eqref{eq:xndef}, and thus consist of quark fields  separated in the transverse direction by $x_\perp^\mu$ as well as along the light cone.

Products of power-suppressed operators may similarly be written as convolutions of higher dimension operators, 
\begin{equation}\begin{aligned}
T_{(i,j)}(q,\uvars)
\!&=
\!\!\!\int\!\!\! \frac{d^d x}{2(2\pi)^d} e^{-i q\cdot x} \,
\!\Phi_\1^{(i)}(x_n,\!\uvars) \Phi_\2^{(j)}(x_\nb,\!\uvars),
\end{aligned} \end{equation}
where we define the relevant subleading 
transverse momentum dependent light-cone distribution operators $\Phi_n^{(i)}(x,\tvars)$ as
\begin{equation}
\begin{aligned}
\Phi_n^{(2_1)}(x_n,\hat{t})\equiv&\,(i\partial^\mu \chibar_n(x_n)) \frac{\slashed{\nb}}{2} \gamma_\mu^\perp \gamma_\nu^\perp \mathcal{B}_n^{\dagger\nu}(-\nb t) \chi_n(0) \\
\Phi_n^{(2_2)}(x_n,\thatt,\thattt)\equiv&-\chibar_n(x_n) \mathcal{B}_n^{\mu}(x_n-\nb t_1) \frac{\slashed{\nb}}{2} \gamma_\mu^\perp \gamma_\nu^\perp\\&\times \mathcal{B}_n^{\dagger\nu}(-\nb t_2) \chi_n(0) \\
\Phi_n^{(2_3)}(x_n,\hat{t})\equiv&  \, 2\pi i \theta(\that) \otimes   \chibar_n(x_n)   \mathcal{B}_n^{\mu\nu}(x_n-\nb t)\\&\times \frac{\slashed{\nb}}{2} \gamma_\nu^\perp \gamma_\mu^\perp \chi_n(0) \\
\Phi_n^{(2_4)}(x_n)\equiv& \, q^+q^-\frac{x^-}{2} \left( n\cdot \partial \, \chibar_n(x_n) \right) \frac{\slashed{\nb}}{2} \chi_n(0) \,.
\end{aligned}\end{equation}

The corresponding $\nb$-sector operators $\Phi_\nb^{(i)}$ are found by taking the Hermitian conjugate and changing $n\leftrightarrow \nb$. The $u$-space Fourier conjugates of these building blocks are defined by the transformation in \eqn{eq:fourier_conventions} for shifts relative to the origin (since these shifts come from an operator), and by the conjugate transformation for shifts relative to $x_{n}$ (since these shifts come from the conjugated operator).

Thus, for example,
\begin{equation}\begin{aligned}
T_{(0,2_1)}(q,u)=&\int \frac{d^d x}{2(2\pi)^d} \, e^{-i q\cdot x} \,
\Phi_\1^{(0)}(x_n) \Phi_\2^{(2_1)}(x_\nb,u)\\=&
\int \frac{d^d x}{2(2\pi)^d} \frac{d\hat{t}}{2\pi} e^{-iu\hat t}\, e^{-i q\cdot x} \,
\left[\chibar_n(x_n) \frac{\slashed{\nb}}{2} \chi_n(0)\right]\\
&\times\left[\chibar_\nbar(0)\mathcal{B}_\nbar^{\nu}(-n t)  \gamma_\nu^\perp \gamma_\mu^\perp\frac{\slashed{n}}{2}   (-i\partial^\mu \chi_\nbar(x_\nb) )\right].
\end{aligned} \end{equation}
In general, we can write
\begin{equation} \begin{aligned} \label{eq:FierzingTransformationFinal}
\int \! \frac{d^d x }{2(2\pi)^d}& \, (-g_{\mu\nu})e^{-iq\cdot x}  O_{2}^{(i)\mu\dagger}(x) O_2^{(j)\nu}(0)\\ =&\sum_{k,\ell}\frac{1}{N_c}\kmat ijk\ell T_{(k,\ell)}+\mbox{spin\,dependent}\,,
\end{aligned} \end{equation}
where the only nonzero elements of $K$ which are relevant at this order are
\begin{equation}\begin{aligned}\label{eq:Kijlk_values}
\kmat {1\perp n}{1A_1}{2_1}{0} &= \kmat{1A_1}{1\perp n}{2_1}{0} 
= \kmat{1A_2}{1A_1}{2_2}{0} = \kmat{1A_1}{1A_2}{2_2}{0} 
\\&
= \kmat{1\perp \nb}{1B_1}{0}{2_1} = \kmat{1B_1}{1\perp\nb}{0}{2_1} 
= \kmat{1B_2}{1B_1}{0}{2_2}  
\\&
= \kmat{1B_1}{1B_2}{0}{2_2} = \kmat{2A_1}{0}{2_3}{0} = \kmat{0}{2A_1}{2_3}{0} 
 \\&
 = \kmat{2\delta^+}{0}{2_4}{0} = \kmat{0}{2B_1}{0}{2_3}= \kmat{2B_1}{0}{0}{2_3} 
\\&=\kmat{2\delta^-}{0}{0}{2_4}=1.\end{aligned}
\end{equation}

%
%
\subsection{Matrix elements of operator products} \label{sec:fierzed_mat_elements}
%
%

Individual $n$- and $\nb$-sector graphs contributing to the matrix elements of each $T_{(i,j)}$ are rapidity divergent and require a regulator to give finite results. We use a version of the pure rapidity regulator introduced in \cite{Ebert:2018gsn}. As discussed in that reference, other commonly used rapidity regulators such as the $\delta$ regulator \cite{Chiu:2009yx} or the $\eta$ regulator \cite{Chiu:2011qc} are not suitable for handling the power-law rapidity divergences that arise at NLP. An explicit example of the $\delta$ regulator failing to regulate rapidity divergences at NLP is given in \secref{sec:T23example}.

In what follows we define the pure rapidity regulator by modifying the integration measure of $n$-sector and $\nb$-sector particles as 
\begin{equation} \begin{aligned} \label{eq:purerapidity}
d^d k_n &\rightarrow  \wn^2 \left( \frac{\QLsqr}{\nun^2}\right)^{\etan/2} 
    \!\! \left(  \frac{q^-}{q^+} \frac{k_\n^+  }{ k_\n^-}\right)^{\etan/2}  d^d k_n 
    \\ 
d^d k_\nb &\rightarrow  \wnb^2 \left( \frac{\QLsqr}{\nunb^2}\right)^{\etanb/2}
    \!\!\left(  \frac{q^+}{q^-} \frac{k_\nb^- }{ k_\nb^+ }\right)^{\etanb/2}   d^d k_\nb  \,.
\end{aligned} \end{equation}
This regulator has the distinct advantage that -- as in dimensional regularization -- scaleless integrals vanish, and as a result all overlap integrals evaluate to zero. This greatly simplifies the calculation since, as is discussed in detail in \secref{sec:overlap_subtractions}, in a scheme where overlap integrals do not vanish,  the overlap subtraction procedure must be carried out to subleading powers.

The regulator in \eqn{eq:purerapidity} is slightly modified from the form presented in \cite{Ebert:2018gsn}: the factors of $q^\pm$ ensure boost invariance, as in \cite{Moult:2019vou}, the dimensionless parameter $\upsilon$ has been replaced by the equivalent dimensionful parameters $\nu_i$, and distinct parameters $\eta_i$, $\nu_i$, and $w_i$ have been introduced for each sector, since the fields in the $n$ and $\nb$ sectors are independent. As discussed in \cite{Inglis:2020rpi}, rapidity logarithms in SCET correspond to a scheme dependence in defining the sum of individually rapidity divergent graphs in the $n$ and $\nb$ sectors. Regulating both sectors (and their corresponding overlap graphs) in the same way and then removing the regulator is equivalent to na\"\i vely adding the individual graphs together before performing the loop integrals, and reproduces the rapidity logarithms of QCD. Since QCD has no rapidity divergences, rapidity divergences cancel for this choice, which corresponds to choosing the parameters $\etan=-\etanb$, $\nun\nunb=q_L^2$ and $\wn=\wnb$.  This was explicitly demonstrated up to NLP in \cite{Ebert:2018gsn}: these authors showed that if QCD diagrams are first rapidity regulated and then expanded in the $n$-collinear, $\nb$-collinear, and soft limits, the leading and subleading power matrix elements reproduce the rapidity-finite QCD results expanded to the same order. A similar cancellation of rapidity divergences will be shown here.

Using different rapidity regulator parameters in the two sectors moves the rapidity logarithm of $\QLsqr$ into the Wilson coefficients of the EFT and allows the scheme dependence of the resulting graphs to be exploited to sum the corresponding rapidity logs. The corresponding rapidity divergences correspond to $1/\eta_i$ singularities which are canceled by introducing the appropriate counterterms into the EFT, and rapidity logarithms are then summed using rapidity renormalization group (RRG) techniques similar to \cite{Chiu:2011qc}.  The bookkeeping constants $w_i$ are taken to formally obey the RRG equation
\begin{equation} \begin{aligned} \label{eq:w12_running}
\frac{d w_i}{d\log \nu_i} = \frac{\eta_i}{2} w_i\,,
\end{aligned} \end{equation}
which cancels the scheme dependence in the measure, keeping the bare theory $\nu_i$ independent and allowing techniques analogous to those in dimensional regularization to be used to extract the rapidity anomalous dimensions. As in \cite{Chiu:2011qc,Ebert:2018gsn}, these bookkeeping constants are set to unity at the end of calculation. Rapidity logarithms are minimized by the appropriate choice of the dimensionless parameters $\nunnb$. 

As noted in \cite{Inglis:2020rpi}, choosing $\nu_n\nu_\nb\neq q_L^2$ requires rapidity counterterms for each $T_{(i,j)}$ which are sensitive to the scale $q_T$. Scale sensitivities in the counterterm generate the same scale dependence in the Wilson coefficient through the RGE, and since Wilson coefficients in an EFT must be independent of infrared physics, this adds the constraint that the theory must first be evolved to $\mu\sim q_T$ before running in rapidity. This will be discussed in more detail in \secref{sec:rap_running}.

At LP the only operator is $T_{(0,0)}$, so its divergences are absorbed by the renormalization constant $Z_{(0,0),(0,0)}$,
\begin{equation} \begin{aligned}\label{eq:renormalization_LP}
    T&_{(0,0)}^B(q^-,q^+,\vecQ)= \int \frac{d\omegan}{\omegan} \frac{d\omeganb}{\omeganb} d^{d-2}\myvec{p}_T \\ & \times Z_{(0,0),(0,0)}(\omegan,\omeganb, \myvec{p}_T)  T_{(0,0)}\left(\frac{q^-}{\omegan},\frac{q^+}{\omeganb},\vecQ -\myvec{p}_T\right) \,,
\end{aligned}\end{equation}
where $T_{(i,j)}^B$  and $T_{(i,j)}$ 
denote the bare and renormalized operators, respectively, and the integral corresponds to summing over the infinite set of operators $T_{(0,0)}(k^-,k^+,\bf{k}_T)$.
At subleading powers the various operators may mix with one another, so we have the general relation
\begin{equation} \begin{aligned} \label{eq:renormalization_NLP}
    T_{(i,j)}^B&(q^-,q^+,\vecQ,\uvars)= \sum_{(k,l)}\int \! \frac{d\omegan}{\omegan} \frac{d\omeganb}{\omeganb} d^{d-2}\myvec{p}_T d\{v\}\\ & \times Z_{(i,j),(k,l)}(\omegan,\omeganb, \myvec{p}_T,\vvars)
  \\ &\times
    T_{(k,l)}\left(\frac{q^+}{\omeganb},\frac{q^-}{\omegan},\vecQ -\myvec{p}_T, \{ u - v\}\right)\,,
\end{aligned} \end{equation}
where the sum over operators includes each subleading $T_{(i,j)}$ as well as the leading operator $T_{(0,0)}$ with a power-suppressed coefficient, as will be discussed in the following sections.

%
\subsubsection{Leading power example} \label{sec:fierzed_mat_elements_LP}
%

The leading power calculation of DY production in this formalism was presented in \cite{Inglis:2020rpi} using the $\delta$ regulator; we repeat the calculation here with the pure rapidity regulator.
At leading power, there is a single bilocal operator contributing to the rate,
\begin{equation}\begin{aligned}
    T&_{(0,0)}(q^-, q^+, \vecQ) = \\& \int \!\!\frac{d^d x}{2(2\pi)^d} \, e^{-iq\cdot x}
    \left[ \chibar_n(x_n) \frac{\slashed{\nb}}{2} \chi_n(0) \right] 
    \left[ \chibar_\nb(0) \frac{\slashed{n}}{2} \chi_\nb(x_\nb) \right]\,.
\end{aligned} \end{equation}
With incoming quark and antiquark states $q(p_1)$ and $\bar{q}(p_2)$ the tree-level matrix element of this operator is
\begin{equation} \begin{aligned}\label{eq:T00mat}
\frac{1}{4}\sum_\mathrm{spins}& \bra{ p_1^n p_2^\nb} T_{(0,0)} \ket{ p_1^n p_2^\nb}
= \frac{1}{4}\sum_\mathrm{spins} \int \!\! \frac{d^d x}{2(2\pi)^d} \, e^{-iq\cdot x}\\&\quad \times 
    \bra{p_1^n} \left[ \chibar_n(x_n)  \ket{0}\frac{\slashed{\nb}}{2} \bra{0} \chi_n(0) \right] \ket{p_1^n} \\ &\quad \times
    \bra{p_2^\nb}\left[ \chibar_\nb(0) \ket{0}\frac{\slashed{n}}{2}\bra{0}  \chi_\nb(x_\nb) \right] \ket{p_2^\nb} \\
&= \delta(\zbar) \delta(\zzbar) \delta^{d-2}(\vecQ) \equiv \delta_1 \delta_2 \deltaT\,,
\end{aligned} \end{equation}
where the superscripts $n$- and $\nb$ in \eqn{eq:T00mat} denote the sector of the corresponding parton. We also use the notation 
\begin{equation}
z_1\equiv \frac{q^-}{p_1^-},\ z_2 \equiv \frac{q^+}{p_2^+},\ \bar{z}_i\equiv1-z_i
\end{equation}
and
\begin{equation}
\delta_i \equiv \delta(\bar{z}_i),\ \delta_i^\prime\equiv \delta^\prime(\bar{z}_i),\ \deltaT\equiv\delta^{d-2}(\vecQ) \,.
\end{equation}

\begin{figure*}
	\includegraphics[width=0.65\textwidth]{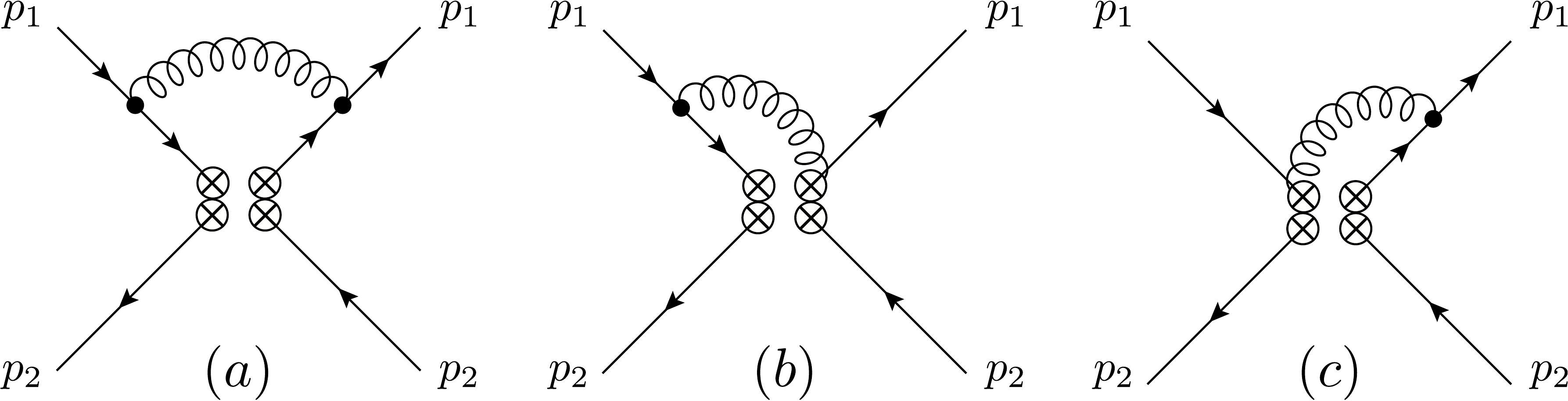}
	\caption{Nonvanishing graphs in the $n$ sector contributing to the matrix element $\bra{ p_1^n p_2^\nb}T_{(0,0)} \ket{ p_1^n p_2^\nb}$ at $\order{\alpha_s}$.}
	 \label{fig:F0_1graphs} 
 \end{figure*}

At $O(\alpha_s)$, the matrix element corresponding to the emission of a real $\1$-sector gluon is given by the three $\1$-sector graphs shown of \figref{fig:F0_1graphs}.  Denoting the spin-averaged one-loop matrix element of $T_{(0,0)}$ by $\mmat{(0,0)}$, we write
\begin{equation}
    \mmat{(0,0)}=\mmatn{(0,0)}{n}+\mmatn{(0,0)}{\nb}-\mmatn{(0,0)}{O}\,,
\end{equation}
where the superscripts $n$ and $\nb$ denote the $O(\alpha_s)$ contribution from a gluon in the corresponding sector, and the $O$ superscript denotes the overlap subtraction.  
Since these matrix elements correspond to a matching calculation at the scale $\mu_S \sim q_T \gg \lqcd$ we use the initial-state kinematics $\pquark^+ = {\pquark}_\perp = 0 = {\panti}_\perp = \panti^-$, and we obtain 
\begin{equation} \begin{aligned} \label{eq:n_graph_calc}
\mmatn{(0,0)}{n}& =
-2\pi g^2 C_F \int \! \frac{d^d k}{(2\pi)^d} \wn^2 \left( \frac{q_L^2}{\nun^2}  \frac{q^-}{q^+} \frac{k^+  }{ k^-}\right)^{\etan/2}  
    \\ &  \times 
    \delta(k^2) \delta(\pquark^- \!- q^- \!- k^-)\delta(\panti^+ - q^+) \delta^{d-2}(\vecQ+\myvec{k}_T)
    \\ &  \times 
    \mathrm{Tr} \left[ \frac{\pquarksl}{2}
    \left( \frac{2\pquark^\alpha-\gamma^\alpha\slashed{k}}{-2\pquark\cdot k} + \frac{\nb^\alpha}{k^-} \right) \frac{\slashed{\nb}}{2}\right. \\&\qquad\left. \times \left( \frac{2{\pquark}_{\alpha}-\slashed{k}\gamma_\alpha}{-2\pquark\cdot k} - \frac{\nb_\alpha}{-k^-} \right)   \right]  \mathrm{Tr} \left[ \frac{\slashed{p}_2}{2} \frac{\slashed{n}}{2} \right] \,,
\end{aligned} \end{equation}
which evaluates to 
\begin{equation}\begin{aligned}  \label{eq:n_graph_final}
\mmatn{(0,0)}{n} =& \frac{\alphabar}{\pi} f_\epsilon \wn^2 \delta_2 
\frac{(\mu^2)^{-\etan/2}}{(\vecQ^2)^{1-\etan/2}} 
\left( \frac{z_1 \mu}{ \nun}\right)^{\!\etan} \\&\times\frac{ (2-2\zbar+(1-\epsilon)\zbar^2) }{\zbar^{1+\etan}} \,,
\end{aligned} \end{equation}
where
\begin{equation}
\alphabar \equiv \frac{\alpha_s C_F}{2\pi},\qquad f_\epsilon\equiv (\pi \mu^2 e^\gamma)^\epsilon 
\end{equation}
and we work in $d=4-2\epsilon$ dimensions.

To extract the singularity structure of this matrix element at $\bar z_1=0$ we use the distributional identity 
\begin{equation}\label{eq:plus_id_1}
\frac{ \theta(\zbar)}{\zbar ^{1+\eta}}= -\frac{\delta(\zbar)}{\eta}+\left[ \frac{\theta(\zbar)}{\zbar}\right]_+ + \cdots 
\end{equation} 
for scalars (see Appendix \ref{sec:plus_distributions} for definitions) as well as the identity \cite{Ebert:2016gcn}
\begin{equation}\label{eq:vecPlusIdentity}
\frac{(\mu^2)^{-\eta/2}}{(\vecQ^2)^{1-\eta/2} } =  \mu^{-2\epsilon}\frac{S_{d-2}}{2} 
\left( 
     \frac{ \deltaT}{\frac{\eta}{2}-\epsilon} 
    + \myvec{\mathcal{L}}_{0T}   + \frac{\eta}{2} \myvec{\mathcal{L}}_{1T}  + \cdots \right)
\end{equation}
for vectors in $(d-2)$ dimensions, where the $\myvec{\mathcal{L}}_{nT}$ are vector plus distributions \cite{Chiu:2012ir,Ebert:2016gcn} defined in Appendix \ref{sec:vector_plus_distributions}, 
\begin{equation}\begin{aligned}
\myvec{\mathcal{L}}_{nT} \equiv \mathcal{L}_n(\vecQ,\mu)= \frac{2\mu^{2\epsilon}}{S_{d-2}}
\left[ 
    \frac{
        \log^n \frac{\vecQ^2}{\mu^2} \theta(\vecQ^2)
        }
        {
        \vecQ^2
    } 
\right]_+^{\mu^2} \ ,
\end{aligned}
\end{equation} 
and $S_{d-2}=2\pi^{\frac{d-2}{2}}/\Gamma \left( \frac{d-2}{2} \right)$. Upon expanding first in $\etan$ then in $\epsilon$, the $n$-sector contribution to the matrix element from a single real emission is
\begin{equation}\begin{aligned}\label{eq:n_final_result}
\mmatn{(0,0)}{n}=&\frac{\alphabar}{2} \wn^2 \delta_2 
\Bigg[  
        \frac{4}{\etan} \delta_1 \left(\frac{\deltaT}{\epsilon} - \myvec{\mathcal{L}}_{0T} \right)
        \\
        &+ \delta_1 \deltaT \left( \frac{2}{\epsilon^2}+\frac{3-2\log \frac{\nun^2}{\mu^2}}{\epsilon} \right) 
     \\
    & + 2 \left(  \myvec{\mathcal{L}}_{0T}  - \frac{\deltaT}{\epsilon}\right) \left[ \frac{1+z_1^2}{\zbar} \right]_+  \deltaT \\& + 2\zbar \deltaT - \zeta_2 \delta_1 \deltaT - 2\delta_1 \myvec{\mathcal{L}}_{1T}
   \\
    &  -\delta_1 \myvec{\mathcal{L}}_{0T}\left( 3-2\log \frac{\nun^2}{\mu^2} \right)  
\Bigg] \, .
\end{aligned}\end{equation}
The $\nbar$-sector contribution is obtained under the replacements $z_1 \rightarrow z_2$, $\wn \rightarrow \wnb$, $\nun\rightarrow \nunb$ and $\etan \rightarrow \etanb$.

As in \cite{Goerke:2017ioi}, the overlap between the two sectors is obtained by taking the opposite-sector gluon limit of the $n$- and $\nb$-sector graphs. As detailed in \secref{sec:overlap_subtractions}, the subtraction prescription corresponds to subtracting half the wrong sector limit for the gluon of each sector,\footnote{In \cite{Goerke:2017ioi}, the limits from either sector were equal.} which we denote
\begin{equation}
    \mmatn{(0,0)}{O}=\frac{1}{2}\left(\mmatn{(0,0)}{\1\rightarrow\2}+\mmatn{(0,0)}{\2\rightarrow\1}
    \right).
    \end{equation}
For example, the wrong sector limit of \eqn{eq:n_graph_calc} is 
\begin{equation} \begin{aligned} \label{eq:n_graph_soft_calc}
\mmatn{(0,0)}{\1\rightarrow\2} =& 
-2\pi g^2 C_F \int  \frac{d^d k}{(2\pi)^d} w_{n}^2 \left( \frac{q_L^2}{\nu_{n}^2}  \frac{q^-}{q^+} \frac{k^+  }{ k^-}\right)^{\eta_{n}/2} 
    \\ &  \times 
     \delta(k^2) \delta(\pquark^- - q^-)\delta(\panti^+ - q^+) \delta^{d-2}(\vecQ+\myvec{k}_T)
    \\ &   \times 
    \mathrm{Tr} \left[ \frac{\slashed{p}_1}{2}
\left( \frac{n^\alpha}{-k^+} + \frac{\nb^\alpha}{k^-} \right) \frac{\slashed{\nb}}{2} \left( \frac{n_\alpha}{-k^+} - \frac{\nb_\alpha}{-k^-} \right)   \right]\\&\times  \mathrm{Tr} \left[ \frac{\slashed{p}_2}{2} \frac{\slashed{n}}{2} \right] +\cdots
\end{aligned} \end{equation}
where the dots indicate terms suppressed by powers of $1/q_L$ relative to leading power.
Integrating with respect to $\myvec{k}_T$ and then $k^+$, we find the $\nb$ limit of the $n$-sector graphs is
\begin{equation} \begin{aligned} \label{eq:n_graph_soft}
\mmatn{(0,0)}{\1\rightarrow\2} =& \frac{2 \alphabar}{\pi} f_\epsilon  w_{n}^2 \delta_1\delta_2\left( \frac{q_L^2}{\nu_{n}^2} \frac{ q^-  }{q^+} \right)^{\eta_{n}/2} \frac{1}{(\qtsq)^{1-\eta_{n}/2}} \\&\times \int_0^\infty \frac{dk^-}{(k^-)^{1+\eta_{n}}}\,,
\end{aligned} \end{equation}
which is a scaleless divergence and vanishes in this regularization scheme. The overlap subtraction between the two sectors is therefore zero when using the pure rapidity regulator at $\order{\alpha_s}$, and this remains true beyond leading power.

Summing the contributions from each sector and subtracting off the (vanishing) overlap, we find the $O(\alpha_s)$ contribution to the matrix element of $ T_{(0,0)}$ 
\begin{equation}\begin{aligned}\label{eq:F0_bare}
\mmat{(0,0)}=&\alphabar
\Bigg[ 
     2 \left( \frac{\wn^2}{\etan} + \frac{\wnb^2}{\etanb}\right)\delta_1 \delta_2 
        \left( \frac{\deltaT}{\epsilon} - \myvec{\mathcal{L}}_{0T} \right) 
    \\&\!+ \delta_1 \delta_2 \myvec{\delta}_{T} \left(\frac{2}{\epsilon^2}
    + \frac{3-2\log \frac{\nun \nunb}{\mu^2}}{\epsilon}\right) \\
    &\! + \left(  \myvec{\mathcal{L}}_{0T} - \frac{ \deltaT }{\epsilon} \right) \!\left( \delta_2 \bigg[ \frac{1+z_1^2}{\zbar}\bigg]_+ 
        \!\!+\delta_1 \bigg[ \frac{1+z_2^2}{\zzbar}\bigg]_+\right) \\
    &\! -\delta_1 \delta_2\left[2\myvec{\mathcal{L}}_{1T} 
    + \myvec{\mathcal{L}}_{0T}\left( 3-2\log \frac{\nun \nunb}{\mu^2} \right)
     \right] \\
    &\!  + (\zbar \delta_2 +\zzbar\delta_1 - \delta_1 \delta_2 \zeta_2 )\deltaT 
\Bigg]\,,
\end{aligned}\end{equation}
where we have set $w_{\thisN,\thisNb}=1$ for all $\eta_i$-independent terms. As discussed earlier, the rapidity divergences appear as the $\eta$-divergent terms in the first line. The rapidity-finiteness of the full theory is reflected in the fact that setting $w_{\thisN} = w_{\thisNb}$ and $\etan=-\etanb$ gives a total rate that is free from $\eta$ poles, which is to be expected since this scheme corresponds to regulating the $n$ and $\nbar$ sectors identically; using different schemes for the two sectors spoils the cancellation of rapidity divergences between the sectors. However, resumming the rapidity logarithms
requires keeping the $w_{\thisN,\thisNb}$ and $\eta_{\thisN,\thisNb}$ scheme dependence. This introduces explicit rapidity divergences in the matrix elements which require rapidity counterterms, from which the RRG may be derived.

In the scheme where $\nun \nunb=q_L^2$, the purely $\epsilon$-divergent terms (ultraviolet divergences) in the first line of \eqn{eq:F0_bare} are canceled by the renormalization constant $Z_{2,(0)}$ for $O_{2}^{(0)}$ 
\begin{equation}
Z_{2,(0)} = 1 + \frac{\alphabar}{2} \left( \frac{2}{\epsilon^2} + \frac{3-2\log\frac{\QLsqr }{\mu^2}}{\epsilon} \right) \ ,
\end{equation}
which follows the product of renormalized operators $O_{2,(0)}^{\dagger\mu} O_{2,(0)}^{\dagger\nu}$ through the Fierz rearrangement. Since $Z_{2,(0)}$ depends only on $\log(\QLsqr /\mu^2)$, the scheme $\nun \nunb=q_L^2 $ is enforced at $\mu\sim q_L$ and throughout the $\mu$ running when $\mu>q_T$. As we later discuss in \secref{sec:rap_running}, when $\mu\sim q_T$ then $q_T$ is no longer an infrared scale, and then $\nunnb$ can be evolved with the RRG, allowing for the resummation of rapidity logarithms.

The IR divergent terms in the third line of \eqn{eq:F0_bare} are the Altarelli-Parisi splitting functions and are reproduced by the infrared divergences in the 
light-cone distribution operators in the low-energy theory. The remaining divergences are rapidity divergences, and are absorbed by the counterterm in \eqn{eq:renormalization_LP}, where 
\begin{equation} \begin{aligned}\label{eq:T_0_cterm}
Z_{(0,0),(0,0)}&(\omegan,\omeganb,\vecQ)=\\&\delta_1 \delta_2 \left( \deltaT + 2 \alphabar  \left( \frac{\wn^2}{\etan} + \frac{\wnb^2}{\etanb}\right)\right.\\&\left.\times
        \left( \frac{\deltaT}{\epsilon} - \myvec{\mathcal{L}}_{0T} \right) 
        + 2\alphabar \frac{ \myvec{\delta}_{T}   }{\epsilon} \log  \frac{q_L^2}{\nun \nunb} \right).
\end{aligned}\end{equation}
Using the running of the fictional coupling $w_{n,\nb}$ in \eqn{eq:w12_running}, we can obtain the rapidity anomalous dimension and rapidity evolution equation for $T_{(0,0)}$, which we further discuss in \secref{sec:rap_running}.

Note that for $\eta_n=-\eta_\nb$ and $\nunnb=q_L^2$ there are no additional ultraviolet (UV) divergences in the matrix element of $T_{(0,0)}$ beyond the renormalization of  $O_2^{(0)}(\mu)$, indicating that the phase space integral in SCET is UV finite at $O(\alpha_s)$. Similarly, one-loop matrix elements at NLP will also be found to be UV finite. Additional UV divergences in matrix elements of the $T_{(i,j)}$'s would indicate phase space integrals which were sensitive to the UV scale $q_L$, in which case the RG running of the corresponding $T_{(i,j)}$ would not simply be given by the running of its constituent SCET operators, but would have additional contributions. 
It is possible that this could complicate the RG running of the $H_{(i,j)}$'s at higher orders in $\alpha_s$, where the final state phase space can include multiple gluons with individually large $\myvec{k}_T$ which largely cancel to contribute at small $q_T$, but this would not affect the one-loop running or the form of the factorization \eqn{eq:final_Cff_factorization}. 

Subtracting the counterterms yields the renormalized matrix element 
\begin{equation}\begin{aligned}\label{eq:F0_renormalized}
\bra{p_1^n p_2^\nb} 
&T_{(0,0)}
\left( q^-, q^+, \vecQ, \frac{\mu}{q_T},\frac{\nunnb}{\mu} \right) 
\ket{p_1^n p_2^\nb}_{\mathrm{1-gluon}}
= \\
&\alphabar 
\left( \!
     \left( \myvec{\mathcal{L}}_{0T} - \frac{ \deltaT }{\epsilon} \right) \!\left( \delta_2 \left[ \frac{1+z_1^2}{\zbar}\right]_+ 
        \!+\delta_1 \left[ \frac{1+z_2^2}{\zzbar}\right]_+\right)\right. \\
    & -\delta_1 \delta_2
    \left[
        2\myvec{\mathcal{L}}_{1T} 
        + \myvec{\mathcal{L}}_{0T}\left( 3-2\log \frac{\nun \nunb}{\mu^2 } \right) 
    \right] \\
    &  + \left.(\zbar \delta_2 +\zzbar\delta_1 - \delta_1 \delta_2 \zeta_2 )\deltaT 
\Bigvert\right) \,,
\end{aligned}\end{equation}
which, with the replacement $\nun \nunb = \nu^2$, also reproduces the result in \cite{Inglis:2020rpi}.

\subsubsection{Next-to-leading power example:  \texorpdfstring{   $T_{(2_4,0)}$   }{  T\_(2\_4,0) }  } \label{sec:fierzed_mat_elements_NLP_27}

Since $\order{\lqcd}$ contributions are not considered, there is no 0-gluon contribution to the matrix element of $T_{(2_4,0)}$, and there is also no $\nb$-sector contribution at $O(\alpha_s)$. Thus at first nontrivial order the graphs $T_{(2_4,0)}$ are those shown in \figref{fig:F0_1graphs}, yielding
\begin{equation} \begin{aligned} \label{eq:T_2_7_calc}
\mmatn{(2_4,0)}{n} 
=& 2\pi g^2 C_F\, q^+ q^-\! \int \! 
    \frac{d^d k}{(2\pi)^d} \wn^2 \left( \frac{q_L^2}{\nun^2} \frac{ q^- k^+}{ q^+ k^-}\right)^{\etan/2} 
    \\
& \times   k^+ \delta(k^2)\delta(\pquark^- - q^- - k^-)\delta^\prime(\panti^+ - q^+) \\&\times \delta^{d-2}(\vecQ+\myvec{k}_T) 
\mathrm{Tr} \left[ \frac{\pquarksl}{2}
\left( \frac{2\pquark^\alpha-\gamma^\alpha\slashed{k}}{-2\pquark\cdot k} + \frac{\nb^\alpha}{k^-} \right) \right.\\&\qquad\times \frac{\slashed{\nb}}{2} \left( \left.\frac{2{\pquark}_{\alpha}-\slashed{k}\gamma_\alpha}{-2\pquark\cdot k} - \frac{\nb_\alpha}{-k^-} \right)   \right]  \mathrm{Tr} \left[ \frac{\pantisl}{2} \frac{\slashed{n}}{2} \right]   \\
=& - \frac{\alphabar}{\pi} f_\epsilon z_1 z_2 \wn^2 \delta^\prime_2
\left( \frac{z_1 q_T}{\nun}\right)^{\etan} \\&\times \frac{ (2-2\zbar+(1-\epsilon)\zbar^2) }{\zbar^{2+\etan}}
 \,.
\end{aligned} \end{equation}
Here, we use the scalar distributional identity
\begin{equation}\label{eq:plus_id_2}
    \frac{\theta(\zbar)}{\zbar^{2+\eta}}=\frac{\delta^\prime(\zbar)}{\eta}-\delta(\zbar)+ \left[         \frac{\theta(\zbar)}{\zbar}\right]_{++} + \cdots
\end{equation}
where the double-plus distribution \cite{Ebert:2018gsn} is defined in Appendix
\ref{sec:plus_distributions}. We also use the 
usual expansion
\begin{equation} \begin{aligned} \label{eq:vecPlusIdentity2}
\frac{(\mu^2)^{-\eta/2}}{(\qtsq)^{-\eta/2} } &= 
    1 - \frac{\eta}{2} \log\frac{\mu^2}{\qtsq} +  \cdots
 \,.
\end{aligned} \end{equation}
Equation \eqref{eq:T_2_7_calc} is finite as $\epsilon\to 0$. Expanding in $\etan$, we find the bare matrix element of $ T_{(2_4,0)}$
\begin{equation} \begin{aligned}\label{eq:T_2_7_bare}
\mmatn{(2_4,0)}{n}=& \frac{\alphabar}{\pi}  z_1 z_2 \delta^\prime_2
\left( 
(\delta_1+ \delta^\prime_1)\left( -\frac{2\wn^2}{\etan} + \log \frac{\nun^2}{\qtsq} \right)\right. \\
& - \left.2\left[ \frac{\theta(\zbar)}{\zbar} \right]_{++} \! + \ 2 \left[ \frac{1}{\zbar}\right]_+ - \, 1 
\vphantom{\log \frac{\QLsqr}{\qtsq \upsn^2}}\right).
\end{aligned}\end{equation}

The $1/\etan$ rapidity divergence in \eqref{eq:T_2_7_bare} is similar in form to that found in the study of NLP jet and soft functions in \cite{Moult:2018jjd, Moult:2019mog,Moult:2019vou, Paz:2009ut}. The divergence is independent of $\vecQ$ and may be absorbed through mixing of $T_{(2_4,0)}$ with the leading-power operator $T_{(0,0)}$, as in \eqn{eq:renormalization_NLP}, with
\begin{equation} \begin{aligned} \label{eq:T_2_7_offdiagonal}
Z_{(2_4,0),(0,0)}&(\omegan,\omeganb,\vecQ)  \\ &= -2 \frac{\alphabar}{\pi} \frac{\wn^2}{\etan} \, \omegan \omeganb  \left( \delta(\omegabarn) + \delta^\prime(\omegabarn) \right) \delta^\prime(\omegabarnb) \,.
\end{aligned} \end{equation}
This rapidity renormalization factor is suppressed by one power of $q_T^2$ relative to the leading term $Z_{(0,0),(0,0)}$ in \eqref{eq:T_0_cterm} since it does not contain a factor of $\delta(q_T^2)$, and so the mixing is consistent with power counting. Equivalently, the divergence may be absorbed by $O(1)$ mixing with the NLP operator
\begin{equation}\label{eq:cumul}
    \begin{aligned}
    \int d^{d-2}\myvec{p}_T  T_{(0,0)}(q^-,q^+,\myvec{p}_T) \,.
    \end{aligned}
\end{equation}
This is similar to the cumulant operators introduced in \cite{Moult:2018jjd, Moult:2019mog,Moult:2019vou, Paz:2009ut} except that \eqn{eq:cumul} has no upper cutoff $|p_T|<\Lambda$ on the integral. 

%
%
\subsubsection{Next-to-leading power example: \texorpdfstring{   $T_{(0,2_3)}$   }{  T\_(0,2\_3) }  }\label{sec:T23example}
%
%

$T_{(0,2_3)}$ provides an example of a matrix element with nontrivial $u$ dependence. Calculating its spin-averaged matrix element,
the only contribution comes from an $\nb$-sector gluon and we find 
\begin{equation} \begin{aligned} \label{eq:T_2_11_calc}
\mmatn{(0,2_3)}{\2} 
=& -2\pi g^2 C_F\, \int  
    \frac{d^d k}{(2\pi)^d} \wnb^2 \left( \frac{q_L^2}{\nunb^2}\frac{q^+ k^-}{ q^- k^+}\right)^{\etanb/2} \delta(k^2)
     \\
& \times \delta(\pquark^- - q^- - k^-)\delta(\panti^+ - q^+) \delta^{d-2}(\vecQ+\myvec{k}_T) 
\\ &\times \mathrm{Tr} \left[ \frac{\pquarksl}{2} \frac{\slashed{n}}{2} \right] \!
\mathrm{Tr} \left[ 
    \frac{1}{u} \vbar(\panti) \! \left( 
        \frac{2p_{2\alpha}-\gamma_\alpha\slashed{k}}{-2\panti\cdot k} - \frac{\nb_\alpha}{-k^-}
    \!\right)  \right.\\&\left.\qquad\times
    \frac{\slashed{n}}{2} \gamma_\nu^\perp \gamma_\mu^\perp   
    \Delta^{\alpha\mu}(k) k^\nu v(\panti) \delta\left(u+\frac{k^+}{q^+}\right)   \right] .
\end{aligned} \end{equation}
After using $\delta(u+k^+/q^+)/u = -q^+ \delta(u+k^+/q^+)/k^+$ and integrating over the gluon's phase space, the bare matrix element is 
\begin{equation} \begin{aligned} \label{eq:T_2_11_bare}
\mmatn{(0,2_3)}{\2} \!&=\!  -\frac{\alphabar}{\pi} f_\epsilon \wnb^2 z_1 z_2 
\left( \frac{ z_2 q_T }{ \nunb}\right)^{\etanb} \!\!\! \frac{1}{\zzbar^{2-\etanb}} \delta_1 \delta\!\left( u+ \frac{\zzbar}{z_2} \right)  .
\end{aligned} \end{equation}
Using distributional identities to extract the pole structure of \eqn{eq:T_2_11_bare}, we expand to find
\begin{equation} \begin{aligned} \label{eq:T_2_11_result}
\mmatn{(0,2_3)}{\2} =&  -\frac{\alphabar}{\pi}  z_1 z_2 \, \delta_1  
\left( \delta^\prime_2 
    \left( 
        \frac{\wnb^2}{\etanb} - \frac{1}{2}\log\frac{\nunb^2}{\qtsq} 
    \right) \right.
    \\ &
    + \left.\left[ \frac{\theta(\zzbar)}{\zzbar^2} \right]_{++} \! 
\right) 
\delta \left( u + \frac{\zzbar}{z_2}\right) \,.
\end{aligned} \end{equation}
where the rapidity divergence is absorbed by \eqn{eq:renormalization_NLP} with the renormalization constant
\begin{equation} \begin{aligned} \label{eq:T_2_11_diag_offdiag}
Z&_{(0,2_3),(0,0)}(\omegan,\omeganb,\vecQ,u)   \\&=-\alphabar \, \tilde{\theta}(\qtsq) \frac{\wnb^2}{\etanb} \, \omegan \omeganb \,  \delta( \omegabarn ) \delta^\prime(\omegabarnb) \delta\left(u + \frac{\omegabarnb}{\omeganb}\right) \,.
\end{aligned} \end{equation}
Note that if the operator definition of $O_2^{(2B_1)}(x,\hat{t})$ had not included the $\theta(\hat{t})$ convolution discussed in \secref{sec:hard_scale_matching}, then its $u$-space matching coefficient would instead be $C_2^{(2B1)}=1/u$, and the corresponding expression in \eqref{eq:T_2_11_bare} would contain only delta functions and single plus distributions in $\zzbar$,
\begin{equation}\begin{aligned}
   \mathcal{M}\sim& \frac{1}{\bar z_2^{1-\eta_\nb}}\delta\left(u+\frac{\bar z_2} {z_2}\right)\\&= \left(\frac{\delta(\bar z_2)}{\eta_\nb}+\left[\frac{\theta(\bar z_2)}{\bar z_2}\right]_++O(\eta_\nb)\right)\delta\left(u+\frac{\bar z_2}{ z_2}\right).
\end{aligned}\end{equation}
Multiplying this by the Wilson coefficient $\sim 1/u$ and integrating over $u$ would then give an unregulated  divergence at $\zzbar=0$. Instead, keeping the singular $1/u$ dependence in the matrix element of the operator gives the properly regulated result in \eqn{eq:T_2_11_result} and \eqn{eq:T_2_11_diag_offdiag} in terms of $\delta'$ and double-plus distributions.

Finally, we can also demonstrate here that the $\delta$ regulator does not regulate all rapidity divergences at NLP. Replacing the previous pure rapidity regulator in Eqs. \eqref{eq:T_2_11_calc} and \eqref{eq:T_2_11_bare} with the $\delta$ regulator, the same expressions read
\begin{equation} \begin{aligned} \label{eq:T_2_11_calc_delta}
\mmatn{(0,2_3)}{\2\delta} 
=& -2\pi g^2 C_F\!\! \int  
    \!\!\frac{d^d k}{(2\pi)^d} \mathrm{Tr} \left[ \frac{\pquarksl}{2} \frac{\slashed{n}}{2} \right] \delta\left(u+\frac{k^+}{q^+}\right)  \delta(k^2)
     \\
& \times \delta(\pquark^- - q^- - k^-)\delta(\panti^+ - q^+) \delta^{d-2}(\vecQ+\myvec{k}_T) 
\\ &\times  \!
\mathrm{Tr} \left[ 
    \frac{1}{u} \vbar(\panti) \! \left( 
        \frac{2p_{2\alpha}-\gamma_\alpha\slashed{k}}{-2\panti\cdot k} - \frac{\nb_\alpha}{-k^--\delta_\nb}
    \!\right)  \right.\\&\left.\qquad\times
    \frac{\slashed{n}}{2} \gamma_\nu^\perp \gamma_\mu^\perp   
    \left(g^{\alpha\mu}-\frac{n^\alpha k^\mu}{k^++\delta_\nb}\right) k^\nu v(\panti)  \right] \\
&=\!  -\frac{\alphabar}{\pi} f_\epsilon \wnb^2 z_1 z_2 
 \frac{1}{\zzbar(\zzbar+\delta_\nb/p_2^+)} \delta_1 \delta\!\left( u+ \frac{\zzbar}{z_2} \right)  
\end{aligned} \end{equation}
which contains an uncontrolled rapidity divergence when integrated over $\zzbar$. Since the unregulated divergence does not originate from a Wilson line propagator, any regulator that only modifies the definition of a Wilson line, such as the $\eta$ regulator of \cite{Chiu:2012ir}, will suffer from similar problems.

%
%
\subsubsection{One-loop results}\label{sec:one_loop_results}
%
%
As shown in previous examples, matrix elements of 
the $T_{(i,j)}$'s are rapidity divergent and require subtractions via rapidity counterterms proportional to the leading order operator $T_{(0,0)}$. The renormalization constants for the rest of the subleading 
$T_{(i,j)}$'s are found to be
\begin{equation} \begin{aligned} \label{eq:off_diagonal_counterterms1a}
Z_{(2_1,0),(0,0)}(\omegan,\omeganb,\vecQ,u) =&  \frac{\alphabar}{\pi}  \frac{\wn^2}{\etan}  \delta( \omegabarn )  \delta(\omegabarnb)\delta(u)\,,  \\
Z_{(0,2_1),(0,0)}(\omegan,\omeganb,\vecQ,u)  =& \frac{\alphabar}{\pi}  \frac{\wnb^2}{\etanb}  \delta( \omegabarn )  \delta(\omegabarnb) \delta(u)\,,   \\
\end{aligned}
\end{equation}
and
\begin{equation} \begin{aligned} \label{eq:off_diagonal_counterterms1b}
Z_{(2_2,0),(0,0)}&(\omegan,\omeganb,\vecQ,u_1,u_2)  \\&=  -\frac{\alphabar}{\pi}  \frac{\wn^2}{\etan}  \delta( \omegabarn )  \delta(\omegabarnb) \delta(u_1)  \delta(u_2)\,,   \\
Z_{(0,2_2),(0,0)}&(\omegan,\omeganb,\vecQ,u_1,u_2)  \\&= -\frac{\alphabar}{\pi}  \frac{\wnb^2}{\etanb}  \delta( \omegabarn )  \delta(\omegabarnb) \delta(u_1)  \delta(u_2)\,,   \\
\end{aligned} \end{equation}
for the operators $T_{(2_1,0)}$ through $T_{(0,2_2)}$, and
\begin{equation} \begin{aligned} \label{eq:off_diagonal_counterterms2}
Z_{(2_3,0),(0,0)}&(\omegan,\omeganb,\vecQ,u)  \\&= -\frac{\alphabar}{\pi}  \frac{\wn^2}{\etan} \, \omegan \omeganb \,  \delta^\prime(\omegabarn) \delta( \omegabarnb )
\delta\left(u + \frac{\omegabarn
}{\omegan}\right)\,, \\
Z_{(0,2_3),(0,0)}&(\omegan,\omeganb,\vecQ,u) \\&=  -\frac{\alphabar}{\pi} \frac{\wnb^2}{\etanb} \, \omegan \omeganb \,  \delta( \omegabarn )\delta^\prime(\omegabarnb)
\delta\left(u + \frac{\omegabarnb}{\omeganb}\right)\,,  \\ 
Z_{(2_4,0),(0,0)}&(\omegan,\omeganb,\vecQ)  \\&=  -2 \frac{\alphabar}{\pi}  \frac{\wn^2}{\etan} \, \omegan \omeganb \,  ( \delta(\omegabarn) + \delta^\prime(\omegabarn) ) \delta^\prime(\omegabarnb)\,, \\
Z_{(0,2_4),(0,0)}&(\omegan,\omeganb,\vecQ,u)  \\&= -2 \frac{\alphabar}{\pi} \frac{\wnb^2}{\etanb} \, \omegan \omeganb \,  ( \delta(\omegabarnb) + \delta^\prime(\omegabarnb) ) \, \delta^\prime(\omegabarn)\,,
\end{aligned} \end{equation}
for the remaining operators $T_{(2_3,0)}$ through $T_{(0,2_4)}$. 

In contrast to the leading power operator, the matrix elements of the power suppressed operators $T_{(i,j)}$ are individually rapidity divergent even when setting $\wn=\wnb=1$ and $\etan=-\etanb$. Nevertheless, these divergences cancel pairwise between $T_{(2_1,0)}$ and $T_{(0,2_1)}$, and $T_{(2_2,0)}$ and $T_{(0,2_2)}$. The divergences also cancel in the sum over the four operators in  \eqn{eq:off_diagonal_counterterms2} when weighted and integrated against their appropriate prefactor $H_{(i,j)}(\{ u \}) \kmat ijk\ell$.
The cancellation of rapidity divergences in the total rate reflects the rapidity finiteness of the total NLP cross section in SCET and is a nontrivial check on the validity of the EFT expansion. In \secref{sec:regulator_free_heuristics} we will show that this cancellation can be understood without an explicit regulator, in which case the correct treatment of overlap subtraction graphs, which vanished here when using the pure rapidity regulator, is critical. 

The hard-scale matching coefficients of all the subleading operators $T_{(i,j)}$ have the same LL anomalous dimensions \cite{Goerke:2017lei}, so these cancellations are manifestly maintained to all orders in the leading-log approximation. Since the rate must be finite beyond leading logarithms, finiteness of the theory will place constraints on rapidity divergences which are beyond the scope of this paper.

As in \eqn{eq:soft_matching}, the 
$T_{(i,j)}$ operators are matched onto a theory solely consisting of light-cone distribution operators, defined as 
\begin{equation}\begin{aligned} \label{eq:Oqdefns}
O_q(\ell^-)&=\frac{1}{2\pi} \int d\xi \, e^{-i \xi \ell^-} \bar \psi_n(\nb\xi) \frac{\slashed{\nb}}{2} W(\nb\xi, 0) \psi_n(0)\,, \\
O_{\bar q}(\ell^+)&=\frac{1}{2\pi} \int d\xi \, e^{-i \xi \ell^+} \bar \psi_\nb(0)\frac{\slashed{n}}{2}  W(0,n\xi) \psi_\nb(n \xi) \,.
\end{aligned}
\end{equation}
Since the renormalized partonic matrix element of the product of these soft theory operators is \cite{Manohar:2003vb,Collins:1981uw}
\begin{equation}\begin{aligned}\label{eq:soft_matrix_element}
&\bra{p_1^\n p_2^\nb} O_q(q^-) O_{\bar{q}}(q^+) \ket{p_1^\n p_2^\nb}
\\&=
\left( \delta_1 - \frac{\alphabar}{\epsilon} \left[ \frac{ 1+z_1^2}{\zbar}\right]_+ 
\right)
\left( \delta_2 - \frac{\alphabar}{\epsilon} \left[ \frac{ 1+z_2^2}{\zzbar}\right]_+ 
\right)+\cdots
\end{aligned}\end{equation}
and since these IR divergences are precisely reproduced in the renormalized matrix element of $T_{(0,0)}$ (see \eqn{eq:F0_bare}), the leading-power soft matching coefficient is then
\begin{equation}\begin{aligned}\label{eq:CS_LP}
C_{S,(0,0)}
&\left( z_1, z_2, \vecQ, \frac{\mu}{q_T},\frac{\nunnb}{\mu} \right)
= \delta_1 \delta_2 \deltaT \\&+  
\alphabar 
\Bigg\{ 
     -\delta_1 \delta_2\left[2\myvec{\mathcal{L}}_{1T} 
    + \myvec{\mathcal{L}}_{0T}\left( 3-2\log \frac{\nun \nunb}{\mu^2 } \right) 
     \right] 
    \\
    & + \left( \delta_2  \left[ \frac{1+z_1^2}{\zbar} \right]_+ 
    + \delta_1  \left[ \frac{1+z_2^2}{\zzbar} \right]_+ \right)  \myvec{\mathcal{L}}_{0T} 
    \\
    & + (\zbar \delta_2 +\zzbar\delta_1 - \delta_1 \delta_2 \zeta_2)\deltaT 
\Bigg\} \,.
\end{aligned}\end{equation}
This also provides the fixed order expansion of $V_{(0,0),(0,0)}$,
\begin{equation}\begin{aligned}\label{eq:V_LP}
V_{(0,0),(0,0)}
&\left( z_1, z_2, \vecQ, \frac{\mu}{q_T},\frac{q_L}{\nunnb} \right)
\\&
=\delta_1 \delta_2 \deltaT 
+2\alphabar 
\myvec{\mathcal{L}}_{0T} \log \frac{q_L^2}{\nun \nunb}+\cdots \ ,
\end{aligned}\end{equation}
where higher order terms can be generated using the running in \secref{sec:rap_running}.

At subleading power the renormalized matrix elements of $T_{(i,j)}$ begin at $\order{\alpha_s}$ and thus match onto the tree-level term $\delta(\zbar) \delta(\zzbar)$ of \eqn{eq:soft_matrix_element}.  The renormalized matrix elements of $T_{(i,j)}$ are thus equal to the soft matching coefficients $C_{S,(i,j)}$. Suppressing their scale dependence, the first four NLP soft matching coefficients are
%
\begin{equation} \begin{aligned} \label{eq:CS_N}
C_{S,(2_1,0)}( z_1, z_2, \vecQ) =& - \frac{\alphabar}{\pi}  \delta_2 \left( \left[ \frac{\theta(\zbar)}{\zbar} \right]_{+} + \frac{1}{2} \delta_1 \log \frac{\nun^2}{\qtsq} \right)\\ &\times \delta\left( u + \frac{\zbar}{z_1}\right)\,,   \\
C_{S,(0,2_1)}( z_1, z_2, \vecQ) =&-\frac{\alphabar}{\pi}  \delta_1 \left( \left[ \frac{\theta(\zzbar)}{\zzbar} \right]_{+} + \frac{1}{2} \delta_2 \log \frac{\nunb^2 }{\qtsq }\right)\\&\times \delta\left( u+ \frac{\zzbar}{z_2}\right)\,,    \\
C_{S,(2_2,0)}( z_1, z_2, \vecQ)=& \frac{\alphabar}{\pi}  \delta_2 \left( \left[ \frac{\theta(\zbar)}{\zbar} \right]_{+} + \frac{1}{2} \delta_1 \log \frac{\nun^2}{\qtsq}  \right) \\&\times\delta\left( u_1 + \frac{\zbar}{z_1}\right) \delta(u_1-u_2) \,,   \\
C_{S,(0,2_2)}( z_1, z_2, \vecQ)=&\frac{\alphabar}{\pi} \delta_1 \left( \left[ \frac{\theta(\zzbar)}{\zzbar} \right]_{+} + \frac{1}{2} \delta_2  \log \frac{\nunb^2 }{\qtsq}  \right) \\&\times\delta\left( u_2 + \frac{\zzbar}{z_2}\right) \delta(u_1-u_2)\,,    \\
\end{aligned} \end{equation}
and the remaining four matching coefficients are
\begin{equation} \begin{aligned} \label{eq:CS_12-N}
C_{S,(2_3,0)}( z_1, z_2, \vecQ) &=\frac{\alphabar}{\pi} z_1 z_2 \, \delta_2
\delta\left( u + \frac{\zbar}{z_1}\right)
\\ &\times 
\Bigg( 
    \frac{1}{2}\delta_1^\prime \log\frac{\nun^2}{\qtsq} - \left[ \frac{\theta(\zbar)}{\zbar^2} \right]_{++}  
\Bigg)\,,\\
C_{S,(0,2_3)}( z_1, z_2, \vecQ) &=  \frac{\alphabar}{\pi} z_1 z_2 \, \delta_1 \delta\left( u + \frac{\zzbar}{z_2}\right)
\\ &\times
\Bigg( \frac{1}{2}\delta^\prime_2 \log\frac{\nunb^2 }{\qtsq} - \left[ \frac{\theta(\zzbar)}{\zzbar^2} \right]_{++}  
\Bigg)\,, \\
C_{S,(2_4,0)}( z_1, z_2, \vecQ) &=\frac{\alphabar}{\pi}  z_1 z_2 \,\delta_2^\prime
\Bigg( 
    (\delta(\zbar)+\delta^\prime(\zbar))\log\frac{\nun^2}{\qtsq} \\
    & -2\left[ \frac{\theta(\zbar)}{\zbar^2} \right]_{++} \!\!+ 2\left[ \frac{\theta(\zbar)}{\zbar} \right]_{+} -1 
\Bigg)\,,  \\
C_{S,(0,2_4)}( z_1, z_2, \vecQ)&=\frac{\alphabar}{\pi}  z_1 z_2 \, \delta^\prime_1 
\Bigg( 
    (\delta(\zzbar)+\delta^\prime(\zzbar))\log\frac{\nunb^2 }{\qtsq} \\
    \!\!&-2\left[ \frac{\theta(\zzbar)}{\zzbar^2} \right]_{++} \!\!+ 2\left[ \frac{\theta(\zzbar)}{\zbar} \right]_{+} -1 
\Bigg)\,.
\end{aligned} \end{equation}

Matching QCD onto SCET at $\mu=q_L$ and $\nunnb=q_L$, these matrix elements have large logarithms of $q_L^2/\qtsq$. We will discuss the resummation of these logarithms using the rapidity renormalization group in \secref{sec:rap_running}.

%
%
\subsection{\texorpdfstring{$\Cff$}{Cff} at fixed order}\label{sec:Cff_fixed_order}
%
%

It is useful at this stage to check the fixed order results for $\Cff$ by comparing with the corresponding QCD calculation. 
At leading power, the $O(\alpha_s)$ expression for $\Cff^{(0)}$ in SCET is given by $C_{S(0,0)}$ in \eqn{eq:CS_LP} with $\nunnb=q_L$ and multiplied by the hard function $H_{(0,0)} = C_{2}^{(0)\dagger} C_{2}^{(0)}$. After integrating $d\Omega_T$, this gives the one-loop expression
\begin{equation}  \begin{aligned} \label{eq:LP_unsummed_eta}
\Cff^{(0)}&(z_1,z_2,q_L^2,\qtsq) \\&=\alphabar \Bigg\{ \delta_1 \delta_2 \delta(\qtsq) \left(  -  \log^2 \frac{\QLsqr}{\mu^2} + 3 \log\frac {\QLsqr}{\mu^2} -8 + 7\zeta_2  \right) \\
& +  \left[ \frac{1}{\qtsq} \right]_+^{\mu^2} 
\left( 
    \delta_1 \left[ \frac{1+z_2^2}{\zzbar} \right]_+
    + \delta_2 \left[ \frac{1+z_1^2}{\zbar} \right]_+
\right) \\
&  - \delta_1 \delta_2 
\left[ 
    2 
     \left[ \frac{\log\frac{\qtsq}{\mu^2}}{\qtsq} \right]_+^{\mu^2} 
    + 
     \left[ \frac{1}{\qtsq} \right]_+^{\mu^2} 
     \left( 3  - 2  \log\frac{\QLsqr}{\mu^2}\right) 
\right]  \\
&  + \delta(\qtsq)
\left( 
    \zbar \delta_2 + \zzbar \delta_1  - \zeta_2 \delta_1 \delta_2  
\right) \Bigg\}\,. 
\end{aligned} \end{equation}
At NLP, adding up the contributions from the unsummed matching coefficients in \eqn{eq:CS_N} and \eqn{eq:CS_12-N} with $\nunnb=q_L$ weighted by the corresponding coefficients $H_{(i,j)}\kmat ijk\ell$, gives the $O(\alpha_s)$ coefficient function
\begin{equation}  \begin{aligned} \label{eq:NLP_unsummed_eta}
\Cff^{\pcorrec}&(z_1,z_2,\vecQ) \\
=&\alphabar \,  z_1 z_2
\left[ \Bigvert
     \left(2\log\frac{\hat s}{\qtsq} -3\right) \delta_1^\prime \delta_2^\prime 
    \right.
    \\& 
        + \left(2\log\frac{\hat s}{ \qtsq}+1\right)                
    (\delta_1^\prime \delta_2 + \delta_1 \delta_2^\prime)+4\delta_1\delta_2 \\
    &  - 
        \delta_1^\prime  \left[ \frac{2-2\zzbar+\zzbar^2}{\zzbar^2} \right]_{++}
        - \delta_2^\prime  \left[ \frac{2-2\zbar+\zbar^2}{\zbar^2} \right]_{++}
         \\    & 
         -2 \left.\left( 
        \delta_1 \left[ \frac{1}{\zzbar^2} \right]_{++}
        + \delta_2 \left[ \frac{1}{\zbar^2} \right]_{++} 
    \right) 
\right] \,,
\end{aligned} \end{equation}
where, along with $q_L^2 = z_1 z_2 \hat{s}$, we have used the identities
\begin{equation} \begin{aligned} \label{eq:identities_for_comparison}
\left[\frac{\theta(\overbar{z})}{\overbar{z}} \right]_{++} &= \left[\frac{\theta(\overbar{z})}{\overbar{z}} \right]_{+} + \delta^\prime(\overbar{z}),
\\
\left[\theta(\overbar{z}) \right]_{++} \, &= 1 + \frac{1}{2}\delta^\prime(\overbar{z}) - \delta(\overbar{z}).
\end{aligned} \end{equation}

These results may be compared with the direct QCD calculation. $\Cff$ is determined in QCD by the partonic rate 
\begin{equation}  \begin{aligned} \label{eq:QCD_rate_definition}
\mathcal{R}_{\qcd}=& - \int\frac{d^dx}{(2\pi)^d} 
\bra{\pquark \panti} 
\bar{\psi}(x) \gamma^\mu \psi(x)  \bar{\psi}(0) \gamma_\mu \psi(0) 
\ket{\pquark \panti}\\
=&  \frac{1}{2} \int\!\frac{d^dx}{(2\pi)^d}
\bra{\pquark \panti} 
\bar{\psi}(x) \gamma^\mu \psi(0)  \bar{\psi}(0) \gamma_\mu \psi(x) 
\ket{\pquark \panti} \,.
\end{aligned} \end{equation}
The single gluon real emission contribution evaluates to
\begin{equation}  \begin{aligned} \label{eq:QCD_rate}
\mathcal{R}_\qcd^{1g} =&  \frac{\alphabar}{\pi} f_\epsilon \frac{\delta(\zbar \zzbar \hat{s} - \vecQ^2)}{\zbar \zzbar}
\left[ 
    2 - 2(\zbar + \zzbar)\right. \\& \left. + (\zbar^2 + \zzbar^2)  -\epsilon(\zbar+\zzbar)^2 
\right].
\end{aligned} \end{equation}
Expanding \eqref{eq:QCD_rate} in powers of $\qtsq/\QLsqr$,
\begin{equation}  \begin{aligned}
\mathcal{R}_\qcd^{1g} = \mathcal{R}_\qcd^{(0)1g} + \mathcal{R}_\qcd^{\pcorrec1g} + \cdots \ ,
\end{aligned} \end{equation}
is straightforward away from $\zbar=\zzbar=0$
\begin{equation}  \begin{aligned} \label{eq:QCD_n_expansion}
&\left. \mathcal{R}_\qcd^{1g} \right\vert_{\bar z1\neq 0} =\frac{\alphabar}{\pi} \frac{f_\epsilon}{\vecQ^2} 
\Bigg[ 
    \delta(\zzbar) \left( \frac{2-2\zbar+\zbar^2}{\zbar} \right)\nonumber\\
    &\qquad -2  \frac{\qtsq}{\hat{s}} \delta_2 
        \left( \frac{1}{\zbar^2} \right)
    - \frac{\qtsq}{\hat{s}} \delta^\prime_2 
        \left( \frac{2-2\zbar+\zbar^2}{\zbar^2} \right)\\&\qquad +O\left( \frac{q_T^4}{\hat{s}^2} \right)\bigg]
\end{aligned} \end{equation}
(with a similar result for $\zzbar\neq 0$), but care is required at the singular points. At leading power, $\mathcal{R}$ may be written
\begin{equation}  \begin{aligned} \label{eq:LP_generic_form_QCD}
\mathcal{R}_\qcd^{(0)1g} \!&=\! \frac{\alphabar}{\pi} \frac{f_\epsilon}{\vecQ^2} 
\!\left(\!
    A^{(0)} \delta_1 \delta_2 
    + \delta_2 \left[ f_n(\zbar) \right]_{+} 
    + \delta_1 \left[ f_\nb(\zzbar) \right]_{+} 
\!\right) .
\end{aligned} \end{equation}
where, from \eqref{eq:QCD_n_expansion},
\begin{equation}\label{eq:fnnbar}
    f_{n,\nbar}(\bar z)=  \left( \frac{2-2\bar z+\bar z^2}{\bar z} \right)
\end{equation}
and $A^{(0)}$ is determined from the integrated rate 
\begin{equation}  \begin{aligned} \label{eq:LP_Aconst}
 A  = \frac{\qtsq}{\frac{\alphabar}{\pi} f_\epsilon} \int_0^1 \!\! d\zbar d\zzbar \ \mathcal{R}_\qcd^{1g}=A^{(0)} + \frac{\qtsq}{\hat{s}} A^{\pcorrec}+ \cdots\,.
\end{aligned} \end{equation}
Similarly at NLP the rate has the general form
\begin{equation}  \begin{aligned} \label{eq:NLP_generic_form_QCD}
\mathcal{R}_\qcd^{\pcorrec1g} =& \frac{\alphabar}{\pi} \frac{f_\epsilon}{\hat{s}} 
\Bigg(
     A^{\pcorrec} \delta_1 \delta_2
    + B^{\pcorrec} \delta_1^\prime \delta_2^\prime 
    + C^{\pcorrec} (\delta_1^\prime \delta_2 + \delta_1 \delta_2^\prime)
    \\ &
    + \delta_2^\prime  \left[ g_\n(\zbar) \right]_{++}
    +  \delta_2  \left[ h_\n(\zbar) \right]_{++}
    + \delta_1^\prime  \left[ g_\nb(\zzbar) \right]_{++}\\&
    +  \delta_1  \left[ h_\nb(\zzbar) \right]_{++}
\Bigg)\,,
\end{aligned} \end{equation}
where
\begin{equation}\label{eq:NLP_ghfns}
    \begin{aligned}
g_{n,\nbar}(\bar z)&=    -  
        \left( \frac{2-2\bar z+\bar z^2}{\bar z^2} \right)
        \\
h_{n,\nbar}(\bar z)&=   
        -2 \left( \frac{1}{\bar z^2} \right) 
\end{aligned} \end{equation}
and the constants $B$ and $C$ are given by the appropriate moments of the rate,
\begin{equation}  \begin{aligned} \label{eq:NLP_BCconstants}
B  =& \frac{\qtsq}{\frac{\alphabar}{\pi} f_\epsilon} \int_0^1 d\zbar d\zzbar 
     \zbar \zzbar \mathcal{R}_\qcd^{1g}=B^{(0)}+\frac{q_T^2}{\hat{s}}B^{(2)}+\dots 
    \\
C  =& -\frac{\qtsq}{\frac{\alphabar}{\pi} f_\epsilon} \int_0^1  d\zbar d\zzbar 
    \zbar \mathcal{R}_\qcd^{1g} 
    \\=& -\frac{\qtsq}{\frac{\alphabar}{\pi} f_\epsilon} \int_0^1 \!\! d\zbar d\zzbar 
    \ \zzbar \mathcal{R}_\qcd^{1g}  
    =C^{(0)} + \frac{q_T^2}{\hat{s}}C^{(2)}+\dots 
    \,.
\end{aligned} \end{equation}
The integrals in Eqs.\ \eqref{eq:LP_Aconst} and \eqref{eq:NLP_BCconstants} give the endpoint constants 
\begin{equation}  \begin{aligned} \label{eq:QCDendpoint_ABC}
A^{(0)} &= 2\log\frac{\hat{s}}{\qtsq} -3 - \epsilon\,, \\
A^{\pcorrec} &= 4\,, \\
B^{\pcorrec} &= 2\log\frac{\hat{s}}{\qtsq} -3\,, \\
C^{\pcorrec} &= 2\log\frac{\hat{s}}{\qtsq} +1\,,
\end{aligned} \end{equation}
where we drop the $\epsilon$ dependence in the NLP terms since, unlike the LP rate, the NLP rate contains no infrared divergences stemming from a $1/\qtsq$ prefactor.

At leading power, applying \eqref{eq:vecPlusIdentityAppendixL1} and \eqref{eq:vecPlusIdentityAppendixLog} gives
\begin{equation}  \begin{aligned} \label{eq:real_rate_in_QCD}
\mathcal{R}_\qcd^{(0)1g} =& \frac{\alphabar}{\pi} \frac{f_\epsilon}{\vecQ^2} 
\left\{
    \left( 2\log\frac{\hat{s}}{\qtsq}-3-\epsilon \right) \delta_1 \delta_2\right.\\& 
    + \left.\delta_2 \left[ \frac{2-2\zbar+\zbar^2}{\zbar} \right]_+
    + \delta_1 \left[ \frac{2-2\zzbar+\zzbar^2}{\zbar} \right]_+
\right\} \\
=& \alphabar
\Bigg[ 
    \delta_1 \delta_2 \myvec{\delta}_{T} \left(\frac{2}{\epsilon^2}
    + \frac{3-2\log \frac{\hat{s}}{\mu^2 }}{\epsilon}\right) 
    \\&\! 
    + \!\left(  \myvec{\mathcal{L}}_{0T} - \frac{ \deltaT }{\epsilon} \right) \!\left( \delta_2 \bigg[ \frac{1+z_1^2}{\zbar}\bigg]_+ 
        \!\!+\delta_1 \bigg[ \frac{1+z_2^2}{\zzbar}\bigg]_+\right) 
    \\ &\! 
    -\delta_1 \delta_2\left[2\myvec{\mathcal{L}}_{1T} 
    + \myvec{\mathcal{L}}_{0T}\left( 3-2\log \frac{\hat{s}}{\mu^2 } \right) 
     \right] 
    \\ &\!
    + (\zbar \delta_2 +\zzbar\delta_1 - \delta_1 \delta_2 \zeta_2 )\deltaT 
\Bigg].
\end{aligned} \end{equation}
The LP vertex correction gives an additional contribution
\begin{equation}  \begin{aligned} \label{eq:virtual_rate_in_QCD}
\mathcal{R}_\qcd^{(0)\mathrm{virt}} 
=& \alphabar\delta_1 \delta_2 \myvec{\delta}_{T}
\Bigg[ 
    - \left(\frac{2}{\epsilon^2}
    + \frac{3-2\log \frac{\hat{s}}{\mu^2 }}{\epsilon}\right)\\ &
     -  \log^2 \frac{\QLsqr}{\mu^2} + 3 \log\frac {\QLsqr}{\mu^2} -8 + 7\zeta_2  
\Bigg] .
\end{aligned} \end{equation}
Combining the finite pieces of \eqn{eq:virtual_rate_in_QCD} and \eqn{eq:real_rate_in_QCD} and then integrating $d\Omega_T$ reproduces the SCET result for $\Cff^{(0)}$ in \eqref{eq:LP_unsummed_eta}. The remaining divergent terms are equal to the infrared divergences of the 
light-cone distribution operator matrix elements and thus cancel in the matching onto the soft theory.

QCD virtual corrections do not contribute to the NLP coefficient function at $O(\alpha_s)$, and so $\Cff^{(2)}$ is determined from Eqs.\ \eqref{eq:NLP_generic_form_QCD}-\eqref{eq:QCDendpoint_ABC}. After integrating $d\Omega_T$, this gives 
\begin{equation}  \begin{aligned} \label{eq:QCD_Cff2}
\Cff^{\pcorrec}&(z_1,z_2,q_L^2,q_T^2) 
= \alphabar z_1 z_2
\left[4\delta_1\delta_2
     +\left(2\log\frac{\hat s}{\qtsq} -3\right) \delta_1^\prime \delta_2^\prime 
    \right.
    \\& 
        + \left(2\log\frac{\hat s}{ \qtsq}+1\right)                
    (\delta_1^\prime \delta_2 + \delta_1 \delta_2^\prime) \\
    &  - 
        \left(\delta_1^\prime  \left[ \frac{2-2\zzbar+\zzbar^2}{\zzbar^2} \right]_{++}
        + \delta_2^\prime  \left[ \frac{2-2\zbar+\zbar^2}{\zbar^2} \right]_{++}
        \right) \\
    &  -2 
    \left.\left( 
        \delta_1 \left[ \frac{1}{\zzbar^2} \right]_{++}
        + \delta_2 \left[ \frac{1}{\zbar^2} \right]_{++} 
    \right) 
\right] \,,
\end{aligned} \end{equation}
in  agreement with the SCET result in \eqn{eq:NLP_unsummed_eta}. Thus, the SCET result and the expanded QCD result agree to NLP, as required.

Our fixed order results may also be compared with those obtained in \cite{Ebert:2018gsn}.\footnote{Similar results, integrated over rapidity, were presented in \cite{Cieri:2019hopc}.} In that reference, the DY rate was determined up to NLP by expanding the QCD matrix element in the $n$ collinear, $\nb$ collinear and soft limits, regulating the ensuing rapidity divergences, and combining the results. The results in that reference are also in agreement with the expanded QCD results in this section, but are presented in different variables which makes the comparison more involved. We have checked that our results are in agreement with theirs; details of this comparison are given in Appendix \ref{app:litcompare}.

%
%
\subsection{Rapidity running}\label{sec:rap_running}
%
%

Rapidity logarithms arise in this formalism as a scheme dependence in summing together the individually divergent contributions from the $n$ and $\nb$ sectors to a given matrix element. It was argued in \cite{Inglis:2020rpi} that in this formalism rapidity renormalization should be performed at the matching scale onto the 
light-cone distribution operators in order to ensure that Wilson coefficients in SCET are independent of infrared physics. 

As discussed in \cite{Inglis:2020rpi}, the rapidity regulators in the two sectors are fixed by matching at the hard scale from QCD onto SCET by the requirement that when $\mu\gg \mu_S$ the Wilson coefficients of SCET are independent of infrared energy scales of order $\mu_S$. In the rapidity regularization scheme used here, this corresponds to choosing $ \nun \nunb = q_L^2$, which, as discussed in \secref{sec:fierzed_mat_elements}, corresponds to using the same rapidity regulator in the $n$ and $\nb$ sectors, and is required for the rapidity divergences to cancel in the EFT.  The necessity of this choice can be seen from the $q_T$ dependence of the leading-power
matrix element of \eqn{eq:F0_renormalized}, which contains the term
\begin{equation}\begin{aligned}\label{eq:F0_soft_rap_term}
\bra{p_1^\n p_2^\nb} 
&T_{(0,0)}
\left( q^-, q^+, \vecQ,\nunnb\right) \ket{p_1^\n p_2^\nb}_{\mathrm{1-gluon}}
\\& = 
2 \alphabar \delta_1 \delta_2
     \myvec{\mathcal{L}}_{0T} \log \frac{\nun \nunb }{\mu^2 } +\dots \ .
\end{aligned}\end{equation}
Since physical quantities are independent of the rapidity regulator, any variation in $\nun\nunb$ in the matrix element of $T_{(0,0)}$ must be compensated by a Wilson coefficient proportional to $\myvec{\mathcal{L}}_{0T}$ in the EFT. This variation would then introduce nonanalytic dependence on the IR scale $q_T$ into the effective Lagrangian through the Wilson coefficient, which is inconsistent with factorization of hard and soft scales.

However, at the soft scale $\mu\sim q_T$ where SCET operators are matched onto 
light-cone distribution operators,  the scale $q_T$ is no longer an infrared scale in the EFT, and the Wilson coefficients are free to have nonanalytic dependence on $q_T$. The operators $T_{(i,j)}$ may therefore be run in $\nunnb$ to minimize rapidity logarithms in the matching coefficients $C_S$ in Eqs.\ \eqref{eq:CS_LP}, \eqref{eq:CS_N}, and \eqref{eq:CS_12-N}. These operators obey the RRG equation
\begin{equation}  \begin{aligned} \label{eq:gamma_kkprime_defn}
\frac{d}{d\log\nunnb} & T_{(i,j)}\left( q^-,q^+, \vecQ,\nunnb \right) 
 \\& =\sum_{k,\ell}
\left(
    \gamma^{n,\nb}_{(i,j),(k,\ell)}\rrgtimes
    T_{(k,\ell)}\right) \left( q^-,q^+, \vecQ,\nunnb \right) \,, 
\end{aligned} \end{equation}
where $\gamma^{\n,\nb}$ is the rapidity anomalous dimension for each sector, and we define the convolution $\rrgtimes$ by
\begin{equation}\begin{aligned}
(f \rrgtimes g)(\lambda_1,\lambda_2,\myvec{k}_T)& \equiv 
\int \frac{d\omegan}{\omegan}\frac{d\omeganb}{\omeganb} d^{d-2}\myvec{p}_T \\
&\!\times f(\omegan,\omeganb,\myvec{p}_T)
g\left(   \frac{\lambda_1}{\omegan},\frac{\lambda_2}{\omeganb},\myvec{k}_T-\myvec{p}_T\right).
\end{aligned}\end{equation}
The solution to \eqn{eq:gamma_kkprime_defn} can be written in the form of \eqn{eq:scheme_matching},
\begin{equation}\begin{aligned} \label{Trunning}
    &T_{(i,j)}(q^-,q^+,\vecQ,\nunnb=q_L)\\&\ =\sum_{k,\ell}\left(V_{(i,j)(k,\ell)}(q_L,\nunnb) \rrgtimes T_{(k,\ell)}(\nunnb)\right)(q^-,q^+,\vecQ) \ .
\end{aligned}\end{equation}
The explicit form of this solution to the RRG in momentum space can be found using the techniques in \cite{Ebert:2016gcn}. 

From the counterterm definitions in \eqn{eq:renormalization_LP} and \eqn{eq:renormalization_NLP} relating the bare and renormalized operators, and using the fact that the bare operators are independent of the parameters $\nunnb$ (as guaranteed by the fictional coupling $w_{\thisN,\thisNb}$), the rapidity anomalous dimensions for the operators $T_{(i,j)}$ may be calculated in terms of the renormalization constants as
\begin{equation}  \begin{aligned} \label{eq:gamma_kkprime_cterm}
\gamma^{n,\nb}_{(i,j),(k,\ell)}  =
& - \sum_{\kappa,\lambda}Z^{-1}_{(i,j),(\kappa,\lambda))}\rrgtimes \frac{d}{d\log\nunnb}  Z_{(\kappa,\lambda),(k,\ell)}  \,.
\end{aligned} \end{equation}
Here, the inverse counterterm satisfies the relation 
\begin{equation}  \begin{aligned} \label{eq:cterm_inverse_relation}
\sum_{\kappa,\lambda}&\left(Z_{(i,j)(\kappa,\lambda)}^{-1}\rrgtimes Z_{(\kappa,\lambda)(k,\ell)}\right)(\omegan,\omeganb,\vecQ)\\&\qquad  =\delta(\omegabarn) \delta(\omegabarnb) \delta(\vecQ) \delta_{ik}\delta_{j\ell} \,.
\end{aligned} \end{equation}

At leading power the rapidity anomalous dimension of $T_{(0,0)}$ is calculated from the renormalization constant in \eqn{eq:T_0_cterm}, which gives 
\begin{equation}  \begin{aligned} \label{eq:gamma_0/0}
\gamma^n_{(0,0),(0,0)} = \gamma^\nb_{(0,0),(0,0)} 
= 2\alphabar \delta(\omegabarn) \delta(\omegabarnb) \myvec{\mathcal{L}}_{0T}.
\end{aligned} \end{equation}

The leading-power operator $T_{(0,0)}$ thus obeys the RRG equation
\begin{equation}  \begin{aligned} \label{eq:RRG_LP}
&\frac{d}{d\log\nunnb}  T_{(0,0)} \left( \omegan,\omeganb, \vecQ,\nunnb \right) 
\\&=2\alphabar  \int  d^2\myvec{p}_T   \myvec{\mathcal{L}}_{0T}(\vecQ-\myvec{p}_T,\mu)  T_{(0,0)} \left( \omegan,\omeganb, \myvec{p}_T,\nunnb \right) \,,
\end{aligned}  \end{equation}
similar to the results in \cite{Inglis:2020rpi}, and with all the complications of running and scale setting of vector distributions described in \cite{Ebert:2016gcn}. Symmetrically, this RRG equation begins in the scheme $\nunnb=q_T$ where the logarithms of $T_{(0,0)}$ are minimized, and runs up to the scheme $\nunnb=q_L$ which, as we have argued, reproduces the QCD result.

At subleading power, the rapidity mixing of each $T_{(i,j)}$ with the leading order $T_{(0,0)}$ may easily be read off from Eqs. (\ref{eq:off_diagonal_counterterms1a})--(\ref{eq:off_diagonal_counterterms2}):
\begin{equation} \begin{aligned} \label{eq:D_N}
\gamma^n_{(2_1,0),(0,0)} &=\gamma^\nb_{(0,2_1),(0,0)}= - \frac{\alphabar}{\pi} \delta(\omegabarn) \delta(\omegabarnb) \,\delta(u)       \\
\gamma^n_{(2_2,0),(0,0)} &= \gamma^\nb_{(0,2_2),(0,0)}=\frac{\alphabar}{\pi}  \delta(\omegabarn) \delta(\omegabarnb) \,\delta(u_1) \delta(u_2)\,,    \\
\gamma^n_{(2_3,0),(0,0)} &= \frac{\alphabar}{\pi} 
\omegan  \omeganb \, \delta^\prime(\omegabarn) \delta(\omegabarnb) \,
\delta\left( u + \frac{\omegabarn}{\omegan}\right)\,,  \\
\gamma^\nb_{(0,2_3),(0,0)}&=\frac{\alphabar}{\pi} 
\omegan  \omeganb \, \delta(\omegabarn) \delta^\prime(\omegabarnb) \,
\delta\left( u + \frac{\omegabarnb}{\omeganb}\right)\,,  \\
\gamma^n_{(2_4,0),(0,0)}& =  2 \frac{\alphabar}{\pi} 
\omegan \omeganb \, \delta^\prime(\omegabarn) 
    \left( \delta(\omegabarnb) +  \delta^\prime(\omegabarnb) \right)\,,\\
\gamma^\nbar_{(0,2_4),(0,0)} &= 2 \frac{\alphabar}{\pi} 
\omegan \omeganb \left( \delta(\omegabarn) +  \delta^\prime(\omegabarn) \right) \delta^\prime(\omegabarnb)\,. 
\end{aligned} \end{equation}

As noted in \cite{Beneke:2019mua}, since each subleading $T_{(i,j)}$ only has a nonvanishing matrix element beginning at $\order{\alpha_s}$, calculating the complete rapidity renormalization for each  $T_{(i,j)}$ requires calculating matrix elements at $\order{\alpha_s^2}$. There will be some constraints on these rapidity anomalous dimensions because of $\mu$ independence of the final result \cite{Chiu:2012ir}, as discussed in this formalism in \cite{Inglis:2020rpi}, but the full calculation is beyond the scope of this paper and will be the subject of future work.

%
%
\section{Overlap subtractions at NLP} \label{sec:overlap_subtractions}
%
%

As discussed in \cite{Goerke:2017ioi,Inglis:2020rpi}, in this formulation of SCET it is necessary to subtract the double-counting of low-energy degrees of freedom which are simultaneously below the cutoff of both the $n$ and $\nb$ sectors, analogous to zero-bin subtraction in SCET \cite{Manohar:2006nz}.  
Rapidity logarithms in this formulation of SCET arise from the scheme dependence in summing the individually rapidity divergent diagrams in each sector and subtracting the corresponding overlap.

In this paper we have used a rapidity renormalization scheme in which overlap subtraction graphs vanish; while this is convenient for calculations, it obscures the cancellations that occur between different operators in different regions of phase space which are required to obtain a rapidity-finite result. In this section we generalize the overlap subtraction prescription to NLP and repeat the calculations without a rapidity regulator in order to explicitly show the cancellation of rapidity divergences due to the overlap subtraction, similar to what was done at LP in \cite{Inglis:2020rpi}.

At LP, the zero-bin prescription of \cite{Manohar:2006nz} has been shown to be equivalent to the nonperturbative subtraction definition of dividing the na\"\i ve matrix element by a vacuum expectation value of Wilson lines \cite{Lee:2006nr,Idilbi:2007on,Idilbi:2007tl}. This equivalence also holds for the overlap prescription of \cite{Goerke:2017ioi,Goerke:2017lei}. At subleading power, however, this simple prescription does not hold: matrix elements of the NLP operators $T_{(i,j)}$ begin at $\order{\alpha_s}$, and thus dividing by a vacuum expectation value of the form $(1+\order{\alpha_s})$ does not provide the necessary $\order{\alpha_s}$ subtraction to regulate their matrix elements. Calculations of probabilities in the effective theory therefore require a systematic way to implement the necessary overlap subtraction.
In this section we describe a simple diagram-based prescription to perform the overlap subtraction at subleading powers, and illustrate in the case of DY at NLP that it is required to obtain the correct, finite, result. This allows us to extend the LP discussion of \cite{Inglis:2020rpi} on the relationship between scheme dependence and rapidity logarithms up to NLP. We show that the previous observation in \secref{sec:one_loop_results} -- that at NLP rapidity divergences do not cancel for matrix elements of individual operators, but instead cancel between distinct operators -- occurs because different linear combinations of operators are required to reproduce the correct rate in different regions of phase space. 

Consider, for example, the process in \figref{fig:F0_1graphs} in which a gluon is produced in DY annihilation in addition to the lepton pair. In SCET this corresponds to two distinct processes in which the gluon is emitted in the $n$ sector or the $\nb$ sector. At NLP, the first receives contributions from the $T_{(2_i,0)}$ operators while the second receives contributions from the corresponding $T_{(0,2_i)}$ operators. Since in loop graphs all momenta are integrated over, the first class of operators will give nonvanishing spurious contributions in the momentum region described by the second, and vice versa. Thus, the overlap subtraction procedure at NLP necessarily
involves cancellations between different operators, and the subtraction required in order to avoid overcounting in each is found by taking the wrong limit of matrix elements in the other sector. In the symmetric process that we are examining in this paper, this may be achieved by subtracting one-half of each of the wrong limits from each sector. Schematically, we have the prescription
\begin{equation}\label{eq:overlap}
P_{\rm SCET}= P_n+P_\nbar 
- \frac{1}{2}\left(P_{n\rightarrow\nbar}+P_{\nbar\rightarrow n}\right)\,,
\end{equation}
where $P_i$ is the probability to produce a gluon in the $i$ sector and the subscripts $i\rightarrow j$ denote the wrong sector limits. 

The power counting of these subtractions follows the power counting of the limit in which the gluon is taken. An $n$-sector gluon has the scaling $k_n^-/q^- \sim O(1)$, $k_n^+/q^+ \sim O(q_T^2/q_L^2) $, while its wrong-sector limit has the scaling $k_{n\rightarrow \nb}^-/q^- \sim O(q_T^2/q_L^2)$ and $k_{n\rightarrow \nb}^+/q^+ \sim O(1)$. 
This definition of overlap subtraction ensures that probabilities in QCD are properly reproduced to the appropriate order by SCET in all regions of phase space.
This prescription is inherently perturbative, and further work is required to determine an operator definition of overlap subtraction which correctly reproduces QCD probabilities at both leading and next-to-leading power. 

In the next subsection we review the discussion of overlap subtraction at LP presented in \cite{Inglis:2020rpi} using the prescription \eqref{eq:overlap}. We then demonstrate that the same prescription may be used to calculate the NLP coefficient function $\Cff^{\pcorrec}$, and discuss the nature of the overlap subtraction in various regions of phase space.

%
%
\subsection{Overlap subtraction and scheme dependence at LP}\label{sec:LP_ambiguity}
%
%

The DY cross section at LP is determined by the spin-averaged matrix element of $T_{(0,0)}$, which takes the general form
\begin{equation}  \begin{aligned} \label{eq:LP_generic_form}
\mmat{(0,0)} = \frac{\alphabar}{\pi} \frac{f_\epsilon}{\qtsq} 
\left( 
    \amat{(0,0)} \delta_1 \delta_2 
    + \delta_2 \left[ f_n(\zbar) \right]_{+} 
    + \delta_1 \left[ f_\nb(\zzbar) \right]_{+} 
\right)
\end{aligned} \end{equation}
where as before we define $\delta_1 \equiv \delta(\zbar)$, $\delta_2\equiv\delta(\zzbar)$.
Away from the singular point $\zbar=\zzbar=0$ the unregulated $n$- and $\nb$-sector contributions to the matrix element of $T_{(0,0)}$ are determined by the graphs in \figref{fig:F0_1graphs} and their $\nb$-sector equivalents, and are given by \eqn{eq:n_graph_final} (and the corresponding expression in the $\nb$ sector) with $\omega_n=1$ and $\eta_n=0$. This immediately gives the functions $f_n(\zbar)$ and $f_\nb(\zzbar)$ in \eqref{eq:fnnbar} which describe the spectrum away from the endpoint. Since each $f_{n,\nb}$ only receives contributions from a single sector, there is no overcounting, and these expressions are finite and well-defined without a rapidity regulator.

The constant $\amat{(0,0)}$ may be most simply obtained by integrating the rate over $\zbar$ and $\zzbar$, which receives contributions from both sectors.
Adding these contributions overcounts the probability of producing a gluon that lies below the cutoff of both sectors and so must be subtracted using the overlap prescription \eqref{eq:overlap}, given by taking the wrong limit of the matrix elements in each sector. These are given by \eqn{eq:n_graph_soft_calc} with the rapidity regulator set to unity,
\begin{equation} \begin{aligned} \label{eq:heuristic_n_graph_soft}
\mmatn{(0,0)}{\1 \rightarrow \2}=& 
-2\pi g^2 C_F \int \frac{d^d k}{(2\pi)^d} \, \delta(\pquark^- - q^-)\delta(\panti^+ - q^+)\\&  \times\delta^{d-2}(\vecQ+\myvec{k}_T)   \delta(k^2) \mathrm{Tr} \left[ \frac{\pquarkslashed}{2}
\left( \frac{n^\alpha}{-k^+} + \frac{\nb^\alpha}{k^-} \right) \right.\\&\times\left. \frac{\slashed{\nb}}{2}\left( \frac{n_\alpha}{-k^+} - \frac{\nb_\alpha}{-k^-} \right)   \right]  \mathrm{Tr} \left[ \frac{\pantislashed}{2} \frac{\slashed{n}}{2} \right] + \cdots \\
 \end{aligned}  \end{equation}
and the corresponding (and identical) wrong limit $\mmatn{(0,0)}{\2 \rightarrow \1}$ of the $\nbar$ matrix element. The dots indicate terms suppressed by powers of $1/q_L$ relative to leading power, which do not contribute at LP but which will be important at NLP.
By integrating these graphs with respect to $\zbar$ and $\zzbar$ before integrating over the gluon momentum the contributions to the endpoint constant $\amat{(0,0)}$ from each sector and their wrong limit subtractions can be obtained. As discussed in \cite{Inglis:2020rpi}, because the individual graphs each have rapidity divergences, the ordering of integration is important; the sum is defined here by performing the $\zbar$, $\zzbar$, $\myvec{k}_T$, and $k^+$ integrals, leaving only a single rapidity-divergent $k^-$ integral\footnote{This is equivalent to the prescription in \cite{Chiu:2009yx} of adding the integrands together before performing any loop integrals.}
\begin{equation}\label{eq:DYambig_A}
    \begin{aligned}
    \amat{(0,0)}=&\int_0^\infty \frac{dk^-}{k^-}\left[\theta(p_1^--k^-) \amatn{(0,0)}{n}(k^-)\right.\\
    &+\theta\left(k^--\frac{\qtsq}{p_2^+}\right) \amatn{(0,0)}{\nb}(k^-) \\&-\left. \frac12 \left(\amatn{(0,0)}{\1 \rightarrow \2}(k^-)+\amatn{(0,0)}{\2 \rightarrow \1}(k^-)\right)\right]\,,
    \end{aligned}
\end{equation}
where
\begin{equation} \begin{aligned}  \label{eq:DYambig_A_explicit}
\amatn{(0,0)}{n}(k^-)&= 
2 - 2 \left( \frac{k^-}{ \pquark^- } \right)  + (1-\epsilon) \left( \frac{k^-}{\pquark^-}\right)^2\,,  \\
\amatn{(0,0)}{\nb}(k^-)&=  
 2  -2 \left(\frac{\qtsq }{ k^- \panti^+}\right) + (1-\epsilon)\left(\frac{ \qtsq }{k^- \panti^+} \right)^2 \,,  \\
 \amatn{(0,0)}{\1 \rightarrow \2}(k^-)&=\amatn{(0,0)}{\2 \rightarrow \1}(k^-)= 2\,.
 \end{aligned}
 \end{equation}
Physically, regions of phase space where $k^-\sim O(q^-)$ are properly described in the EFT by $n$-sector gluons. Regions where $k^+=k_T^2/k^-\sim O(q^+)$ give spurious contributions in the $n$ sector, producing the unphysical divergence in $\amatn{(0,0)}{n}(k^-)$ as $k^-\to 0$. Similarly, the divergence in the $\nb$ sector as $k^-\to\infty$ corresponds to the large $k^-$ region which is not properly described by the $\nb$ sector.  Both of these spurious divergent contributions are canceled by the overlap terms, leaving the finite result
\begin{equation} 
\amat{(0,0)}=2\log{ \frac{\hat{s}}{\qtsq} }-3-\epsilon \,.
\end{equation}
This is the same endpoint constant we determined from QCD in \eqref{eq:QCDendpoint_ABC}, and so we find the same LP coefficient function $\Cff^{(0)}$. Equivalently, in \eqn{eq:DYambig_A_explicit}, the constant terms in $A_n^{(0,0)}$ and $A_\nb^{(0,0)}$ are common to both sectors, and so the double-counting is removed by subtracting the overlap on the third line.

As discussed in \cite{Inglis:2020rpi}, however, the $\hat{s}$ dependence in $\amat{(0,0)}$ is actually a scheme dependence in the EFT, which allows rapidity divergences to be resummed in SCET. Since each integral represents the momentum of a distinct particle in each sector, the momentum in each integral can be independently rescaled, which changes the term in the rapidity logarithm. For example, rescaling $k^-\to k^- \zeta^2/\hat s$ in the $A_\nb$ integral of \eqn{eq:DYambig_A_explicit} gives the manifestly scheme-dependent result
\begin{equation} \label{eq:endLP}
\amat{(0,0)}(\zeta)=2\log{ \frac{\scale^2}{\qtsq} }-3-\epsilon \,.
\end{equation}
In the remainder of this section we will demonstrate a similar origin of rapidity logarithms at NLP.
%
%
%
\subsection{Overlap subtraction and scheme dependence at NLP} \label{sec:regulator_free_heuristics}
%
%

At NLP the overlap subtraction follows the same procedure as at LP, but here more terms are kept in the wrong limit expansion of each operator's matrix elements. The NLP cancellation of divergences is also slightly more involved, since rapidity divergences cancel between different operators, as may be seen in Eqs.\ (\ref{eq:off_diagonal_counterterms1a})--(\ref{eq:off_diagonal_counterterms2}). Similar cancellations between different operators in SCET were also discussed in detail in \cite{Liu:2020wbn}. There, the endpoint divergences are regulated by explicit hard cutoffs, and expressed in a refactorized form that makes obvious the cancellation  between different NLP operators contributing to the observable. Overcounting of hard regions arises from the convolutional structure of the operators with a hard cutoff and thus an ``infinity-bin" prescription, distinct from the usual zero-bin prescription, is introduced to correct for this double counting.  In this section we will show that the same overlap subtraction required to correct for overcounting in the soft region also properly regulates endpoint divergences. This uniform treatment of divergences is possible because all spurious terms have a common origin, arising from an overcounting of probabilities induced by wrong limit contributions in each individual sector. 

The operator products $T_{(2_1,0)}$ through $T_{(0,2_2)}$ come from products of scattering operators $O_{2,{(1_i})}^{\dagger\mu} O_{2,{(1_j)}}^{\nu}$ whose definitions pick out the longitudinal Lorentz structure $\nb^\mu n^\nu$ or $\nb^\nu n^\mu$, while the remaining operators $T_{(2_3,0)}$ through $T_{(0,2_4)}$, along with the leading order $T_{(0,0)}$, come from products of operators that are proportional to $g_\perp^{\mu\nu}$.
It is therefore convenient to classify each $T_{(i,j)}$ according to its Lorentz structure, as either transverse or longitudinal. 
We consider these two classes of operators in turn. 

%
%
\subsubsection{Longitudinal class}
%
%

From Eqs. \eqref{eq:off_diagonal_counterterms1a} and \eqref{eq:off_diagonal_counterterms1b}, matrix elements of $T_{(2_1,0)}$ and $T_{(0,2_1)}$ are individually rapidity divergent, but the divergences cancel in the sum (and hence in the cross section, since their Wilson coefficients are equal). The same is true for  $T_{(2_2,0)}$ and $T_{(0,2_2)}$, and in both cases the cancellation may be understood by examining the unregulated diagrams and corresponding overlaps, as in the previous section.

Taking $T_{(2_1,0)}$ as an example, its unregulated spin-averaged matrix element  is
\begin{equation} \begin{aligned} \label{eq:T_2_1_mat_elem}
\mmatn{(2_1,0)}{n} =&-2\pi g^2 C_F \int \!\!\frac{d^d k}{(2\pi)^d} 
 \delta(\pquark^- \!- q^- \!- k^-)\delta(\panti^+ - q^+)\\&\times \delta^{d-2}(\vecQ+\myvec{k}_T)  \delta(k^2)  \delta\left(u+\frac{k^-} {q^-}\right)\mathrm{Tr} \left[ \frac{\pantislashed}{2} \frac{\slashed{n}}{2} \right]\\
& \times    \mathrm{Tr} \left[ \frac{\pquarkslashed}{2} \!
\left( \frac{2\pquark^\alpha-\gamma^\alpha\slashed{k}}{-2\pquark\cdot k} + \frac{\nb^\alpha}{k^-} \right) \! \frac{\slashed{\nb}}{2}  \slashed{k}_\perp \gamma^\perp_\mu \Deltabar^{\alpha\mu}(k)   \right]
\\ =& -\frac{\alphabar}{\pi} \delta_2 \int_0^\infty \!\!\frac{dk^-}{k^-}  
p_1^- \delta(p_1^- \!- q^- \!- k^-)  \delta\left(\!u+\frac{k^-}{q^-}\!\right)
\\ =& - \frac{\alphabar}{\pi} \frac{\delta_2}{\zbar} \delta\left(u+\frac{\zbar}{z_1} \right)\,,
\end{aligned}   \end{equation}
where $\Deltabar$ is defined in Appendix \ref{sec:appendix_matrix_elements} and its wrong-sector limit is
\begin{equation} \begin{aligned} \label{eq:T_2_1_mat_elem_soft}
\mmatn{(2_1,0)}{\1 \rightarrow \2} =& 
-2\pi g^2 C_F \int  \frac{d^d k}{(2\pi)^d} \,
 \delta(\pquark^- - q^- )\delta(\panti^+ - q^+) \\&\times \delta^{d-2}(\vecQ+\myvec{k}_T)  \delta(k^2)  \delta\left(u+\frac{k^-}{q^-}\right) \mathrm{Tr} \left[ \frac{\pantislashed}{2} \frac{\slashed{n}}{2} \right]  \\
&  \times  \mathrm{Tr} \left[ \frac{\pquarkslashed}{2}
\left( \frac{\n^\alpha}{-k^+} + \frac{\nb^\alpha}{k^-} \right) \frac{\slashed{\nb}}{2} \slashed{k}_\perp \gamma^\perp_\mu \Deltabar^{\alpha\mu}(k)  \right] 
\\ =& -\frac{\alphabar}{\pi} \delta(\zbar) \delta(\zzbar) \int_0^\infty \frac{dk^-}{k^-}  
 \delta\left(u+\frac{k^-}{q^-}\right).
\end{aligned}   \end{equation} 
Away from $\zbar=\zzbar=0$ the overlap does not contribute and \eqref{eq:T_2_1_mat_elem} gives a well-defined result; however, it is rapidity divergent at $\zbar=0$. Following the LP approach, the matrix element may be written in the general form 
\begin{equation} \begin{aligned} 
\mmat{(2_1,0)} &= 
\frac{\alphabar}{\pi} \delta_2
\left( 
 \amat{(2_1,0)} \delta_1 \delta(u)
    - \left[\frac{1}{\zbar}\right]_+ \delta\left( u + \frac{\zbar}{z_1}\right) 
\right),
\end{aligned}   \end{equation}
in accordance with \eqn{eq:CS_N}. The constant $A_{(2_1,0)}$ is determined by integrating with respect to $u$, $\zbar$, and $\zzbar$, which gives 
\begin{equation}  \begin{aligned} \label{eq:T21}
\amat{(2_1,0)}=&\int_0^\infty \frac{dk^-}{k^-}
\left[\theta(p_1^--k^-) \amatn{(2_1,0)}{n}(k^-)\right. \\&-\left.\frac12\amatn{(2_1,0)}{\n \rightarrow \nb}(k^-)\right]\,,
\end{aligned}\end{equation}
where
\begin{equation}\begin{aligned}
\amatn{(2_1,0)}{n}(k^-) &= \amatn{(2_1,0)}{\n \rightarrow \nb}(k^-)= -1.
\end{aligned} \end{equation}
The integral in \eqn{eq:T21} is divergent: matrix elements of $T_{(2_1,0)}$ alone are not rapidity finite, in agreement with the result \eqref{eq:off_diagonal_counterterms1a} using the pure rapidity regular. This is to be expected, since gluons in both the $n$ and $\nb$ sectors are required to reproduce the QCD rate, and the corresponding $\nb$-sector gluon is emitted from the operator $T_{(0,2_1)}$. Including this operator and its corresponding subtraction gives 
\begin{equation} \begin{aligned} 
\mmat{(2_1,0)} &+ \mmat{(0,2_1)} = 
\frac{\alphabar}{\pi}
\left\{ \Bigvert
 \amat{2_1} \delta_1\delta_2 \delta(u)\right. \\ &\left. 
    -  \delta_2 \left[\frac{1}{\zbar}\right]_+  \delta\left( u + \frac{\zbar}{z_1}\right) 
     - \delta_1\left[\frac{1}{\zzbar}\right]_+ \delta\left( u + \frac{\zzbar}{z_2}\right) 
\right\}
\end{aligned} \end{equation}
where
\begin{equation}\begin{aligned}\label{eq:T21tot}
\amat{2_1}=&\int_0^\infty \frac{dk^-}{k^-}\left[\theta(p_1^--k^-) \amatn{(2_1,0)}{n}(k^-)\right.\\&\left. 
+\theta\left(k^--\frac{\qtsq}{p_2^+}\right) \amatn{(0,2_1)}{\nb}(k^-)\right.\\
&-\left.\frac12 \left(\amatn{(2_1,0)}{\n \rightarrow \nb}(k^-)+\amatn{(0,2_1)}{\nb \rightarrow n}(k^-)\right)\right]
  \end{aligned}\end{equation}
and 
  \begin{equation}\begin{aligned}
\amatn{(0,2_1)}{\nb}(k^-) &= \amatn{(0,2_1)}{\nb \rightarrow \n}(k^-)=
 -1 \,.
\end{aligned} \end{equation}
The integral in \eqn{eq:T21tot} is finite; as at LP, the spurious divergences from the $n$ sector as $k^-\to 0$ and the $\nb$ sector as $k^-\to\infty$ have been canceled by the overlap subtraction to give the finite result
\begin{equation}
    \amat{2_1}=-\log\frac{\hat{s}}{\qtsq} \ ,
\end{equation}
which, by a similar rescaling argument as at leading power, gives a scheme-dependent rapidity logarithm reproducing that in \eqn{eq:CS_N}.

A similar argument holds for $T_{(2_2,0)}$ and $T_{(0,2_2)}$. Explicitly, we find
\begin{equation} \begin{aligned} 
\mmat{(2_2,0)} &+ \mmat{(0,2_2)} = 
\frac{\alphabar}{\pi}
\left\{\vphantom{\frac{zzbar}{zbar}}
 \amat{2_2} \delta_1\delta_2 \delta(u_1)\right.
    \\ & +  \delta_2 \left[\frac{1}{\zbar}\right]_+\delta\left( u_1 + \frac{\zbar}{z_1}\right)
    \\ & 
     + \delta_1\left[\frac{1}{\zzbar}\right]_+ \delta\left( u_1 + \frac{\zzbar}{z_2}\right) 
\!\bigg\}\delta(u_1 - u_2) \,,
\end{aligned}   \end{equation}
where
\begin{equation}\begin{aligned}
\amatn{(2_2,0)}{n}(k^-) 
=& \amatn{(2_2,0)}{\n \rightarrow \nb}(k^-)
=\amatn{(0,2_2)}{\nb}(k^-)\\ =&
\amatn{(0,2_2)}{\nb \rightarrow \n}(k^-)
= 1\,,
\end{aligned}\end{equation}
and so
\begin{equation}
A_{2_2}=\log\frac{\hat{s}}{\qtsq}\,,
\end{equation}
again in agreement with \eqn{eq:CS_N}. The total fixed-order contribution to the cross section therefore cancels between the four longitudinal operators.

%
%
\subsubsection{Transverse class}
%
%
 
Matrix elements of the transverse class of operators $T_{(2_3,0)}$ through $T_{(0,2_4)}$ are more complicated because they originate from operator products having the same Lorentz structure as those that produce the leading-power operator $T_{(0,0)}$, and power corrections to the overlap subtraction of $T_{(0,0)}$ must also be included to achieve a rapidity-finite combination. Thus, while in the longitudinal case rapidity divergences canceled between the corresponding $n$- and $\nb$-sector operators, here they only cancel in the particular linear combination of transverse operators that contribute to the DY cross section. 

The contribution of the transverse operators to the coefficient function $\Cff^{(2)}$ is calculated from \eqn{eq:Cff_defn} and has the general form
\begin{equation}  \begin{aligned} \label{eq:trans_generic_form}
\Cff^{(2)T} =& \alphabar 
z_1z_2
\left( 
     A^{(2)}_T \delta_1 \delta_2   
    + B^{(2)}_T \delta_1^\prime \delta_2^\prime 
     +C^{(2)}_T (\delta_1^\prime \delta_2 + \delta_1 \delta_2^\prime) \right.
    \\&+ \delta_2^\prime  \left[ g_\n^T(\zbar) \right]_{++}
    +  \delta_2  \left[ h_\n^T(\zbar) \right]_{++}\\
    &\left.
    + \delta_1^\prime  \left[ g_\nb^T(\zzbar) \right]_{++}+  \delta_1  \left[ h_\nb^T(\zzbar) \right]_{++}
\right) \,.
\end{aligned} \end{equation}

Away from the endpoint $\zbar=\zzbar=0$ there are no rapidity divergences, so the contribution from each operator to $g_{n,\nb}^T$ and $h_{n,\nb}^T$ are the same as in Eqs. \eqref{eq:CS_N} and \eqref{eq:CS_12-N}. After summing and integrating over $u$'s, these combine to give the functions $g_{n,\nb}^T$, $h_{n,\nb}^T$:
\begin{equation}  \begin{aligned} \label{eq:gh_7-12}
g_{n,\nb}^{T}(\overbar{z}) &= \ -\left( \frac{2-2\overbar{z}+\overbar{z}^2}{\overbar{z}^2} \right)\,,
\\
h_{n,\nb}^{T}(\overbar{z}) &= - \frac{2}{\overbar{z}^2} \,.
\end{aligned} \end{equation}

The endpoint region is overcounted in the sum of the two sectors, and must be compensated by subtracting away half the wrong limit of each sector. In contrast with the previous cases, the power counting of the required overlap subtractions is more subtle because the overlap graphs must subtract not only logarithmic, but also linear rapidity divergences.

Consider first the various contributions to $A_T^{(2)}$, which are found by integrating unweighted matrix elements over $\{u\}$, $\zbar$, and $\zzbar$. The na\"\i ve contributions from $T_{(2_3,0)}$ and $T_{(0,2_3)}$ are calculated to be
\begin{equation}\label{eq:DYambig_A_T23}
    \begin{aligned}
    \int_0^\infty &\frac{dk^-}{k^-}\left[\vphantom{\frac{zzbar}{zzbar}}\theta(p_1^--k^-) \amatn{(2_3,0)}{n}(k^-)\right.\\&\left. +\theta\left(k^--\frac{\qtsq}{p_2^+}\right) \amatn{(0,2_3)}{\nb}(k^-)\right]\,,
\end{aligned}
\end{equation}
where 
\begin{equation}\begin{aligned}\label{eq:DYambig_A_T_explicit02}
\amatn{(2_3,0)}{n}(k^-) 
&= -2 \left( \frac{p_1^-}{k^-} \right)\,,
\\
\amatn{(0,2_3)}{\nb}(k^-) 
&=  -2 \left( \frac{k^- p_2^+}{q_T^2} \right)\,.
\end{aligned} \end{equation}
The integral in \eqref{eq:DYambig_A_T23} is rapidity divergent.
Both terms in \eqn{eq:DYambig_A_T_explicit02} are $O(1)$ in their respective correct regions, $k^-\sim O(q^-)$ in the $n$ sector and $k^+ = q_T^2/k^- \sim O(q^+)$ in the $\nb$ sector, but are enhanced and give rise to linear rapidity divergences in the regions where this correct momentum scaling is no longer valid. The contributions to $\Cff$ from these spurious regions are subtracted away by the overlap. There are two sources of overlap subtraction for $A_T^{(2)}$: the wrong limits $\amatn{(2_3,0)}{n\to\nb}$ and $\amatn{(0,2_3)}{\nb\to n}$, and also the subleading wrong limits from the leading power operator $T_{(0,0)}$. 

Expanding the Feynman diagrams of $T_{(2_3,0)}$ and $T_{(0,2_3)}$
in their wrong limits gives the same functions as in \eqn{eq:DYambig_A_T_explicit02} but are integrated over the region $0<k^-<\infty$. 
Explicitly, there are two nonvanishing terms from the wrong limit of the $T_{(2_3,0)}$ matrix element, 
\begin{equation} \begin{aligned} \label{eq:T_2_3_mat_elem_soft1}
\mmatn{(2_3,0)}{\1\to\2,\mathrm{I}} =&g^2 C_F \int \frac{d^d k}{(2\pi)^d} \,
\delta(\pquark^- - q^-)\delta(\panti^+ - q^+)\\&\times \delta^{d-2}(\vecQ+\myvec{k}_T)  2\pi \delta(k^2) \mathrm{Tr} \left[ \frac{\pantislashed}{2} \frac{\slashed{n}}{2} \right]
\\&\times
\mathrm{Tr} \left[ \frac{\pquarkslashed}{2} \frac{1}{u}k^\mu \Deltabar(k)^{\alpha\nu}
 \frac{\slashed{\nb}}{2}\gamma_\nu^\perp \gamma^\perp_\mu\right. \\&\qquad\times \left.
 \left( 
    \frac{\n_\alpha}{-k^+} - \frac{\nb_\alpha}{-k^-} 
\right)
 \delta\left(u+\frac{k^-}{q^-}\right)  \right]
\\ =& \frac{\alphabar}{\pi} \delta_1\delta_2 \int_0^\infty \frac{dk^-}{k^-}
\frac{1}{u} \delta\left(u+\frac{k^-}{q^-}\right)\,,
\end{aligned}   \end{equation}
and
\begin{equation} \begin{aligned} \label{eq:T_2_3_mat_elem_soft2}
\mmatn{(2_3,0)}{\1\to\2,\mathrm{II}} =&g^2 C_F \int \frac{d^d k}{(2\pi)^d} \,
(-k^-)\delta^\prime(\pquark^- - q^-)\delta(\panti^+ - q^+) \\ &\times \delta^{d-2}(\vecQ+\myvec{k}_T)  2\pi  \delta(k^2)\mathrm{Tr} \left[ \frac{\pantislashed}{2} \frac{\slashed{n}}{2} \right] 
\\&\times
\mathrm{Tr} \left[ \frac{\pquarkslashed}{2} \frac{1}{u}k^\mu \Deltabar(k)^{\alpha\nu}
 \frac{\slashed{\nb}}{2}\gamma_\nu^\perp \gamma^\perp_\mu\right.\\ &\qquad\times\left.
 \left( 
    \frac{\n_\alpha}{-k^+} - \frac{\nb_\alpha}{-k^-} 
\right)
 \delta\left(u+\frac{k^-}{q^-}\right)  \right] 
\\ =&- \frac{\alphabar}{\pi} \delta^\prime(\zbar) \delta(\zzbar) \int_0^\infty \frac{dk^-}{p_1^-}  
\frac{1}{u} \delta\left(u+\frac{k^-}{q^-}\right) \,.
\end{aligned}   \end{equation}
The wrong-limit expansion is truncated after the terms reach an $O(q_T^2/q_L^2)$ suppression relative to the leading-power operator in the wrong-limit momentum scaling $p_1^- \sim O(q^-)$ and $k^+\!,\,p_2^+ \sim O(q^+)$. The subtraction term in \eqn{eq:T_2_3_mat_elem_soft1} contributes to $A_T^{(2)}$ while the term in \eqn{eq:T_2_3_mat_elem_soft2} contributes to $C_T^{(2)}$.

Similarly, expanding the $\nb$- and $n$-sector graphs  of $T_{(0,0)}$ up to NLP gives the $O(1/k^-)$ term in $A^n_{(2_3,0)}$ and the $O(k^-)$ term in $A^\nb_{(0,2_3)}$ in \eqn{eq:DYambig_A_T_explicit02}, respectively, which are again integrated over all values of $k^-$. Explicitly, expanding $\mmatn{(0,0)}{n}$ gives two contributions to the subleading overlap,
\begin{equation} \begin{aligned} \label{eq:T00_NLP_subtractions}
\mmatn{(0,0)}{\1 \rightarrow \2,\mathrm{NLP}_1} &= 
-2\pi g^2 C_F \int \! \frac{d^d k}{(2\pi)^d} \, (-k^-)\delta^\prime(\pquark^- - q^-)\\&\times\delta(\panti^+ - q^+) \delta^{d-2}(\vecQ+\myvec{k}_T)  \delta(k^2) \mathrm{Tr}\left[ \frac{\pantislashed}{2} \frac{\slashed{n}}{2} \right]\\
&  \times \mathrm{Tr} \left[ \frac{\pquarkslashed}{2} \!
\left( \!\frac{n^\alpha}{-k^+} + \frac{\nb^\alpha}{k^-} \right) \!\frac{\slashed{\nb}}{2} \!\left( \frac{n_\alpha}{-k^+} - \frac{\nb_\alpha}{-k^-} \! \right)   \right]
\\&= -2 \delta^\prime(\zbar) \delta(\zzbar) \frac{\alphabar}{\pi} \int_0^\infty \frac{dk^-}{k^-} \frac{k^- p_2^+}{\qtsq}\,,
\end{aligned}\end{equation}
and
\begin{equation}\begin{aligned} \label{eq:T00_NLP_subtractions_2}
\mmatn{(0,0)}{\1 \rightarrow \2,\mathrm{NLP}_2}& = 
-2\pi g^2 C_F \int \! \frac{d^d k}{(2\pi)^d} \, \delta(\pquark^- - q^-)\delta(\panti^+ - q^+) \\ &\times \delta^{d-2}(\vecQ+\myvec{k}_T)  2\pi \delta(k^2)\mathrm{Tr}\left[ \frac{\pantislashed}{2} \frac{\slashed{n}}{2} \right] \\
&  \times \mathrm{Tr} \left[ \frac{\pquarkslashed}{2}
\left( \frac{n^\alpha}{-k^+} + \frac{\nb^\alpha}{k^-} \right) \frac{\slashed{\nb}}{2} 
\left(  \frac{ \slashed{k}_\perp\gamma^\mu_\perp}{\pquark^- k^+}\Delta_{\alpha\mu}(k) \right)   \right. 
\\&  +
\left. \frac{\pquarkslashed}{2}
\left(  \Delta^{\alpha\mu}(k) \frac{\gamma_\mu^\perp \slashed{k}_\perp}{\pquark^- k^+} \right)  \frac{\slashed{\nb}}{2} \left( \frac{n_\alpha}{-k^+} - \frac{\nb_\alpha}{-k^-} \right)   \right]
\\&= -2\frac{\alphabar}{\pi} \delta(\zbar) \delta(\zzbar) \int_0^\infty \frac{dk^-}{k^-} \frac{k^- p_2^+}{\qtsq}\,,
 \end{aligned}  \end{equation}
the first of which comes from higher corrections to the momentum-conserving delta function, while the second comes from higher corrections to the quark propagator expansions. Only the second term contributes here; the first contributes to $C_T^{(2)}$.

Putting these together, we obtain the expression for $A_T^{(2)}$,
\begin{equation}\label{eq:DYambig_A_T}
    \begin{aligned}
    A_T^{(2)}=&\int_0^\infty \frac{dk^-}{k^-}\left[\theta(p_1^--k^-) \amatn{(2_3,0)}{n}(k^-)\right.\\ &\left.+\theta\left(k^--\frac{\qtsq}{ p_2^+}\right) \amatn{(0,2_3)}{\nb}(k^-)\right.
    \\ &  
    -\frac12\left(
            \amatn{(0,0)}{\1 \rightarrow \2 ,\NLP}(k^-)
            +\amatn{(0,0)}{\2 \rightarrow \1 ,\NLP}(k^-)
    \right)
    \\ &\left.  
    -\frac12\left(
            \amatn{(2_3,0)}{\1 \rightarrow \2}(k^-)
            +\amatn{(0,2_3)}{\2 \rightarrow \1}(k^-)
    \right)\right]\,,
    \end{aligned}
\end{equation}
where the contributions from $T_{(2_4,0)}$ and $T_{(0,2_4)}$ all vanish, and explicitly
\begin{equation}\begin{aligned}\label{eq:DYambig_A_T_explicit}
\amatn{(2_3,0)}{\1 \rightarrow \2}(k^-) 
=& \amatn{(0,0)}{\2 \rightarrow \1 ,\NLP}(k^-)  
= -2 \left( \frac{p_1^-}{k^-} \right)\,,
\\
\amatn{(0,2_3)}{\2 \rightarrow \1}(k^-) 
=& \amatn{(0,0)}{\1 \rightarrow \2 ,\NLP}(k^-) 
=  -2 \left( \frac{k^- p_2^+}{q_T^2} \right) \,.
\end{aligned} \end{equation}
This gives the finite result
\begin{equation} \label{eq:A_T_2}
A_T^{\pcorrec} = 4 \,.
\end{equation}

Next consider the contributions to the endpoint constant $C_T^{(2)}$, which are obtained by integrating the various matrix elements weighted with $\zbar$ (or equivalently $\zzbar$). The na\"\i ve contributions to the $\zbar$ moment give
\begin{equation}\label{eq:DYambig_C_T2324}
    \begin{aligned}
\int_0^\infty \frac{dk^-}{k^-}&\left[\theta(p_1^--k^-) \cmatn{(2_3,0)}{n}(k^-)\right.\\&\left. +\theta\left(k^--\frac{\qtsq}{p_2^+}\right) \cmatn{(0,2_4)}{\nb}(k^-)\right]\,,
    \end{aligned}
\end{equation}
where 
\begin{equation}\begin{aligned}\label{eq:DYambig_C_T_explicitnaive}
\cmatn{(2_3,0)}{n}(k^-) 
&= 2 \,,
\\
\cmatn{(0,2_4)}{\nb}(k^-) &= -2 \left( \frac{k^- p_2^+}{q_T^2}\right) + 2 
                                    - \left( \frac{q_T^2}{k^- p_2^+} \right). 
\end{aligned} \end{equation}
This is again is rapidity divergent: $\cmatn{(2_3,0)}{n}(k^-)$ gives a logarithmically divergent contribution as $k^-\to 0$, while $\cmatn{(0,2_4)}{\nb}(k^-)$ gives contributions that are both logarithmically and linearly divergent as $k^-\to\infty$. As with $A_T^{(2)}$, taking the wrong limit of the Feynman diagrams contributing to \eqn{eq:DYambig_C_T_explicitnaive} gives the $k^-\to 0$ and $k^-\to\infty$ expansions of these terms. For example, the wrong-limit expansion of $\mmatn{(0,2_4)}{\nb}$ gives three terms which correspond almost exactly to the overlaps of $T_{(0,0)}$ in Eqs. (\ref{eq:heuristic_n_graph_soft}), (\ref{eq:T00_NLP_subtractions}), and (\ref{eq:T00_NLP_subtractions_2}), except they have a different momentum-conserving delta function structure. These give the contributions
\begin{equation}\begin{aligned}\label{eq:DYambig_C_T_explicit}
\cmatn{(0,2_4)}{\2 \rightarrow \1}(k^-) =&  -2 \left( \frac{k^- p_2^+}{q_T^2}\right) + 2 \ ,
\end{aligned} \end{equation}
while from the $T_{(2_3,0)}$ overlap given in  \eqn{eq:T_2_3_mat_elem_soft2} is the contribution
\begin{equation}\begin{aligned}\label{eq:DYambig_C_T_explicit_overlap}
\cmatn{(2_3,0)}{\1 \rightarrow \2}(k^-)   
=  2 \,.
\end{aligned} \end{equation}
The overlap term from $T_{(0,2_4)}$ contains two terms: the leading term is proportional to $k^-$ and cancels a linear rapidity divergence, while the $O(1)$ term contributes to the cancellation of a logarithmic divergence.

We also have the contribution from the NLP overlap of $T_{(0,0)}$ in \eqn{eq:T00_NLP_subtractions},
\begin{equation}
\cmat{(0,0)}^{\1 \rightarrow \2,\NLP}(k^-) =  -2 \left( \frac{k^- p_2^+}{q_T^2} \right) \ ,
\end{equation}
with the sum of all contributions giving the result
\begin{equation}\label{eq:DYambig_C_T}
    \begin{aligned}
    C_T^{(2)}=&\int_0^\infty \frac{dk^-}{k^-}\left[\theta(p_1^--k^-) \cmatn{(2_3,0)}{n}(k^-)\right.\\&\left. +\theta\left(k^--\frac{\qtsq}{ p_2^+}\right) \cmatn{(0,2_4)}{\nb}(k^-)\right.
    \\ & 
    -\frac12\left(
            \cmatn{(0,0)}{\1 \rightarrow \2,\NLP)}(k^-)
    \right)
    \\ &\left.  
    -\frac12\left(
            \cmatn{(2_3,0)}{\1 \rightarrow \2}(k^-)
            +\cmatn{(0,2_4)}{\2 \rightarrow \1}(k^-)
    \right)\right]\\
    =&  2\log\frac{\hat{s}}{\qtsq} + 1 \,.
    \end{aligned}
\end{equation}
Once again there is a precise interplay between na\"\i ve matrix elements and overlap subtractions required to obtain the same finite result as using the pure rapidity regulator. 
Rescaling the integrals for
$\cmatn{(0,2_4)}{\2}(k^-)$, $\cmatn{(0,2_4)}{\2 \rightarrow \1}(k^-)$, and $\cmatn{(0,0)}{\1 \rightarrow \2,\NLP}(k^-)$ as $k^- \rightarrow k^- \zeta^2/\hat{s} $ replaces the $\hat{s}$ dependence in the result of \eqn{eq:DYambig_C_T} with $\zeta^2$ scheme dependence. This correlation between the rescaling of individual integrals is necessary to maintain a finite result, and is a general feature of power-law divergences.

Finally, the endpoint constant $B_T$ is found by weighting the integrals by $\zbar \zzbar$, giving
\begin{equation}\label{eq:DYambig_B_T} 
\begin{aligned}
    B_T^{(2)}=&\int_0^\infty \frac{dk^-}{k^-}\left[\theta(p_1^--k^-) \bmatn{(2_4,0)}{n}(k^-)\right.\\&\left. +\theta\left(k^--\frac{\qtsq}{ p_2^+}\right) \bmatn{(0,2_4)}{\nb}(k^-)\right.
    \\ &\left.  
    -\frac12\left(
            \bmatn{(2_4,0)}{\1 \rightarrow \2}(k^-)
            +\bmatn{(0,2_4)}{\2 \rightarrow \1}(k^-)
    \right)\right]
    \end{aligned}
\end{equation}
where explicitly
\begin{equation}\begin{aligned}\label{eq:DYambig_B_T_explicit}
\bmatn{(2_4,0)}{n}(k^-) 
&= 2  - 2 \left( \frac{k^-}{p_1^-}\right) + \left( \frac{k^-}{p_1^-}\right)^2\,,
\\
\bmatn{(0,2_4)}{\nb}(k^-) &=  2  - 2 \left( \frac{\qtsq}{k^- p_2^+}\right) 
                                    + \left( \frac{\qtsq}{k^- p_2^+}\right)^2 \,,
\end{aligned} \end{equation}
and the overlap terms cancel just the logarithmic divergences,
\begin{equation}
\bmatn{(2_4,0)}{\1 \rightarrow \2}(k^-) = \bmatn{(0,2_4)}{\2 \rightarrow \1}(k^-) = 2 \,.
\end{equation}
This gives
\begin{equation} \begin{aligned} \label{eq:DYambig_B_T_final}
B_T^{(2)} &= 2\log\frac{\hat{s}}{\qtsq} - 3 \ ,
\end{aligned}\end{equation}
where rescaling the second line of \eqn{eq:DYambig_B_T_explicit} as $k^- \rightarrow k^- \zeta^2/\hat{s} $ replaces the $\hat{s}$ dependence in \eqn{eq:DYambig_B_T_final} with $\zeta^2$ scheme dependence. 

This concludes the calculation of all the endpoint constants $A_{L,T}$, $B_{L,T}$, and $C_{L,T}$. In each case, these constants agree with those of QCD as calculated in \secref{sec:Cff_fixed_order}. We have thus demonstrated an overlap subtraction prescription that allows us to properly calculate probabilities at NLP without an explicit rapidity regulator, providing a nontrivial cross-check of our results using different rapidity regularization schemes.

%
%
\section{Conclusion}\label{sec:conc}
%
%

In this paper we have shown that factorization of the Drell-Yan production cross section into hard matching coefficients, rapidity evolution factors, soft matching coefficients and PDFs occurs naturally in a formulation of SCET in which the low energy degrees of freedom are not separated into distinct fields for each mode relevant to the process. 
The DY rate is given by the matrix element of the nonlocal product of two external currents in SCET. Usually in SCET observables are factorized into jet and soft factors which are separately renormalized and run to the appropriate scales; here, the EFT is first run in $\mu$ down to the soft matching scale $\mu\sim q_T$, at which point the product of currents is renormalized in rapidity space. After resumming the rapidity logs at the soft matching scale, the 
operator products are then matched onto 
a product of light-cone distribution operators, whose hadronic matrix elements are the usual PDFs.  At $O(\alpha_s)$, our EFT cross section reproduces the fixed-order QCD cross section at NLP, as well as the equivalent fixed-order cross section calculated using the pure rapidity regulator in \cite{Ebert:2018gsn}. Off-diagonal rapidity anomalous dimensions were calculated and rapidity divergences were shown to cancel in the cross section. The resummation of rapidity logarithms at NLP requires the complete rapidity anomalous dimension matrix for the subleading operators $T_{(i,j)}$, which is beyond the scope of this paper, and will be the subject of future work.

The factorization and resummation of the DY process is particularly simple in this approach: it does not depend on proving factorization at a given order in the SCET expansion or in the leading-log approximation, but instead is a straightforward consequence of the usual EFT approach of matching and running.  By not explicitly factorizing modes in the Lagrangian, the complication of power corrections coupling different modes in the Lagrangian is avoided, as is the necessity to refactorize the result to make individual jet and soft functions well-defined. Divergences analogous to the endpoint divergences arising at NLP in other approaches arise, but are regulated by the rapidity regulator and systematically canceled by the same overlap subtraction procedure required to avoid double counting at leading power.

Rapidity divergences were considered in detail, and the cancellation of rapidity divergences in the rate was shown in two ways. Using the pure rapidity regulator, it was shown that all rapidity poles canceled between the different linear combinations of subleading operators arising in the expression for the differential rate, as was found in previous analyses \cite{Moult:2019vou,Liu:2019oav,Liu:2020tzd,Liu:2020wbn}. In \secref{sec:overlap_subtractions} it was shown that even without an explicit rapidity regulator, rapidity divergences in the DY cross section cancel between particular linear combinations of operators, and that these linear combinations could be understood by requiring that SCET reproduce the correct differential rate in different regions of phase space. A consistent treatment of subleading overlap subtractions from the leading order operator was shown to be necessary for this cancellation.

\section*{Acknowledgements}

We thank Christian Bauer for useful comments on the manuscript. This work was supported in part by the National Science and Engineering Research Council of Canada.

\appendix
%
%
\section{Summary of matrix elements of hard scattering operators} \label{sec:appendix_matrix_elements}
%
%

In the following equations we list all relevant $u$-space matrix elements of scattering operators which contribute to the quark-induced DY process through the emission of an $n$-sector gluon. We use the soft-scale matching kinematics $p_1^+ = p_{1\perp} = 0 = p_2^- = p_{2\perp}$, and define the noncommon factor $\mathcal{A}_{n,\nb}^{(i)}$ of these matrix elements through the relation
\begin{equation}\begin{aligned} \label{eq:Amunu_defn}
\int \! \frac{d^dx}{2(2\pi)^d} e^{-iq\cdot x} &\bra{ k^{n,\nb} } O_2^{(i)\mu}(x,\{u\}) \ket{p_1^{n} p_2^{\nb}} \\ & \equiv g \vbar(p_2^\nb)  T^{a}\mathcal{A}_{n,\nb}^{(i)} u(p_1^n)\epsilon^\ast_\nu \,,
\end{aligned}\end{equation}
where $\mathcal{A}$ is tensor valued with implied Lorentz indices $\mu$ and $\nu$. We find for the $n$-gluon emissions
\begin{equation}\begin{aligned} \label{eq:Amunu_n}
\mathcal{A}^{(0)}_\n &= - \Phat_n \gamma^\mu \Phat_n \left(\frac{2 \pquark^\nu - \slashed{k}\gamma^\nu}{-2 \pquark \cdot k} - \frac{\nb^\nu}{-k^-} \right)  \delta_n^- \delta_n^+ \deltaP 
\\
\mathcal{A}^{(1_{\perp\1})}_\n &= -\Phat_n \gamma^\mu \frac{\slashed{\etabar}}{2}  \slashed{k}_\perp \left(\frac{2 \pquark^\nu - \slashed{k}\gamma^\nu}{-2 \pquark \cdot k} - \frac{\nb^\nu}{-k^-} \right) \delta_n^- \delta_n^+ \deltaP\,,
\\
\mathcal{A}^{(1A_1)}_\n &=\gamma_\alpha^\perp \frac{\slashed{\eta}}{2}\gamma^\mu\Phat_n \Deltabar^{\nu\alpha}(k) \delta_n^- \delta_n^+ \deltaP \delta(u+\hat{k}^-)\,,
\\
\mathcal{A}^{(1A_2)}_\n &= - \Phat_n \gamma^\mu \frac{\slashed{\etabar}}{2} \gamma_\alpha^\perp  \Deltabar^{\nu\alpha}(k) \delta_n^- \delta_n^+ \deltaP \delta(u+\hat{k}^-)\,,
\\
\mathcal{A}^{(2\delta^+)}_\n &= q^+ q^- \Phat_n \gamma^\mu \Phat_n \! \left(\frac{2 \pquark^\nu - \slashed{k}\gamma^\nu}{-2 \pquark \cdot k} -  \frac{\nb^\nu}{-k^-} \! \right)\!k^+ \delta_n^- \delta_n^{+\prime} \deltaP,
\\
\mathcal{A}^{(2A_1)}_\n &=-\frac{1}{u}  \gamma_\alpha^\perp \gamma_\beta^\perp \Phat_n \gamma^\mu \Phat_n  \Deltabar^{\nu\alpha}(k) k^\beta \delta_n^- \delta_n^+ \deltaP \delta(u+\hat{k}^-)\,,
\end{aligned}\end{equation}
while for the $\nb$-gluon emissions we find
\begin{equation}\begin{aligned} \label{eq:Amunu_nbar}
\mathcal{A}^{(0)}_\nb &=  \left(\frac{2 \panti^\nu - \gamma^\nu\slashed{k}}{-2 \panti \cdot k} - \frac{n^\nu}{-k^+} \right) \Phat_n \gamma^\mu \Phat_n   \delta_\nb^- \delta_\nb^+ \deltaP \,,
\\
\mathcal{A}^{(1_{\perp\2})}_\nb &= \left(\frac{2 \panti^\nu - \gamma^\nu\slashed{k}}{-2 \panti \cdot k} - \frac{n^\nu}{-k^+} \right) \slashed{k}_\perp  \frac{\slashed{\eta}}{2} \gamma^\mu \Phat_n  \delta_\nb^- \delta_\nb^+ \deltaP\,,
\\
\mathcal{A}^{(1B_1)}_\nb &= - \Phat_n \gamma^\mu \frac{\slashed{\etabar}}{2} \gamma_\alpha^\perp  \Delta^{\nu\alpha}(k) \delta_\nb^- \delta_\nb^+ \deltaP \delta(u+\hat{k}^+)\,,
\\
\mathcal{A}^{(1B_2)}_\nb &= \gamma_\alpha^\perp \frac{\slashed{\eta}}{2}\gamma^\mu\Phat_n \Delta^{\nu\alpha}(k) \delta_\nb^- \delta_\nb^+ \deltaP \delta(u+\hat{k}^+)\,,
\\
\mathcal{A}^{(2\delta^-)}_\nb &= \! q^+ q^- \!\!\left( \frac{n^\nu}{-k^+} - \frac{2 \panti^\nu - \gamma^\nu\slashed{k}}{-2 \panti \cdot k}\right) \Phat_n \gamma^\mu \Phat_n   \, k^- \delta_\nb^{-\prime} \delta_\nb^+ \deltaP,
\\
\mathcal{A}^{(2B_1)}_\nb &=\frac{1}{u}   \Phat_n \gamma^\mu \Phat_n \gamma_\alpha^\perp \gamma_\beta^\perp k^\alpha \Deltabar^{\nu\beta}(k)  \delta_\nb^- \delta_\nb^+ \deltaP \delta(u+\hat{k}^+)\,.
\end{aligned}\end{equation}

The one-gluon matrix elements of the scattering operators defined in Eqs. \eqref{eq:Amunu_n} and \eqref{eq:Amunu_nbar} use the following definitions:
\begin{equation} \begin{aligned} \label{eq:DeltabarDefn}
\Deltabar^{\alpha\mu}(k) = g^{\alpha\mu} - \frac{\nb^\alpha k^\mu}{\nb\cdot k} \ , \quad \Delta^{\alpha\mu}(k) = g^{\alpha\mu} - \frac{n^\alpha k^\mu}{n\cdot k} \,.
\end{aligned} \end{equation}
These are common structures associated with the covariant derivative. We also define the dimensionless quantities
\begin{equation} \begin{aligned} \label{eq:HatsDefn}
\hat{\ell}^- = \frac{\ell^-}{q^-} , \quad  \hat{\ell}^+ = \frac{\ell^+}{q^+} 
\end{aligned} \end{equation}
and we have we used the shorthand notation $\delta_n^- = \delta(\pquark^- - k^- - q^-)$,
$\delta_n^+=\delta(\panti^+ - q^+)$, $\delta_\nb^- = \delta(\pquark^- - q^-)$,  $\delta_\nb^+=\delta(\panti^+ -k^+ - q^+)$, and $\deltaP=\delta^{(d-2)}( k_\perp + q_\perp)$.

There are additional operators that are present from the hard-scale matching \cite{Freedman:2011kj,Freedman:2013vya,Freedman:2014uta,Goerke:2017lei,Feige:2017zci}, but that do not contribute to the quark-initiated DY process to the order at which we are working. Up to a $1/q_L^2$ suppression, these include an operator with two perpendicular derivatives
\begin{equation}
O_2^{(2_{\perp\perp})\mu}(x)= [ i\partial^\alpha \bar{\chi}_\nb(\xnb)]  \gamma_\alpha^\perp \frac{\slashed{\eta}}{2} \gamma^\mu  \frac{\slashed{\etabar}}{2} \gamma_\beta^\perp [i\partial^\beta  \chi_{n}(\xn)] \ ,
\end{equation}
the $A$-type operators
\begin{equation} \begin{aligned} \label{eq:AtypeDefnsApp}
O_2^{(2A_2)\mu}(x,\hat{t}\,)&=  2\pi i\,\theta(\hat{t}) \otimes [ \bar{\chi}_\nb(x)  ]
 \gamma_\alpha^\perp \frac{\slashed{\eta}}{2} \gamma^\mu  \frac{\slashed{\etabar}}{2} \gamma_\beta^\perp \\ 
&\times [\mathcal{B}_n^{\dagger\alpha\beta}(x) \chi_{n}(x-\nb t)]\\
O_2^{(2A_3)\mu}(x,\hat{t}\,)&=  2\pi i\,\theta(\hat{t}) \otimes [\bar{\chi}_\nb (x)  ]
\gamma_\beta^\perp \frac{\slashed{\eta}}{2} \gamma^\mu  \frac{\slashed{\etabar}}{2} \gamma_\alpha^\perp \\
&\times [ i\partial^\alpha  \mathcal{B}_n^{\dagger\beta}(x) \chi_{n}(x-\nb t)] \\
O_2^{(2A_4)\mu}(x,\hat{t}\,)&=    -2\pi i\,\theta(\hat{t}) \otimes [i\partial^\beta  \bar{\chi}_\nb (x-n t)  ]
 \gamma_\beta^\perp \frac{\slashed{\eta}}{2} \gamma^\mu  \frac{\slashed{\etabar}}{2} \gamma_\alpha^\perp \\ 
&\times  [\mathcal{B}_n^{\dagger\alpha}(x)\chi_{n}(x)] \\
O_2^{(2A_5)\mu}(x,\hat{t}\,)&=  2\pi i \,\theta(\hat{t}) \otimes [i\partial^\alpha \bar{\chi}_{\nb}(x) ]
\gamma^\mu \{ \gamma_\alpha^\perp , \gamma_\beta^\perp \} \\ 
&\times [ \mathcal{B}_n^{\dagger\beta}(x-\nb t) \chi_{\n}(x)] \ ,
\end{aligned} \end{equation}
and the corresponding $B$-type operators
\begin{equation} \begin{aligned} \label{eq:BtypeDefnsApp}
O_2^{(2B_2)\mu}(x,\hat{t}\,)&=  2\pi i\,\theta(\hat{t}) \otimes [ \bar{\chi}_\nb(x-n t) \mathcal{B}_\nb^{\alpha\beta}(x)  ] \\&\times
 \gamma_\alpha^\perp \frac{\slashed{\eta}}{2} \gamma^\mu  \frac{\slashed{\etabar}}{2} \gamma_\beta^\perp 
 [ \chi_{n}(x)]\\
O_2^{(2B_3)\mu}(x,\hat{t}\,)&=  2\pi i\,\theta(\hat{t}) \otimes [ i\partial^\alpha   \bar{\chi}_\nb (x-n t) \mathcal{B}_\nb^{\beta}(x) ]
\\&\times
\gamma_\alpha^\perp \frac{\slashed{\eta}}{2} \gamma^\mu  \frac{\slashed{\etabar}}{2} \gamma_\beta^\perp 
[  \chi_{n}(x)] \\
O_2^{(2B_4)\mu}(x,\hat{t}\,)&=  -2\pi i\theta(\hat{t})\otimes  [ \bar{\chi}_\nb (x) \mathcal{B}_\nb^{\alpha}(x-\nb t)  ]\\&\times
 \gamma_\alpha^\perp \frac{\slashed{\eta}}{2} \gamma^\mu  \frac{\slashed{\etabar}}{2} \gamma_\beta^\perp 
  [i\partial^\beta  \chi_{n}(x)] \\
O_2^{(2B_5)\mu}(x,\hat{t}\,)&=  2\pi i \,\theta(\hat{t}) \otimes [ \bar{\chi}_{\nb}(x) \mathcal{B}_\nb^{\alpha}(x-\nb t) ]\\&\times
\gamma^\mu \{ \gamma_\alpha^\perp , \gamma_\beta^\perp \}  
[ i\partial^\beta \chi_{n}(x)] \ .
\end{aligned} \end{equation}
There are also the $C$-type operators, which are only relevant for gluon-induced Drell-Yan
\begin{equation} \begin{split} \label{eq:1a11a2Defnsapp}
O_2^{(1C_1)\mu}(x,\hat{t}\,)&=  -2\pi i\theta(\hat{t}) \otimes [\mathcal{B}^{\alpha cc^\prime}_n(x)] 
\\&\times[ \bar{\chi}_\nb^c(x) \gamma^\mu \frac{\slashed{\eta}}{2} \gamma_\alpha^\perp   \chi^{c^\prime}_\nb(x-nt) ]    \\
O_2^{(1C_2)\mu}(x,\hat{t}\,)&=  2\pi i\theta(\hat{t}) \otimes [\mathcal{B}^{\alpha cc^\prime}_n(x)]\\&\times [ \bar{\chi}_\nb^c(x-nt)  \gamma_\alpha^\perp   \frac{\slashed{\eta}}{2}  \gamma^\mu \chi^{c^\prime}_\nb(x) ] \ .
\end{split} \end{equation}

\section{Plus distribution identities} \label{sec:plus_distributions}
\subsection{Single variable plus distributions} \label{sec:single_plus_distributions}

The familiar plus distribution may be written as
\begin{equation} \begin{aligned}
[ \theta(x) f(x)]_+ =& \lim_{\beta\rightarrow 0} \left[F(\beta)-F(1)\right] \delta(x-\beta) \\&+ \theta(x-\beta) f(x)\,,
\end{aligned} \end{equation}
where $f(x)=dF(x)/dx$,
and has the properties
\begin{equation} \begin{aligned}
\int_0^1\! dx\, [ \theta(x) f(x)]_+ &= 0 \\
[ \theta(x) f(x)]_+ &= f(x) \ ,\quad x>0 \,.
\end{aligned} \end{equation}
Rearranging these equations gives the differential relation
\begin{equation} \begin{aligned} \label{eq:x_diff_relation}
\frac{d}{dx} \left[ \theta(x) F(x) \right]= \left[ \theta(x) \frac{dF(x)}{dx} \right]_+ + \delta(x) F(1)\,
\end{aligned} \end{equation}
which is useful for expanding rapidity divergent integrals in terms of plus functions. For example, taking $f(x)=x^{-1-\eta}$ gives $F(x)=-x^{-\eta}/\eta$, and so
\begin{equation}
\frac{d}{dx}\left[\frac{-\theta(x)}{\eta x^\eta}\right]=\frac{\theta(x)}{ x^{1+\eta}}-\frac{\delta(x)}{\eta x^\eta}=\left[\frac{\theta(x)}{ x^{1+\eta}}\right]_+ -\frac{\delta(x)}{\eta}.
\end{equation}
The factor of $\delta(x) x^{-\eta}$ vanishes by analytic continuation, so expanding about $\eta=0$ gives
\begin{equation}
\begin{aligned}
\frac{\theta(x)}{x^{1+\eta}}&= -\frac{\delta(x)}{\eta}+\left[ \frac{\theta(x)}{x} - \eta \frac{\log(x)\theta(x)}{x}  + \cdots \right]_+\\
&= -\frac{\delta(x)}{\eta}+\left[ \frac{\theta(x)}{x}\right]_+ - \eta \left[\frac{\log(x)\theta(x)}{x}\right]_++\dots\,.
\end{aligned}
\end{equation}

Matrix elements at next-to-leading power involve higher-order poles that are more singular than the usual plus distributions. As in \cite{Ebert:2018gsn}, we define double-plus distributions that satisfy
\begin{equation} \begin{aligned}
\int_0^1\! dx\, [ \theta(x) f(x)]_{++} &= 0\,, \\
\int_0^1\! dx\, x [ \theta(x) f(x)]_{++} &= 0\,, \\
[ \theta(x) f(x)]_{++} &= f(x) \ ,\quad x>0.
\end{aligned} \end{equation}
They are related to the single-plus distributions by
\begin {equation}\begin {aligned}\label{eq:single_double_translation}
[\theta (x) f(x)]_{++} &- [\theta (x) f (x)]_{+} 
\\ &
= \lim_{\beta\to0}\delta^\prime (x - \beta)
\int_\beta^1 \!dy\, (y - \beta) f (y)\,.
\end {aligned}\end {equation}
For example, taking $f(x)=x^{-2-\eta} $, then $F(x)=-x^{-1-\eta}/(1+\eta)$, and we obtain
\begin{equation}
\begin{aligned}
\frac{\theta(x)}{x^{2+\eta}}&=\left[\frac{\theta(x)}{ x^{2+\eta}}\right]_+-\frac{\delta(x)}{1+\eta}.
\end{aligned}
\end{equation}
Since $\left[\frac{\theta(x)}{x^2}\right]_+$ is not well-defined, we convert to a double-plus distribution before expanding in $\eta$,
\begin{equation}\begin{aligned}
 \left[\frac{\theta(x)}{x^{2+\eta}}\right]_+&=\left[ \frac{\theta(x)}{x^{2+\eta}}\right]_{++} - \delta^\prime(x) \int_0^1 \! dx  \frac{1}{x^{1+\eta}} 
 \\&
 =\left[ \frac{\theta(x)}{x^{2+\eta}}\right]_{++} +  \frac{\delta^\prime(x)}{\eta}
\end{aligned}\end{equation}
to obtain the expansion
\begin {equation}\begin {aligned}
\frac{1}{x^{2+\eta}} = \frac{\delta^\prime(x)}{\eta} - \frac{\delta (x)}{1+\eta} +  \left[ \frac{\theta(x)}{x^2}\right]_{++} \!\!
- \ \eta  \left[ \frac{\log(x)\theta(x)}{x^2}\right]_{++}
\end {aligned}\end {equation}
as in Eq. (2.40) of \cite{Ebert:2018gsn}.

%
%
\subsection{Vector plus distributions}  \label{sec:vector_plus_distributions}
%
%

The same techniques may be applied to divergent vector-valued functions. Since our operators $T_{(i,j)}$ live in $d\neq4$ spacetime dimensions, we define the vector plus distribution (also known as the $\xi^2$-distribution) by the relations
\begin{equation} \begin{aligned}\label{eq:vplusdefn}
\int_{\qtsq < \xi^2} d^{d-2} \vecQ \, [ \theta(\qtsq)(f(\vecQ)]_+^{\xi^2} &= 0\,, \\
 [ \theta(\qtsq) f(\vecQ)]_+^{\xi^2} &=  f(\vecQ) \ , \quad \qtsq > 0 \,.
\end{aligned} \end{equation}
When $f(\vecQ)=f(\qtsq)$ is a rotationally symmetric function, we have
\begin{equation} \begin{aligned}
\int  d^{d-2} \vecQ f(q_T^2) =   \int \! J_{q_T}^\epsilon f(q_T^2) \, d\qtsq \,,
\end{aligned} \end{equation}
where 
\begin{equation} J_{q_T}^\epsilon\equiv \frac{S_{2-2\epsilon}}{2} q_T^{-2\epsilon}
\end{equation}
and $S_{d-2}=2\pi^{\frac{d-2}{2}}/\Gamma \left( \frac{d-2}{2} \right)$, e.g. $S_2=2\pi$. We also note that a $(d-2)$-dimensional delta function at the origin may be written as 
\begin{equation} \begin{aligned}
\delta(\vecQ) = \frac{\delta(\qtsq)}{J_{q_T}^\epsilon} \,.
\end{aligned} \end{equation}
Therefore if
\begin{equation}
    g(q_T^2)= \int_{\myvec{p}_T^2 < \qtsq} d^{d-2} \myvec{p}_T  \, f(\myvec{p}_T) 
\end{equation}
for some rotationally invariant function $f(\myvec{p}_T)$, then
\begin{equation}
    f(q_T^2)=\frac{1}{J_{q_T}^\epsilon}\frac{d}{dq_T^2}g(q_T^2)\, ,
\end{equation}
which is useful for converting between distributions and their cumulants.

The vector plus distribution may be written as the limit
\begin{equation} \begin{aligned}
[ f(\vecQ)]_+^{\xi^2} = \lim_{\beta\rightarrow 0}   A(\beta,\epsilon,\xi) \delta(\vecQ) + \theta( \qtsq - \beta^2)f(\vecQ) \ ,
\end{aligned} \end{equation}
where, from \eqref{eq:vplusdefn},
\begin{equation} \begin{aligned}
 A(\beta,\epsilon,\xi) = -  \int_{\beta^2 < \qtsq < \xi^2} d^{d-2} \vecQ \ f(\vecQ) \,.
\end{aligned} \end{equation}
For example, we have the explicit form of the $\xi^2$ distribution
\begin{equation} \begin{aligned}
\left[ \frac{1}{\vecQ^2} \right]_+^{\xi^2} =& \xi^{-2\epsilon} \left( 1-\left( \frac{\beta^2}{\xi^2} \right)^{-\epsilon} \right) \frac{S_{d-2}}{2\epsilon}\delta(\vecQ) \\ & +  \frac{\theta( \qtsq - \beta^2)}{\vecQ^2} \,,
\end{aligned} \end{equation}
(where the limit $\beta\to 0$ is implicit). We can also derive the analog of \eqref{eq:x_diff_relation} \cite{Ebert:2016gcn},
\begin{equation} \begin{aligned} \label{eq:qT_diff_relation}
\frac{1}{J_{q_T}^\epsilon}\frac{d}{d\qtsq} \theta(\qtsq) F(\qtsq) =& \left[ \theta(\qtsq) \frac{1}{J_{q_T}^\epsilon}\frac{dF(\qtsq)}{d\qtsq} \right]_+^{\xi^2} \\ & + \delta(\vecQ) F(\xi^2)\,.
\end{aligned} \end{equation}
As in Appendix \ref{sec:single_plus_distributions}, we may then derive the expansion
 \begin{equation} \begin{aligned} \label{eq:vecPlusIdentityAppendix}
\frac{(\nu^2)^{-\eta/2}}{(\vecQ^2)^{1-\eta/2} } =& J_\xi^\epsilon \left( \frac{\nu^2}{\xi^2} \right)^{-\eta/2} \frac{ \delta(\vecQ)}{\frac{\eta}{2}-\epsilon} + \left[ \frac{\theta(\qtsq)}{\qtsq} \right]_+^{\xi^2} \\ & + \frac{\eta}{2}\left[ \frac{\log\frac{\qtsq}{\nu^2}\theta(\qtsq)}{\qtsq} \right]_+^{\xi^2}  + \cdots \ .
\end{aligned} \end{equation}
The choice of $\xi$ in these identities is entirely arbitrary. However, since each diagram comes with an overall $\mu^{2\epsilon}$, and since these identities put all the $\epsilon$ dependence into the delta-function prefactor $J_\xi^\epsilon \propto \xi^{-2\epsilon}$, the canonical choice that avoids spurious logarithms is $\xi=\mu$.

It is convenient to rescale the vector plus distributions to have the same scaling dimensions and $\pi$ counting as $\delta(\vecQ)$. Borrowing from the generalized-log notation of \cite{Ebert:2016gcn}, we define
\begin{equation} \begin{aligned} \label{eq:genLogDefn}
\mathcal{L}_n(\vecQ,\mu) &= \frac{1}{J_\mu^\epsilon}
\left[ 
    \frac{
        \log^n \frac{\vecQ^2}{\mu^2} \theta(\qtsq)
        }
        {
        \vecQ^2
    } 
\right]_+^{\mu^2} \,.
\end{aligned} \end{equation}
With these definitions, and taking $\nu=\mu=\xi$, we have
 \begin{equation} \begin{aligned} \label{eq:vecPlusIdentityAppendixL1}
\frac{(\mu^2)^{-\eta/2}}{(\vecQ^2)^{1-\eta/2} }=&  J_\mu^\epsilon 
\left( 
     \frac{ \delta(\vecQ)}{\frac{\eta}{2}-\epsilon} 
     + \mathcal{L}_0^{(0)}(\vecQ,\mu)   \right.\\& \left.+ \frac{\eta}{2} \mathcal{L}_1^{(0)}(\vecQ,\mu)   + \cdots \right) \,.
\end{aligned} \end{equation}
Finally, we also need the identity
\begin{equation} \begin{aligned} \label{eq:vecPlusIdentityAppendixLog}
\frac{\log\frac{\vecQ^2}{\mu^2}}{\vecQ^2 }  &=  J_\mu^\epsilon \left( - \frac{ \delta(\vecQ)}{\epsilon^2} + \mathcal{L}_1^{(0)}(\vecQ,\mu)  \right) \ ,
\end{aligned} \end{equation}
which appears in the context of calculations without a regulator.

%
%
\section{Fixed-Order Comparison}\label{app:litcompare}
%
%

In this section we compare our results to that of \cite{Ebert:2018gsn}. In that reference, the QCD cross section for the process $N_1N_2 \rightarrow V+X$ up to NLP is decomposed into a sum of convolutions of coefficient functions multiplied by PDFs and their first derivatives, so that
\begin{equation} \begin{aligned} \label{eq:Ebert2018_eq_3.5}
\frac{1}{\sigma_0}&\frac{d\sigma}{dq^2 dy\, d^2\vecQ}  =  \int\frac{dz_a}{z_a}\frac{dz_b}{z_b} \\& \times 
\left[
    C_{f_q\!f_{\bar{q}}}^{(0)}(z_a,z_b,q^2,\qtsq) f\left( \frac{x_a}{z_a} \right)f\left( \frac{x_b}{z_b} \right)\right.
    \\  & \quad
    + \frac{1}{q^2} C_{f_q\!f_{\bar{q}}}^{\pcorrec}(z_a,z_b,q^2,\qtsq) f\!\left( \frac{x_a}{z_a} \right)f\left( \frac{x_b}{z_b} \right)\
    \\ & \quad
    + \frac{1}{q^2}C_{f_q^\prime\!f_{\bar{q}}}^{\pcorrec}(z_a,z_b,q^2,\qtsq) \frac{x_a}{z_a}f^\prime\left( \frac{x_a}{z_a} \right) f\!\left( \frac{x_b}{z_b} \right) 
    \\ & \quad
    + \frac{1}{q^2} C_{f_q\!f_{\bar{q}}^\prime}^{\pcorrec}(z_a,z_b,q^2,\qtsq) f\left( \frac{x_a}{z_a} \right) \frac{x_b}{z_b}f^\prime\left( \frac{x_b}{z_b} \right)
    \\ & \quad
    + \frac{1}{q^2} C_{f_q^\prime\!f_{\bar{q}}^\prime}^{\pcorrec}(z_a,z_b,q^2,\qtsq) \frac{x_a}{z_a} f^\prime\left( \frac{x_a}{z_a} \right) \frac{x_b}{z_b}f^\prime\left( \frac{x_b}{z_b} \right) \bigg]\,,
\end{aligned} \end{equation}
where at one-loop
\begin{equation} \begin{aligned} \label{eq:Ebert2018_Cff}
C_{f_q\!f_{\bar{q}}}^{(0)} =&  \alphabar \left\{\bigvert \delta(\zabar) \delta(\zbbar) \delta(\qtsq) 
\right.\\&\qqquad\times\left.
\left(  -  \log^2 \frac{q^2}{\mu^2} + 3 \log\frac {q^2}{\mu^2} -8 + 7\zeta_2  \right) \right. 
\\&+  
\left[ \frac{1}{\qtsq} \right]_+^{\mu^2}
\!\!\left( 
    \delta(\zabar) \left[ \frac{1+z_b^2}{\zbbar} \right]_+
    \!+ \delta(\zbbar) \left[ \frac{1+z_a^2}{\zabar} \right]_+
\right) 
\\&- \delta(\zabar) \delta(\zbbar) 
\left( 
    2 
    \left[ \frac{\log \qtsq/q^2}{\qtsq} 
    \right]_+^{\mu^2}
    \!\!+ 
    3\left[ \frac{1}{\qtsq} \right]_+^{\mu^2} 
\right)  \\
& \left. + \delta(\qtsq) 
\left( 
    \zabar \delta(\zbbar) + \zbbar \delta(\zabar)  - \zeta_2 \delta(\zabar) \delta(\zbbar)  
\right) \vphantom{\frac{q^2}{mu^2}}\right\}\,,
\end{aligned} \end{equation}
at LP, and 
\begin{equation} \begin{aligned} 
C_{f_q\!f_{\bar{q}}}^{\pcorrec} =& \alphabar \left[ 
-4\delta(\zabar)\delta(\zbbar) 
- \delta(\zabar) \frac{1+z_b^2-4z_b^3}{z_b}
\right.\\& \left. -\frac{1+z_a^2-4z_a^3}{z_a}\delta(\zbbar) 
\right] \,,
\end{aligned}\end{equation}
\begin{equation}\begin{aligned}
C_{f_q^\prime\! f_{\bar{q}}}^{\pcorrec} =& \alphabar 
\left[ 
    \left( -\log\frac{q^2}{\qtsq} -1 \right)\delta(\zabar)\delta(\zbbar) \right.
    \\ &
    + \delta(\zabar) 
    \left( 
        \frac{1+3z_b+2z_b^2}{2z_b} - \left[\frac{1}{\zbbar} \right]_+ 
    \right)\\ &\left.
    - \left( 
        \frac{1+z_a+2z_a^3}{2z_a} + \left[\frac{1}{\zabar} \right]_+
    \right)\delta(\zbbar)
\right]\,,
\end{aligned}\end{equation}
\begin{equation}\begin{aligned}
C_{f_q\!f_{\bar{q}}^\prime}^{\pcorrec} =& \alphabar
\left[ 
    \left( -\log\frac{q^2}{\qtsq} -1 \right)\delta(\zabar)\delta(\zbbar) \right.
    \\ &
    - \delta(\zabar) \left( 
        \frac{1+z_b+2z_b^3}{2z_b} + \left[\frac{1}{\zbbar} \right]_+
    \right)\\ &\left.
    +  
    \left( 
        \frac{1+3z_a+2z_a^2}{2z_a} - \left[\frac{1}{\zabar} \right]_+ 
    \right)\delta(\zbbar)
\right]\,,
\end{aligned}\end{equation}
\begin{equation}\begin{aligned}\label{eq:Ebert2018_Cff1}
C_{f_q^\prime\!f_{\bar{q}}^\prime}^{\pcorrec} =& \alphabar
\left[ 
    \left( 2\log\frac{q^2}{\qtsq} +4 \right)\delta(\zabar)\delta(\zbbar) \right.
    \\ &
    - \delta(\zabar) \left( 
        \frac{1-2z_b-z_b^2}{2z_b} + 2\left[\frac{1}{\zbbar} \right]_+
    \right)\\ & \left.
    +  
    \left( 
        \frac{1-2z_a-z_a^2}{2z_a} + 2\left[\frac{1}{\zabar} \right]_+ 
    \right)\delta(\zbbar)
\right]
\end{aligned} \end{equation}
at NLP.

Since the $x_{a,b}$ in \eqn{eq:Ebert2018_eq_3.5} differ from the $\xi_{1,2}$ used in \eqn{eq:factorization_x_sec} at $O(\qtsq/\QLsqr)$, and since our results are expressed entirely in terms of PDFs instead of PDFs and their first derivatives, the results in \eqn{eq:Ebert2018_Cff} and \eqn{eq:Ebert2018_Cff1} are related to \eqn{eq:LP_unsummed_eta} and \eqn{eq:NLP_unsummed_eta} by a change of variables, integration by parts and a few distributional identities.
Working in the hadronic center-of-mass frame for simplicity, where $P_1^- = \sqrt{s} = P_2^+$, the variables $x_{a,b}$ may be written in terms of $\xi_{1,2}$ as
\begin{equation} \begin{aligned} \label{eq:xab_to_xi12}
x_a &= 
\xi_1\left( 1 - \frac{1}{2}\frac{\qtsq}{q_L^2} + \cdots \right)\,,
\\ 
x_b &= 
\xi_2 \left( 1 - \frac{1}{2}\frac{\qtsq}{q_L^2} + \cdots\right) \,.
\end{aligned} \end{equation}
Expanding \eqref{eq:Ebert2018_eq_3.5}--\eqref{eq:Ebert2018_Cff1} up to $O(\qtsq/\QLsqr)$ gives
\begin{equation} \begin{aligned} \label{eq:transformed_3.5}
\frac{1}{\sigma_0}&\frac{d\sigma}{dq^2 dy\, d^2\vecQ}  = \commonPrefac \int\frac{dz_1}{z_1}\frac{dz_2}{z_2}\\ & \times 
\bigg[
    C_{f_q\!f_{\bar{q}}}^{(0)}(z_1,z_2,q_L^2,\qtsq) f \left( \frac{\xi_1}{z_1} \right)f\!\left( \frac{\xi_2}{z_2} \right)
    \\ & \quad  
    + \frac{1}{q_L^2}\left( C_{f_q\!f_{\bar{q}}}^{\pcorrec}(z_1,z_2,q_L^2,\qtsq) \right. \\ &\left.
        \qquad + \delta C_{f_q\! f_{\bar{q}}}^{(0)}(z_1,z_2,q_L^2,\qtsq) \right) 
        f\left( \frac{\xi_1}{z_1} \right)f\left( \frac{\xi_2}{z_2}\vphantom{\frac{qsq}{\mu^2}}\right)
    \\ & \quad
    + \frac{1}{q_L^2}\left( C_{f_q^\prime\!f_{\bar{q}}}^{\pcorrec}(z_1,z_2,q_L^2,\qtsq) 
        - \frac{1}{2} \frac{\qtsq}{q_L^2}C_{f\!f}^{(0)}(z_1,z_2)  \right)\\ &\qquad \times
    \frac{\xi_1}{z_1}f^\prime\left( \frac{\xi_1}{z_1} \right)
    f \left( \frac{\xi_2}{z_2} \right) 
    \\ &\quad 
    + \frac{1}{q_L^2}\left( C_{f_q f_{\bar{q}}^\prime}^{\pcorrec}(z_1,z_2,q_L^2,\qtsq) 
        - \frac{1}{2} \frac{\qtsq}{q_L^2}C_{f\!f}^{(0)}(z_1,z_2)\right) \\ &\qquad \times
    f \left( \frac{\xi_1}{z_1} \right) 
    \frac{\xi_2}{z_2}f^\prime\left( \frac{\xi_2}{z_2} \right)
    \\ &\quad
    +\frac{1}{q_L^2} C_{f_q^\prime\!f_{\bar{q}}^\prime}^{\pcorrec}(z_1,z_2,q_L^2,\qtsq) \frac{\xi_1}{z_1} f^\prime \left( \frac{\xi_1}{z_1} \right) \frac{\xi_2}{z_2}f^\prime\!\left( \frac{\xi_2}{z_2} \right) \bigg]\,,
\end{aligned} \end{equation}
where $C_{f_q\!f_{\bar{q}}}^{(0)}(z_1,z_2,q_L^2,\qtsq)=\Cff^{(0)}(z_1,z_2,q_L^2;\qtsq)$ in \eqn{eq:LP_unsummed_eta}, and
\begin{equation} \begin{aligned} \label{eq:FF_after_transform}
C_{f_q\!f_{\bar{q}}}^{\pcorrec}& + \delta C_{f_q\!f_{\bar{q}}}^{(0)}   = \alphabar \left[ 
-6\delta_1\delta_2 
- \delta_1 \frac{1+z_2^2-4z_2^3}{z_2}\right.\\& \qquad\left.
- \frac{1+z_1^2-4z_1^3}{z_1}\delta_2 
\right]  \,,
\end{aligned}\end{equation}
\begin{equation}\begin{aligned}
C_{f_q^\prime\!f_{\bar{q}}}^{\pcorrec} 
        &- \frac{1}{2} \frac{\qtsq}{q_L^2}C_{f_q\!f_{\bar{q}}}^{(0)}  = \alphabar
\left[ 
    \left( -2\log\frac{\QLsqr}{\qtsq} -1 \right)\delta_1\delta_2 
     \right.\\ &\qquad+
    \delta_1\left( 
        \frac{1+4z_2+3z_2^2}{2z_2} - 2\left[\frac{1}{\zzbar} \right]_+ 
    \right)
    \\ & \qquad
    - \left.\left( 
        \frac{1-z_1^2 + 2z_1^3}{2z_1} + 2\left[\frac{1}{\zbar} \right]_+
    \right)\delta_2
\right]\,,
\end{aligned}\end{equation}
\begin{equation}\begin{aligned}
\\ 
C_{f_q\!f_{\bar{q}}^\prime}^{\pcorrec} &
        - \frac{1}{2} \frac{\qtsq}{q_L^2}C_{f_q\!f_{\bar{q}}}^{(0)} = \alphabar 
\left[ 
    \left( -2\log\frac{\QLsqr}{\qtsq} -1 \right)\delta_1\delta_2 \right.\\ &\qquad
    - \delta_1 \left( 
        \frac{1-z_2^2+2z_2^3}{2z_2} + 2\left[\frac{1}{\zzbar} \right]_+
    \right)
    \\& \qquad
    + \left.\left( 
        \frac{1+4z_1+3z_1^2}{2z_1} - 2\left[\frac{1}{\zbar} \right]_+ 
    \right)\delta_2
\right]\,,
\end{aligned}\end{equation}
\newpage
\begin{equation}\begin{aligned}
\\ 
C_{f_q^\prime\!f_{\bar{q}}^\prime}^{\pcorrec}& = \alphabar 
\left[ 
    \left( 2\log\frac{\QLsqr}{\qtsq} +4 \right)\delta_1\delta_2\right. \\ &\qquad
    - \delta_1 \left( 
        \frac{1-2z_2-z_2^2}{2z_2} + 2\left[\frac{1}{\zzbar} \right]_+
    \right)
    \\& \qquad
    +  \left.\left( 
        \frac{1-2z_1-z_1^2}{2z_1} + 2\left[\frac{1}{\zbar} \right]_+ 
    \right)\delta_2
\right]\,.
\end{aligned} \end{equation}
Finally, the comparison is completed by applying the following integration by parts identities, valid when $f(x/z)=0$ for $x\ge z$,
\begin{equation} \begin{aligned} \label{eq:Mikes_IBP}
\int \frac{dz}{z}  \delta(\bar{z})  \frac{x}{z}f^\prime \!\left( \frac{\xi}{z} \right) =& \int \frac{dz}{z} \left[ -z \delta^\prime (\bar{z}) \right] f \left( \frac{\xi}{z} \right)  
\\ 
\int \frac{dz}{z} z^n  \frac{x}{z}f^\prime \!\left( \frac{\xi}{z} \right) =& \int \frac{dz}{z} \left[ nz^n - z\delta(\bar{z}) \right] f\!\left( \frac{\xi}{z} \right) 
\\ 
\int \frac{dz}{z}  \left[ \frac{1}{\bar{z}} \right]_+  \frac{x}{z}f^\prime \left( \frac{\xi}{z} \right) =& \int \frac{dz}{z} \left( z \left[ \frac{1}{\bar{z}^2} \right]_{++} + z\delta^\prime(\bar{z})\right.\\ & \left. - \bigvert z\delta(\bar{z}) \right) f\left( \frac{\xi}{z} \right) \,.
\end{aligned} \end{equation}
These identities transform the coefficient functions in \eqn{eq:FF_after_transform} from acting on derivatives of PDFs to the equivalent form of coefficient functions acting only on PDFs, and 
in doing so reproduces the NLP coefficient function in \eqn{eq:NLP_unsummed_eta}.

\bibliographystyle{JHEP}
\bibliography{dy_nlp_prd_rev}

\end{document}